\documentclass[12pt]{article}
\title{
{\bf String Nature of Confinement in}\\
{\bf (Non-)Abelian Gauge Theories}
\thanks{Based on the Ph.D. thesises at the Humboldt University of 
Berlin (1999) (available under {\tt http://dochost.rz.hu-berlin.de}) 
and the Institute of Theoretical and Experimental Physics, Moscow (2000).}}  
\author{Dmitri Antonov 
\thanks{E-mail addresses: {\tt antonov@difi.unipi.it},  
{\tt antonov@heron.itep.ru}.}{\,} 
\thanks {Permanent address: 
Institute of Theoretical and Experimental Physics, 
B. Cheremushkinskaya 25, RU-117 218 Moscow, Russia.}
\\
{\it INFN-Sezione di Pisa, Universit\'a degli studi di Pisa,}\\ 
{\it Dipartimento di Fisica, Via Buonarroti, 
2 - Ed. B - 56127 Pisa, Italy}} 
\date{}
\begin{document}
\maketitle
\vspace{1mm}
\centerline{\bf Abstract}
\vspace{3mm}
\noindent
Recent progress achieved 
in the solution of the problem of confinement in various
(non-)Abelian gauge theories by virtue of a derivation of their 
string representation is reviewed. The theories under study include 
QCD within the so-called Method of Field Correlators, QCD-inspired   
Abelian-projected theories, and compact QED in three and four 
space-time dimensions. Various nonperturbative 
properties of the vacua of the 
above mentioned theories are discussed. The relevance of the 
Method of Field Correlators to the study of 
confinement in Abelian models, allowing for an analytical 
description of this 
phenomenon, is illustrated by an evaluation 
of field correlators in these models.

\newpage

\tableofcontents

\newpage

\section{Introduction}

\subsection{General Ideas and Motivations}

Nowadays, there is no doubt that strong interactions of 
elementary particles are adequately 
described by Quantum Chromodynamics (QCD)~\cite{qcd} (see 
Ref.~\cite{qcdrev} for recent monographs). 
Unfortunately, usual field-theoretical methods are not adequate 
to this theory itself. That is because in the infrared (IR) region, 
the QCD coupling constant becomes large, which makes the standard 
Feynman diagrammatic technique in this region unapplicable. However, it 
is the region of the strong coupling, which deals with the physically 
observable  
colourless objects (hadrons), whereas the standard perturbation theory 
is formulated in terms of coloured (unphysical) 
objects: quarks, gluons, and 
ghosts. This makes it necessary to develop special techniques, applicable 
for the evaluation of effects beyond the scope of 
perturbation theory. The latter are usually referred to as 
{\it nonperturbative phenomena}. Up to now, those are best of all 
studied in the framework of the approach based on lattice gauge 
theory~\cite{montvay}, 
which provides us with a natural 
nonperturbative regularization scheme. Various ideas and methods 
elaborated on in the lattice field theories during the two last decades,  
together with the development of 
algorithms for numerical calculations and progress 
in the computer technology, have made these theories one of the most 
powerful tools for evaluation of nonperturbative characteristica of 
QCD (see Ref.~\cite{lat} for a recent review). However, despite 
obvious progress of this approach, there still remain several problems.  
Those include {\it e.g.} 
the problem of simultaneous reaching the continuum and 
thermodynamic limits. Indeed, physically relevant length scales lie 
deeply inside the region between the lattice spacing and the size of the 
lattice. However, due to the asymptotic freedom of QCD, 
in the weak coupling limit, not only the lattice spacing, but 
also the size of the lattice (for a fixed number of sites) becomes 
small, as well 
as the region between them. However, in order to achieve 
the thermodynamic limit, the size of the lattice 
should increase. 
This makes it 
necessary to construct large lattices, which in particular leads to 
the technical problem of slowdown of simulations on them. 
As far as the problem of reaching the continuum limit alone is concerned, 
recently some progress in the solution of this problem has been achieved 
by making use of the conception of improved lattice actions~\cite{sym} 
in Ref.~\cite{improve}. Another problem of the lattice formulation of 
QCD is the  
appearance of so-called fermion 
doublers ({\it i.e.}, additional modes appearing as relevant dynamical 
degrees of freedom)  
in the definition of the fermionic action on the lattice due 
to the Nielsen-Ninomiya theorem~\cite{nielsen}.  
According to this 
No-Go theorem, would we demand simultaneously 
hermiticity, locality, and 
chiral symmetry (which will be discussed later on) 
of the lattice fermionic action, the doublers 
unavoidably appear, which means that all these three physical requirements 
cannot be achieved together. This makes it necessary to introduce 
fermionic 
actions which violate one of these properties ({\it e.g.} Wilson 
fermions~\cite{wilferm}, violating chiral symmetry for a finite 
lattice), checking afterwards 
lattice artefacts associated with 
a particular choice of the action. Notice however, that recently 
a significant progress 
in the solution of this problem has been achieved (for a review 
see~\cite{nieder} and Refs. therein).
Finally, there remains the important problem of 
reaching the chiral limit, which becomes especially hard 
if one accounts for dynamical fermions.   
That was just one of the reasons 
why the main QCD calculations on large lattices 
have been performed in the quenched approximation, {\it i.e.}, when the 
creation/annihilation of dynamical quark pairs is neglected.

All these problems together with the necessity of getting deeper 
theoretical insights into nonperturbative phenomena require 
to develop 
{\it analytical} nonperturbative 
techniques in QCD and other theories displaying such  
phenomena. This is the main motivation for the present work. 

It is worth realizing that unfortunately up to now a systematic  
way of analytical investigation of nonperturbative phenomena in QCD is 
still lacking. Instead of that, there exist various 
approaches, enabling one 
to take them into account 
phenomenologically for describing hadron interactions. 
Those approaches 
include {\it e.g.} the potential and bag models~\cite{cea1, bag}, the 
large-$N$ expansion methods~\cite{largen}, the effective Lagrangian 
approach~\cite{efflagr}, the 
QCD sum rule approach~\cite{vain}, and later on  
its generalization, the so-called 
Method of Field Correlators~\cite{strten, svm, 
doschov, ufn}. 

The most fundamental problem associated with the 
IR dynamics of QCD, which is 
known to be one of the most important problems of modern Quantum 
Field Theory,  
is the problem of explanation and description of 
{\it confinement} (for a review see {\it e.g.}~\cite{polbook, ufn}).
In general, by confinement one  
implies the phenomenon of absence in the spectrum   
of a certain field theory of the physical $\left|{\rm in}\right>$ and  
$\left|{\rm out}\right>$ states of  
some particles, whose fields are 
however present in the fundamental Lagrangian. 
It is this phenomenon in QCD, which forces quarks and antiquarks 
to combine into colourless hadronic bound states. 
The important characteristic of confinement 
is the existence of string-like field configurations leading to a 
linearly rising quark-antiquark potential as expressed by the 
Wilson's area law (see the next Subsection). 
Such a field configuration 
emerging between external quarks is usually referred to as QCD string. 
In the present review, we shall 
demonstrate that the properties of this string can be naturally studied 
in the framework of the Method of Field Correlators 
and derive by virtue of 
this method the corresponding string Lagrangian. Notice, that 
the advantage of the Method of Field Correlators 
is that it deals directly with QCD, and therefore enables us to 
express the coupling constants of this Lagrangian 
in terms of the  
fundamental QCD quantities, which are the gluonic condensate and the 
so-called correlation length of the vacuum. 
This approach will also allow us 
to incorporate quarks and derive a Hamiltonian of the QCD string with 
quarks in the confining QCD vacuum.

Another fundamental 
phenomenon of nonperturbative QCD is the {\it spontaneous 
breaking of chiral symmetry}, {\it i.e.}, the $U\left(N_f\right)\times
U\left(N_f\right)$-symmetry of the massless QCD action. 
Indeed, though one could 
expect this symmetry to be observed on the level of a 
few MeV, it does not exhibit itself in the hadronic spectra. Were this 
symmetry exact, one would expect parity degeneracy of all hadrons, 
whereas in reality parity partners are generally split by a few 
hundred MeV. Such a spontaneous symmetry breaking has far-reaching 
consequences. In particular, it implies that 
there exist massless Goldstone bosons, 
which are identified with pions. The signal for chiral symmetry 
breaking is the appearance of a nonvanishing quark condensate, which plays 
an important r\^ole in many nonperturbative approaches~\cite{gell, leu82}.

The non-Abelian character 
of the gauge group $SU(3)$ makes it especially difficult to study 
the problems of confinement and 
chiral symmetry breaking  
in the QCD case. To explain the mechanisms of these phenomena  
microscopically, a 
vast amount of models of the QCD vacuum has been proposed 
(see~\cite{ufn} for a recent review). Those are based either on  
an ensemble of classical field 
configurations ({\it e.g.} instantons~\cite{mitya0, 
mitya}, see~\cite{diakrev} for recent reviews) 
or on quantum background fields~\cite{temp, spring}.
The most general demand made on 
all of them was to reproduce two characteristic quantities of the 
QCD vacuum, which are nonzero quark~\cite{gell} and gluon~\cite{vain} 
condensates, related to the chiral symmetry breaking and confinement, 
respectively. However, 
at least the semiclassical scenario 
possesses several weak points. First of all, since the topological charge 
of the QCD vacuum as a whole 
is known to vanish, this vacuum cannot be described 
by a certain unique classical configuration, but should be rather 
built out of a 
superposition of various configurations, {\it e.g.} instantons and 
antiinstantons. However, such a superposition already does not 
satisfy classical equations of motion and is, moreover, unstable 
{\it w.r.t.} 
annihilation of the objects with the opposite topological charge. 
Secondly, in order to reproduce the phenomenological gluon 
condensate~\cite{vain}, classical configurations should be dense packed 
(about one configuration per $({\rm fm})^4$), which leads to a 
significant distorsion 
of the solutions corresponding to these configurations within the 
original superposition ansatz~\cite{shur1}~\footnote{Recently, some 
progress in the solution of this problem has been achieved in 
Ref.~\cite{thur}.}. And last but not least, a further 
counterargument 
against semiclassical models of the QCD vacuum is that not
all of them, once being simulated in the lattice experiments, 
yield the property of confinement 
(see discussion in Ref.~\cite{ufn}).

Another natural way of investigation of the nonperturbative 
phenomena in QCD might lie in the  
simplification of the problem 
under study by considering some solvable theories displaying 
the same type of phenomena. 
In this way, the problem of chiral symmetry breaking 
is best of all analytically studied in the so-called 
Nambu-Jona-Lasinio (NJL) type models, which are models containing 
local four-quark interactions~\cite{vlnjl, njl2, njl3, etal} 
(see Ref.~\cite{dietmar} for a recent review). These models 
lead to a gap equation for the dynamical quark mass, signalling 
spontaneous breaking of chiral symmetry. After applying the so-called 
bosonization procedure (which can be performed either by making use of the 
standard Hubbard-Stratonovich transformation or within the field 
strength approach~\cite{fieldstr}) as well as a derivative expansion of 
the resulting quark determinant 
at low energies, this leads 
to the construction of nonlinear chiral meson 
Lagrangians~\cite{efflagr}. The advantage of the latter ones is that 
they summarize QCD low-energy theorems, which is the reason why 
these Lagrangians 
are intensively used in the modern hadronic physics~\cite{ga85, dietmar}. 
The techniques developed for NJL type models have been in particular 
applied to the 
evaluation of higher-order derivative terms in meson 
fields~\cite{njl3, etal}, 
which enabled one to 
estimate the structure constants of the effective chiral Lagrangians 
introduced in Ref.~\cite{ga85}. Furthermore, in this way in 
Refs.~\cite{njl2, njl3, etal, dietmar} it has been demonstrated that the 
low-energy properties of light pseudoscalar, vector, and axial-vector 
mesons are well described by effective chiral Lagrangians following 
from the QCD-motivated NJL models. In addition, the path-integral 
bosonization of an extended NJL model including chiral symmetry 
breaking of light quarks and heavy quark symmetries of heavy quarks 
has been performed~\cite{dietmarnjl} (see the second paper 
of Ref.~\cite{jurke} for a review), which yielded the effective 
Lagrangians of pseudoscalar, vector, and axial-vector $D$ or $B$ mesons, 
interacting with light $\pi$, $\rho$, and $a_1$ mesons. 
 
As far as 
the theories possessing 
the property of confinement 
are concerned, 
those firstly include compact QED and the 
3D Georgi-Glashow model~\cite{polbook} and, secondly, the so-called 
Abelian-projected theories~\cite{th}~\footnote{In what follows, 
we shall not consider recently 
discovered supersymmetric theories, also possessing the property of 
confinement~\cite{seib}.}. In the present review, 
we shall concentrate ourselves 
on the confining properties of the above mentioned
non-supersymmetric theories. In all of them, confinement occurs due to 
the expected 
condensation of Abelian magnetic monopoles, after which the vacuum 
structure of these theories becomes similar to that of the dual 
superconductor (the so-called dual Meissner scenario of 
confinement)~\cite{mand}. Such a vacuum then leads to the 
formation of 
strings (flux tubes) connecting external electric charges, 
immersed into it.
These strings are dual 
to the 
(magnetic) Abrikosov-Nielsen-Olesen strings~\cite{ano}. The latter ones  
emerge as 
classical field configurations in  
the Abelian Higgs Model, 
which is the standard relativistic version of the 
Ginzburg-Landau theory of 
superconductivity. It turns out that the properties of electric strings 
in the dual superconductor are 
similar 
to the ones of the realistic strings in QCD, which connect quarks with 
antiquarks and ensure confinement.
It is worth noting that this analogy 
based on the 't Hooft-Mandelstam scenario led to several 
phenomenological dual models of QCD (see Ref.~\cite{ball} for a review).
Thus, the properties of the QCD string can be naturally studied in the 
framework of the Abelian projection method. Moreover, this approach turns 
out to provide us 
with the representations for the partition functions of effective    
Abelian-projected theories of the 
$SU(2)$- and $SU(3)$-gluodynamics 
in terms of 
the integrals over string world-sheets. 
Such an integration, which is absent 
in the approach to the QCD string based on the 
Method of Field Correlators, 
appears now  
from the integration over the singular 
part of the phase of the magnetic Higgs field. The reformulation of 
the integral over the singularities of this field into the integral 
over string world-sheets is possible due to the fact 
that such singularities  
just take place at the world-sheets. 
In particular, an interesting string picture emerges in the 
$SU(3)$-gluodynamics, where after the Abelian projection there arise 
three types of magnetic Higgs fields, leading to three types of strings, 
which (self)interact via the exchanges of two massive dual gauge bosons.   
An exact procedure of the derivation of the string representations for the 
partition functions of Abelian-projected theories in the 
language of the path-integral, the so-called path-integral duality 
transformation, will be described in details below. 
In the framework of 
this approach, we shall also investigate field correlators in the 
Abelian-projected theories and find them to parallel those of 
QCD, predicted by the Method of Field Correlators and measured  
in the lattice simulations. Topological properties of Abelian-projected 
theories will be also discussed. Furthermore, the effects brought 
about by the summation over the grand canonical ensembles of 
small vortex loops, built out of the paired electric 
Abrikosov-Nielsen-Olesen  
strings, in these theories will be studied. In particular in the 
dilute gas approximation, the 
effective potential of such vortex loops will be derived and employed 
for the evaluation of their correlators.
Besides that, we shall study the string 
representation and field correlators of compact QED in 3D and 4D.
Notice that due to the absence of the Higgs field in this theory,  
the integration over the string world-sheets 
is realized there 
in another way than in Abelian-projected theories. Namely, it results from 
the summation over the branches 
of the multivalued effective monopole potential, which turns out to 
have the same form as the 3D version of the above discussed potential 
of vortex loops. Finally, similar 
forms of the string effective actions in QCD, Abelian-projected theories, 
and compact QED will enable us to elaborate for all these theories a 
unified method of description of the string world-sheet 
excitations, based on 
the methods of nonlinear sigma models, known from the standard string 
theory.

It is worth realizing, that the string 
theories, to be derived below, should be treated as effective, 
rather than fundamental ones. The actions of all of them turn out to 
have the form of an interaction between the elements of the string  
world-sheet, mediated by certain (nonperturbative) gauge field 
propagators. Being 
expanded in powers of the derivatives {\it w.r.t.} world-sheet 
coordinates, these actions yield as a first term of such an expansion 
the usual Nambu-Goto action. 
The latter one is known to suffer from the problem of conformal anomaly 
in $D\ne 26$ appearing during its quantization, which will not 
be discussed below. It this sense, 
throughout the present review, 
we shall treat the obtained string theories as effective 4D 
ones. It is also worth noting  
that within our approach only pure bosonic strings without 
supersymmetric extensions appear. As far as superstrings are concerned, 
during the last fifteen years, a great progress has been achieved in 
their development (see {\it e.g.}~\cite{gsw} 
for comprehensive monographs). 
Among the achievements of the superstring theory it is worth mentioning 
such ones as the calculation of the critical dimension 
of the space-time, inclusion of gravity in a common scheme, and, 
presumably, the absence of divergencies for some of these theories.  
The aim of all the superstring theories  
is the unification 
of all the four fundamental interactions. In another language, one should 
eventually be able to derive from them both the 
Standard Model and gravity,  
whereas all the auxiliary heavy modes should become irrelevant.  
Therefore, the final strategy 
of superstring theories is a derivation of the known field 
theories out of them. Contrary to this ideology, the aim of the 
present review is the derivation of effective string theories 
from gauge field theories possessing string-like excitations. 
As it has been discussed above, such string-like field configurations 
naturally 
appear in the confining phases of gauge theories.

Another possible direction of investigation of confinement and 
chiral symmetry breaking in QCD 
is based on a derivation of self-coupled equations 
for gauge-invariant vacuum amplitudes starting directly 
from the QCD Lagrangian 
and seeking for solutions allowing for these 
properties~\cite{looprev, 
stoch1, gluoeq, chiral}. Recently, this approach turned our to be quite 
useful for the investigation of the problem of interrelation between 
these two phenomena~\cite{chiral}. Once such an interrelation 
takes place, 
there should exist a relation between quark and gluon condensates 
as well, 
which 
has just been established in Ref.~\cite{chiral}. We shall briefly 
demonstrate the method of derivation of such a relation later on.

The organization of the review is as follows. In the next 
Subsection of the Introduction, 
we shall introduce the main quantitative parameters 
for the description of confinement in gauge theories and quote the 
criterion of this phenomenon in the sense of Wilson's area law. 
This criterion will then 
serve as our starting point in a derivation of certain string effective 
actions in various gauge 
theories. In the last Subsection, 
we shall consider theoretical foundations of the 
Method of Field Correlators.
Section 2 is devoted to a derivation and investigation of the QCD string 
effective action within this method. In Section 3, we investigate the 
problem of string representation of QCD from the point of view of 
Abelian-projected theories and demonstrate a correspondence between 
the Abelian projection method and the Method of Field Correlators. 
In Section 4, 
we study the string representation and vacuum correlators 
of compact QED in 3D and 4D. 
All these investigations eventually bring us to the conclusion 
that both 
QCD within the Method of Field Correlators, 
Abelian-projected theories, and 
compact QED have 
similar string representations. Such an observation then  
enables us to consider 
strings in these theories from the same point of view and 
elaborate for them 
a unified 
mechanism of description of string excitations. Besides that 
in Section 4, by virtue of 
the techniques developed for the investigation of the grand 
canonical ensemble of monopoles in compact QED, collective effects 
in similar ensembles of topological defects emerging in the 
Abelian-projected theories will be studied. This will enable us 
to improve on the calculations of the field strength correlators,
performed in Section 3.
Finally, we discuss the main results summarized in the review 
as well as 
possible future developments in the 
Conclusion and Outlook. In five Appendices, 
some technical details of transformations performed in the main text 
are outlined.

\subsection{Wilson's Criterion of Confinement and the Problem of 
String Representation of Gauge Theories}

As a most natural characteristic quantity for the description 
of confinement one usually considers the so-called Wilson loop. 
For example in the case of QCD, 
this object has the following 
form  

\begin{equation}
\label{wils}
\left<W(C)\right>=\frac{1}{N_c}\left<{\rm tr}{\,} P{\,} \exp\left(ig\oint 
\limits_C^{}A_\mu dx_\mu\right)\right>, 
\end{equation}
which is nothing else, 
but an averaged amplitude of the process of creation, propagation, 
and annihilation of a quark-antiquark pair. In Eq.~(\ref{wils}), 
$A_\mu$ stands for the vector-potential of the gluonic 
field~\footnote{From now on in the non-Abelian case, $A_\mu\equiv A_\mu^a 
t^a, a=1,\ldots, N_c^2-1$, where $t^a=\left(t^a\right)^{ij}$ is the 
Hermitean generator 
of the colour group in the fundamental 
representation, whereas in the Abelian case $A_\mu$ is simply a vector 
potential.}, $g$ is the QCD coupling 
constant, $C$ is a closed 
contour, along which the quark-antiquark pair propagates, $P$ stands 
for the path-ordering prescription, which is present only in the 
non-Abelian case, 
and the 
average on both sides is performed with the QCD 
action~\footnote{In what follows, we call the object defined by 
Eq.~(\ref{wils}) for brevity a ``Wilson loop'', 
whereas in the literature 
it is sometimes referred to as a ``Wilson loop average''.}. 

In order to 
understand why this object really serves as a characteristic 
quantity in QCD, 
let us consider the case when 
the contour $C$ is a rectangular one and lies for concreteness in the 
$(x_1,t)$-plane. Let us also denote the size of $C$ 
along the $t$-axis as $T$, 
and its size along the $x_1$-axis as $R$. Then such a Wilson loop 
in the case $T\gg R$ is related to the energy of the static 
({\it i.e.}, infinitely heavy) quark and antiquark, which are separated 
from each other by the distance $R$, by the formula

\begin{equation}
\label{wils1}
\left<W_{R\times T}\right>\sim {\rm e}^{-E_0(R)\cdot T},~ T\gg R.
\end{equation}
In order to get Eq.~(\ref{wils1}), let us fix the axial gauge $A_4=0$ 
\footnote{Throughout the present review, all the investigations 
will be performed in the Euclidean space-time.}, so that only the segments 
of the rectangular contour $C$ parallel to the $x_1$-axis, 
contribute to $\left<W_{R\times T}\right>$. 
Denoting 

\begin{equation}
\Psi_{ij}(t)\equiv\left[ P{\,} \exp\left(ig 
\int\limits_0^R dx_1A_1\left({\bf x}, 
t\right)\right)\right]_{ij},
\end{equation}
where we have omitted for shortness the dependence of $\Psi_{ij}$ on 
$x_2$ and $x_3$, we are not interested in, 
we get

\begin{equation}
\label{wils2}
\left<W_{R\times T}\right>=\frac{1}{N_c}\left<\Psi_{ij}(0)
\Psi_{ji}^\dagger(T)\right>.
\end{equation}
Inserting into Eq.~(\ref{wils2}) a sum over a complete set of 
intermediate states $\sum\limits_n^{}\left|n\right>\left<n\right|=1$, 
we get 

\begin{equation}
\label{wils3}
\left<W_{R\times T}\right>=\frac{1}{N_c}\sum\limits_n^{} 
\left<\Psi_{ij}(0)\left|n\left>\right<n\right|\Psi_{ji}^\dagger(T)
\right>=\frac{1}{N_c}\sum\limits_n^{}\left|\left<\Psi_{ij}(0)|n
\right>\right|^2{\rm e}^{-E_nT}, 
\end{equation}
where $E_n$ is the energy of the state $\left|n\right>$. At $T\to
\infty$, only the ground state with the lowest energy survives 
in the sum over states standing in Eq.~(\ref{wils3}), and we finally 
arrive at Eq.~(\ref{wils1}).    

The energy $E_0(R)$ in Eq.~(\ref{wils1}) includes a $R$-independent 
renormalization 
of the mass of a heavy (anti)quark due to its 
interaction with the gauge field. To the first order in $g^2$, 
up to a colour factor, it 
is the same as in QED~\cite{zinn} and reads 

\begin{equation}
\label{renorm}
\Delta E_{\rm mass}=C_2\frac{g^2}{4\pi a},
\end{equation}
where $C_2=\frac{N_c^2-1}{2N_c}$ stands for the Casimir operator of the 
fundamental representation, and 
$a\to 0$ is a cutoff parameter ({\it e.g.} lattice spacing). 
The difference $E(R)=E_0(R)-\Delta E_{\rm mass}$ therefore defines 
the potential energy of the interaction between a static quark and 
antiquark. In particular, the exponential dependence of the Wilson 
loop on the area 
of the minimal surface $\Sigma_{\rm min.}[C]$
encircled by the contour $C$, which we shall denote by 
$\left|\Sigma_{\rm min.}[C]\right|$, 

\begin{equation}
\label{area}
\left<W(C)\right>\to {\rm e}^{-\sigma\left|\Sigma_{\rm min.}[C]\right|}
\end{equation}
(the so-called area law behaviour of the Wilson loop) corresponds 
to the linearly rising potential between a quark and an antiquark, 

\begin{equation}
\label{energy}
E(R)=\sigma R.
\end{equation} 
This is the essence of the Wilson's criterion of 
confinement~\cite{wilson}.  

The coefficient $\sigma$ 
entering Eqs.~(\ref{area}) and (\ref{energy})
is called string tension. This is because 
the gluonic field between 
a quark and an antiquark is contracted to a tube or a string 
(the so-called QCD string), whose 
energy is proportional to its length, and $\sigma$ is the energy of 
such a string per unit length. This string plays the central 
r\^ole in the Wilson's picture of confinement, since with the increase of 
the distance $R$ between a quark and an antiquark it stretches and 
prevents them from moving apart to macroscopic distances. 

In order to get an idea of numbers, notice that according 
to the lattice data~\cite{ford} 
the distance $R$, 
at which Wilson's criterion of confinement becomes valid, is 
of the order of $1.0{\,} {\rm fm}$, and the 
string tension is of the order of $0.2{\,} {\rm GeV}^2$ (see 
{\it e.g.}~\cite{ufn}). 
It is worth realizing that the classical QCD 
Lagrangian does not contain 
a dimensionful parameter of such an order ({\it i.e.}, of hundreds MeV)
\footnote{{\it E.g.} the masses of the light quarks are of the order of 
a few MeV.}. However, in quantum theory, there always exists a 
dimensionful cutoff (like the lattice spacing $a$ in 
Eq.~(\ref{renorm})), which is related to the QCD coupling constant 
$g$ through the Gell-Mann--Low equation

\begin{equation}
\label{low}
-a^2\frac{dg^2\left(\frac{1}{a^2}\right)}{da^2}=
g^2\beta_{\rm QCD}\left(g^2\right).
\end{equation}
Here $\beta_{\rm QCD}\left(g^2\right)$ stands for the QCD Gell-Mann--Low 
function, which at the one-loop level reads 
 
\begin{equation}
\label{function}
\beta_{\rm QCD}\left(g^2\right)=-\left(\frac{11}{3}N_c-\frac23N_f\right)
\frac{g^2}{16\pi^2},
\end{equation}
where $N_f$ is the number of light quarks flavours, whose masses are 
smaller than $1/a$. It is Eq.~(\ref{function}), which tells 
us that QCD is an asymptotically free theory, provided that for 
$N_c=3$,  
$N_f\le 16$, which indeed holds in the real world. Consequently, the 
high-energy limit of QCD (the so-called perturbative QCD) is 
similar to the low-energy limit of QED, and the scale parameter 
following from the integration of Eq.~(\ref{low}), 

\begin{equation}
\label{lambda}
\Lambda_{\rm QCD}^2=\frac{1}{a^2}
\exp\left[-\int\limits_{}^{g^2\left(\frac{1}{a^2}
\right)} \frac{dg'^2}{g'^2\beta\left(g'^2
\right)}\right],
\end{equation}
is measurable in QCD as well as the QED fine-structure constant 
(=1/137) with the result $100{\,}{\rm MeV}\le 
\Lambda_{\rm QCD}\le 300{\,} 
{\rm MeV}$~\cite{lambda}. 
The phenomenon of the appearance of a dimensionful 
parameter in QCD, which remains finite in the limit of vanishing 
cutoff, is usually referred to as 
dimensional transmutation. All observable 
dimensionful quantities in QCD ({\it e.g.} widths of hadronic decays), 
and, in particular, the string tension  
are proportional to the corresponding power of 
$\Lambda_{\rm QCD}$\footnote{
The aim of all the nonperturbative phenomenological 
approaches to QCD, such as the 
Method of Field Correlators which will be described in the 
next Subsection, is to calculate the $g$-independent dimensionless 
ratios of these quantities, but not 
$\Lambda_{\rm QCD}$ itself.}. Then according to  
Eqs.~(\ref{function}) and (\ref{lambda}), we get 

\begin{equation}
\label{sigqcd}
\sigma\propto\Lambda_{\rm QCD}^2=\frac{1}{a^2}\exp\left[
-\frac{16\pi^2}{\left(\frac{11}{3}N_c-\frac23N_f\right)g^2\left(
\frac{1}{a^2}\right)}\right],
\end{equation}
which means that all the coefficients in the expansion of the 
string tension in powers of $g^2$ vanish. This conclusion tells us 
that the QCD string has a pure 
nonperturbative origin, as well 
as the phenomenon of confinement, which leads to the process of 
formation of such strings in the vacuum itself.

Throughout the present 
review, we shall be mostly interested in the models and 
properties of the QCD 
string and strings in other gauge theories, possessing a 
confining phase. In the literature, this problem is usually referred to 
as a {\it problem of string representation of gauge theories}.

Contrary to the linearly rising quark-antiquark potential, 
the Coulomb potential, $E(R)\propto -\frac{g^2}{R}$, 
cannot confine quarks, since in this case 
the gauge field between them is distributed over the whole space. 
For such a potential,  
the Wilson loop for the large contour $C$ has the following asymptotic 
behaviour 

\begin{equation}
\label{perim}
\left<W(C)\right>\to {\rm e}^{-{\rm const}\cdot L(C)},
\end{equation}
where $L(C)\equiv\int\limits_0^1 ds\sqrt{\dot x^2(s)}$ 
stands for the length of the contour $C$, parametrized by the 
vector-function $x_\mu (s),~ 0\le s\le 1,~ x_\mu(0)=x_\mu(1)$. 
Such a behaviour is called the perimeter law. 
It is dominant at small distances, $R\le 0.25{\,} {\rm fm}$, where quarks 
can with a good accuracy be 
considered as separate particles not connected by strings,  
{\it i.e.}, in the framework 
of perturbative QCD. 

It is worth noting that 
to each order of perturbation theory, it is the perimeter 
law~(\ref{perim}), rather than the area law~(\ref{area}), that holds 
for the Wilson loops. Because of the ultraviolet divergencies, 
for such a perturbative expansion of 
Eq.~(\ref{wils}) in powers of $g$, one needs a (gauge invariant) 
regularization. When such a regularization is introduced, the 
Wilson loop for a smooth contour $C$ ({\it i.e.}, a contour without 
cusps) takes the form 

\begin{equation}
\label{regul}
\left<W(C)\right>=\exp\left[-C_2\frac{g^2}{4\pi}
\frac{L(C)}{a}\right]\left<W_{\rm ren.}(C)\right>,
\end{equation}
where $\left<W_{\rm ren.}(C)\right>$ is finite when expressed 
via the renormalized charge $g_{\rm ren.}$. The exponential factor 
in Eq.~(\ref{regul}) is due to the renormalization of the mass 
of a heavy (anti)quark, described by 
Eq.~(\ref{renorm}). Such a multiplicative renormalization of the smooth 
Wilson loop has been proved in Refs.~\cite{neveu, rings, dot}. 
Notice also that if the contour $C$ has a cusp 
but no self-intersections, then $\left<W(C)\right>$ is still 
multiplicatively renormalizable~\cite{brandt}. Namely, in that case 
$\left<W(C)\right>=Z(\gamma)\left<W_{\rm ren.}(C)\right>$, where 
the diverging factor $Z(\gamma)$ depends on the cusp angle $\gamma$.

Thus, we conclude that the Wilson loop indeed plays the r\^ole of 
the quantity relevant to the description of confinement. 
Its area law behaviour means that the 
gauge theory under study ({\it e.g.} QCD) is in the confining phase, 
whereas the perimeter law behaviour means that the theory 
is in the Coulomb phase.
We also see that due to Eq.~(\ref{wils1}), the change of  
$-\ln\left<W(C)\right>$ determines the 
change of the action due to the interaction of an (anti)quark with 
the gauge field.

\subsection{Method of Field Correlators in QCD: Theoretical Foundations}

The nonlinear character of the QCD action makes it impossible to 
calculate in this theory the vacuum averages of 
gauge-invariant quantities 
in a closed form, {\it e.g.} 
by making use of 
the path-integral techniques. On the other hand, 
the standard perturbation theory also becomes 
unapplicable at large distances, where the running coupling constant 
is around 0.5-1. This makes it necessary to develop special 
approaches to nonperturbative QCD. In the present Subsection, 
we shall briefly describe one of them, 
the so-called Method of Field Correlators (MFC) in QCD. 
The idea, which lies behind this approach, 
is that a large class of observables in QCD can be expressed in terms of 
the Wilson 
loops~\cite{wilson, loopeqs1}. In this respect, the 
calculation of some QCD observable consists of two steps: 
the calculation of the Wilson loop for an arbitrary contour and 
summation of the Wilson loops over contours with a certain weight, 
which is determined by the observable~\footnote{It should be mentioned  
that contrary to the classical theory (where only the electric and 
magnetic field strengths are observable, while the vector potential 
plays only an auxiliary r\^ole for their determination), 
in the quantum theory Wilson 
loops are observable themselves owing to the 
Aharonov-Bohm effect~\cite{kobe}.}. 
For example, the Green 
function of a system consisting of a scalar quark and antiquark   
with equal masses $m$, which propagates in the QCD vacuum 
from the initial state $\left(y, 
\bar y\right)$ to the final state $\left(x,\bar x\right)$,  

\begin{equation}
\label{fourpoint}
G\left(x,\bar x;y,\bar y\right)\equiv\left<{\rm tr}{\,}
\left(\bar\psi\left(\bar x
\right)\Phi\left(\bar x, x\right)\psi(x)\right)
\left(\bar\psi(y)\Phi\left(y,\bar y\right)\psi\left(\bar y
\right)\right)\right>,
\end{equation}
where 

$$
\Phi(x,y)=\frac{1}{N_c}{\,}
P{\,}\exp\left[ig\int\limits_y^x A_\mu(u)
du_\mu\right]
$$
stands for the parallel transporter factor along the 
straingt line~\footnote{From now on, for quantities including  
$P$-ordering along open paths, we adopt the convention that the 
matrices are ordered from 
the second argument to the first one.}, 
in the quenched approximation reads 
\cite{npb, yadfiz, tjon}

\begin{equation}
\label{green}
G\left(x,\bar x;y,\bar y\right)=\int\limits_0^{+\infty}ds
\int\limits_0^{+\infty}d\bar s{\rm e}^{-m^2\left(s+\bar s\right)}
\int Dz\int D\bar z\exp\left(-\frac14\int\limits_0^s
\dot z^2 d\lambda-\frac14\int\limits_0^{\bar s}\dot{\bar z}^2
d\lambda\right)\left<W(C)\right>.
\end{equation}
In Eq.~(\ref{green}), the dot stands for 
$\frac{d}{d\lambda}$, and 
the closed contour $C$ consists of the trajectories 
$z_\mu$ and $\bar z_\mu$ of a quark and antiquark, 
$z(0)=y,{\,}z(s)=x,{\,}\bar z(0)=\bar y,{\,}\bar z(\bar s)=\bar x$, 
and straight-line 
pieces, which form the initial and final states. One can see that 
the weight factor in the path-integral standing in Eq.~(\ref{green}) 
is completely determined by 
the free theory, and all the dependence on the gauge field 
factors out in the form of the Wilson loop.  

After that, 
the main idea of the MFC is not to directly evaluate the 
Wilson loop, but to express 
it via gauge-invariant irreducible correlators (the so-called 
cumulants) of the gauge field strength tensors. Such correlators 
have been measured in lattice experiments both at large and 
small distances~\cite{di, digiac}, which enables one to use them  
in practical calculations of various physical quantities. 
In order to express the Wilson loop~(\ref{wils}) 
via cumulants, one should first 
make use of the non-Abelian Stokes theorem, derived in 
Refs.~\cite{stok, stokes1}, which yields~\footnote{In the following 
equation, the summation 
is performed over $\mu<\nu$ ({\it cf.} also Eq.~(\ref{expans}) below).}

\begin{equation}
\label{sto}
\left<W(C)\right>=
\frac{1}{N_c}\left<{\rm tr}{\,}P \exp\left(ig\int\limits_\Sigma^{} 
d\sigma_{\mu\nu}(x(\xi))F_{\mu\nu}(x(\xi),x_0)\right)\right>.
\end{equation}
On the R.H.S. of Eq.~(\ref{sto}), the integration is performed over 
an arbitrary surface $\Sigma$ bounded by the contour $C$ and parametrized 
by the vector $x_\mu(\xi)$. Next, $\xi\equiv
\left(\xi^1,\xi^2\right)$ is a two-dimensional coordinate, and   

\begin{equation}
\label{element}
d\sigma_{\mu\nu}(x(\xi))=\sqrt{g(\xi)}t_{\mu\nu}(\xi)d^2\xi
\end{equation}
stands for the infinitesimal surface element, where 
$g(\xi)$ is the determinant of the so-called induced metric tensor of the 
surface, defined as  

\begin{equation}
\label{metric}
g_{ab}(\xi)=
(\partial_a x_\mu(\xi))(\partial_b x_\mu(\xi)),
\end{equation}
where $\partial_a\equiv 
\frac{\partial}{\partial\xi^a},~ a,b=1,2$. 
Next, 

\begin{equation}
\label{curvature}
t_{\mu\nu}(\xi)=\frac{1}{\sqrt{g(\xi)}}\varepsilon^{ab}(\partial_a 
x_\mu(\xi))(\partial_b x_\nu(\xi)) 
\end{equation}
is the extrinsic curvature tensor of the surface, 
$t_{\mu\nu}^2(\xi)=2$, 
$F_{\mu\nu}(x,x_0)\equiv\Phi(x_0,x)F_{\mu\nu}(x)\Phi(x,x_0)$ is the 
covariantly shifted non-Abelian field strength tensor, 
$F_{\mu\nu}=\partial_\mu A_\nu-\partial_\nu A_\mu-ig\left[A_\mu, 
A_\nu\right]$,
and $x_0$ is an arbitrary but fixed reference point, the dependence 
on which actually drops out. 

The second step in expressing the Wilson loop 
in terms of the cumulants is 
to rewrite Eq.~(\ref{sto}) via the cumulant expansion 
theorem~\cite{van}. 
The aim of this theorem is to express cumulants, which are 
usually denoted as 
$\left<\left<F_{\mu_1\nu_1}(x_1,x_0)\cdots F_{\mu_n\nu_n}(x_n,x_0)
\right>\right>$,
through the usual averages 
$\left<F_{\mu_1\nu_1}(x_1,x_0)\cdots F_{\mu_n\nu_n}(x_n,x_0)
\right>$.
To this end, let us first consider for simplicity the Abelian case, 
where

$$
\left<F_{\mu_1\nu_1}(x_1,x_0)\cdots F_{\mu_n\nu_n}(x_n,x_0)
\right>=\left<F_{\mu_1\nu_1}(x_1)\cdots F_{\mu_n\nu_n}(x_n)\right>
\equiv
$$

\begin{equation}
\label{eta}
\equiv \int DA_\mu\eta\left(A_\mu\right)F_{\mu_1\nu_1}
(x_1)\cdots F_{\mu_n\nu_n}(x_n),
\end{equation}
with $\eta\left(A_\mu\right)$ standing for a certain $O(4)$-invariant 
integration measure, and introduce the following generating 
functional for field correlators

\begin{equation}
\label{abel}
{\cal Z}\left[J_{\mu\nu}\right]=
\left<\exp\left(ig\int d^4x J_{\mu\nu}(x)F_{\mu\nu}(x) 
\right)\right>.
\end{equation}
The field correlators then obviously have the form 

\begin{equation}
\label{gen}
\left.\left<F_{\mu_1\nu_1}(x_1)\cdots 
F_{\mu_n\nu_n}(x_n)\right>=(ig)^{-n}
\frac{\delta^n{\cal Z}\left[J_{\mu\nu}\right]}{\delta
J_{\mu_1\nu_1}(x_1)\cdots 
\delta J_{\mu_n\nu_n}(x_n)}\right|_{J=0}.
\end{equation}
By virtue of the Taylor expansion,  

$$
\left.{\cal Z}\left[J_{\mu\nu}\right]=
1+\sum\limits_{n=1}^{\infty}\frac{1}{n!}
\int d^4x_1\cdots\int d^4x_n J_{\mu_1\nu_1}(x_1)\cdots J_{\mu_n\nu_n}(x_n)
\left(\frac{\delta^n{\cal Z}\left[J_{\mu\nu}\right]}{\delta
J_{\mu_1\nu_1}(x_1)\cdots 
\delta J_{\mu_n\nu_n}(x_n)}\right|_{J=0}\right),
$$
where it has been used that ${\cal Z}[0]=1$, 
we obtain for the generating functional 

\begin{equation}
\label{moment}
{\cal Z}\left[J_{\mu\nu}\right]=
1+\sum\limits_{n=1}^{\infty}\frac{(ig)^n}{n!}\int d^4x_1
\cdots\int d^4x_n J_{\mu_1\nu_1}(x_1)\cdots J_{\mu_n\nu_n}(x_n)
\left<F_{\mu_1\nu_1}(x_1)\cdots F_{\mu_n\nu_n}(x_n)\right>.
\end{equation}
By definition, an $n$-th order cumulant is defined as follows 

\begin{equation}
\label{cum}
\left.\left<\left<F_{\mu_1\nu_1}(x_1)\cdots 
F_{\mu_n\nu_n}(x_n)\right>\right>=
(ig)^{-n}\frac{\delta^n\ln {\cal Z}\left[J_{\mu\nu}\right]}{
\delta J_{\mu_1\nu_1}(x_1)\cdots 
\delta J_{\mu_n\nu_n}(x_n)}\right|_{J=0}.
\end{equation}
In particular, if we specify a current to the form 
$J_{\mu\nu}(x)=
\int\limits_{\Sigma}^{}d\sigma_{\mu\nu}(x(\xi))\delta(x-x(\xi))$, 
where $\Sigma$ is a certain surface 
bounded by the contour $C$, then, by virtue of the 
Stokes theorem, the partition 
function~(\ref{abel}) (where the summation is performed over $\mu<\nu$) 
is nothing else, but the Wilson loop. 
The definition~(\ref{cum}) 
means that the cumulants of the field strength tensors are the analogues 
of usual 
irreducible (connected) Green functions, which is quite in line 
with the interpretation of $-\ln \left<W(C)\right>$ as a correction to the 
free energy due to the interaction of the gauge field with 
an external charged particle ({\it cf.} the end of the 
previous Subsection). 
Next, following the same steps which led from Eq.~(\ref{gen}) to 
Eq.~(\ref{moment}), we obtain from Eq.~(\ref{cum}) 

\begin{equation}
\label{expans}
\left<W(C)\right>=\exp\left(\sum\limits_{k=1}^{\infty}\frac{(ig)^k}{k!}
\int\limits_\Sigma^{}d\sigma_{\mu_1\nu_1}(x_1)\cdots\int\limits_\Sigma^{}
d\sigma_{\mu_k\nu_k}(x_k)
\left<\left<F_{\mu_1\nu_1}(x_1)\cdots F_{\mu_k\nu_k}(x_k)\right>\right>
\right).
\end{equation}
Eq.~(\ref{expans}) is usually referred to as the cumulant expansion. 

Notice that in Eqs.~(\ref{moment}) and (\ref{expans}), 
it has been assumed that the   
series on their R.H.S. converge, which is true for non-pathological 
models. To those belongs in particular  
QCD even at large distances, {\it i.e.}, at large 
$g$ (see 
discussion in Ref.~\cite{ufn}). This fact means that the 
cumulant expansion in QCD is indeed a nonperturbative expansion. 
The important property of the cumulants in such models, 
which distinguishes them 
from the usual correlators (thus associated with the usual Green 
functions), is that any cumulant decreases {\it vs.} the distance between 
any two points, in which the fields in the cumulant are defined.   
An example of the model where the cumulant expansion diverges is the 
instanton gas~\cite{gas}. 

One can now write down the following generating equation for 
cumulants

$$
1+\sum\limits_{n=1}^{\infty}\frac{(ig)^n}{n!}\int d^4x_1\cdots 
\int d^4x_n J_{\mu_1\nu_1}(x_1)\cdots J_{\mu_n\nu_n}(x_n)
\left<F_{\mu_1\nu_1}(x_1)\cdots F_{\mu_n\nu_n}(x_n)\right>=
$$

\begin{equation}
\label{gener}
=\exp\left(\sum\limits_{k=1}^{\infty}\frac{(ig)^k}{k!}
\int d^4x_1\cdots \int d^4x_k J_{\mu_1\nu_1}(x_1)\cdots 
J_{\mu_k\nu_k}(x_k)
\left<\left<F_{\mu_1\nu_1}(x_1)\cdots F_{\mu_k\nu_k}(x_k)\right>\right>
\right), 
\end{equation}
which means that varying Eq.~(\ref{gener}) several times {\it w.r.t.} 
$J_{\mu\nu}$ and setting then $J_{\mu\nu}=0$, one can get relations 
between the cumulants and correlators of various orders. 

In this way, the one-fold 
variation and setting $J_{\mu\nu}=0$ yield $\left<\left<F_{\mu\nu}(x)
\right>\right>=\left<F_{\mu\nu}(x)\right>$. 
Notice that due to the $O(4)$-invariance of the Euclidean integration 
measure $\eta\left(A_\mu\right)$, this average vanishes.
Next, the two- and three-fold variations 
and setting $J_{\mu\nu}=0$ yield the following relations

\begin{equation}
\label{biloc}
\left<F_{\mu_1\nu_1}(x_1)F_{\mu_2\nu_2}(x_2)\right>=\left<\left<
F_{\mu_1\nu_1}(x_1)F_{\mu_2\nu_2}(x_2)\right>\right>+
\left<F_{\mu_1\nu_1}(x_1)\right>\left<F_{\mu_2\nu_2}(x_2)\right>
\end{equation}
and

$$
\left<F_{\mu_1\nu_1}(x_1)F_{\mu_2\nu_2}(x_2)F_{\mu_3\nu_3}(x_3)
\right>=\left<\left<
F_{\mu_1\nu_1}(x_1)F_{\mu_2\nu_2}(x_2)F_{\mu_3\nu_3}(x_3)
\right>\right>+$$

$$+\left<\left<F_{\mu_1\nu_1}(x_1)F_{\mu_2\nu_2}(x_2)
\right>\right>\left<F_{\mu_3\nu_3}(x_3)\right>
+\left<\left<F_{\mu_1\nu_1}(x_1)F_{\mu_3\nu_3}(x_3)\right>\right>
\left<F_{\mu_2\nu_2}(x_2)\right>+$$

\begin{equation}
\label{threeloc}
+\left<F_{\mu_1\nu_1}(x_1)\right>
\left<\left<F_{\mu_2\nu_2}(x_2)F_{\mu_3\nu_3}(x_3)\right>\right>+
\left<F_{\mu_1\nu_1}(x_1)\right>
\left<F_{\mu_2\nu_2}(x_2)\right>
\left<F_{\mu_3\nu_3}(x_3)\right>,
\end{equation}
respectively. 
Eq.~(\ref{threeloc}) can be symbolically written as follows 

$$
\left<F_{\mu_1\nu_1}(x_1)F_{\mu_2\nu_2}(x_2)F_{\mu_3\nu_3}(x_3)
\right>=\left<\left<
F_{\mu_1\nu_1}(x_1)F_{\mu_2\nu_2}(x_2)F_{\mu_3\nu_3}(x_3)
\right>\right>+$$

\begin{equation}
\label{symb1}
+(3)\left<\left<F_{\mu_1\nu_1}(x_1)F_{\mu_2\nu_2}(x_2)
\right>\right>\left<F_{\mu_3\nu_3}(x_3)\right>
+\left<F_{\mu_1\nu_1}(x_1)\right>
\left<F_{\mu_2\nu_2}(x_2)\right>
\left<F_{\mu_3\nu_3}(x_3)\right>,
\end{equation}
where the coefficient in curly brackets denotes the number of the 
terms of the same type, which differ from each other only by the 
order of the arguments (and indices, respectively). Making use of 
this notation, we  
can next write {\it e.g.} the following equation  

$$
\left<F_{\mu_1\nu_1}(x_1)F_{\mu_2\nu_2}(x_2)F_{\mu_3\nu_3}(x_3)
F_{\mu_4\nu_4}(x_4)\right>=\left<\left<
F_{\mu_1\nu_1}(x_1)F_{\mu_2\nu_2}(x_2)F_{\mu_3\nu_3}(x_3)
F_{\mu_4\nu_4}(x_4)\right>\right>+$$

$$
+(3)\left<\left<F_{\mu_1\nu_1}(x_1)F_{\mu_2\nu_2}(x_2)\right>\right>
\left<\left<F_{\mu_3\nu_3}(x_3)F_{\mu_4\nu_4}(x_4)\right>\right>+$$

$$
+(4)\left<\left<F_{\mu_1\nu_1}(x_1)F_{\mu_2\nu_2}(x_2)
F_{\mu_3\nu_3}(x_3)\right>\right>\left<F_{\mu_4\nu_4}(x_4)\right>+$$

\begin{equation}
\label{symb2}
+(6)\left<\left<F_{\mu_1\nu_1}(x_1)F_{\mu_2\nu_2}(x_2)\right>\right>
\left<F_{\mu_3\nu_3}(x_3)\right>\left<F_{\mu_4\nu_4}(x_4)\right>+
\left<F_{\mu_1\nu_1}(x_1)\right>
\left<F_{\mu_2\nu_2}(x_2)\right>
\left<F_{\mu_3\nu_3}(x_3)\right>
\left<F_{\mu_4\nu_4}(x_4)\right>.
\end{equation}
Now one can see that on the R.H.S. of Eqs.~(\ref{symb1}) 
and~(\ref{symb2}) stand with the coefficient equal to unity 
all possible terms, which correspond to various splittings 
of the set $\left\{F_{\mu_1\nu_1}(x_1),\ldots,F_{\mu_n\nu_n}(x_n)
\right\}$ into subsets, so that to every subset corresponds a 
cumulant. This property holds also for the higher correlators, 
which gives a simple mnemonic rule for calculation of the higher 
cumulants without using Eq.~(\ref{gener}). This rule is the 
essence of the cumulant expansion theorem. 

For the non-Abelian case, the cumulant expansion theorem remains 
the same except 
for excluding the terms which violate the path-ordering 
prescription~\cite{stokes1}. {\it E.g.} in the non-Abelian version of 
Eq.~(\ref{threeloc}), the term 

$$
\left<\left<F_{\mu_1\nu_1}(x_1,x_0)F_{\mu_3\nu_3}(x_3,x_0)\right>\right>
\left<F_{\mu_2\nu_2}(x_2,x_0)\right>$$
will be absent for a given order of points $x_1, x_2, x_3$.  

It looks natural to address the question on whether cumulants of 
various orders are independent of each other 
or there exist any relations 
between them. Such equations relating cumulants of different orders 
indeed take place. They have been derived by making use of the 
stochastic quantization method~\cite{parisi, zinn} in Ref.~\cite{stoch1}  
and further investigated in Refs.~\cite{stoch3, stoch4, stoch5}. 
Alternative equations for  
cumulants following from the non-Abelian Bianchi identities have 
been proposed in Refs.~\cite{stokes1, stoch1}.
Notice that quite recently, there has also been 
proposed one more approach to a derivation of a system of 
self-coupled equations for cumulants. This has been done in 
Ref.~\cite{gluoeq} by considering a field of a certain colour in the 
gluodynamics Lagrangian as the one propagating 
in the background of the 
fields of all other colours 
by making use of the background field methods. Finally, 
besides equations for cumulants, there have recently been proposed 
new equations for Wilson loops~\cite{stochloop} 
derived by virtue of the stochastic 
quantization method. Several techniques of the loop space approach,  
exploited in a derivation of these equations, have been then applied 
to the solution of the Cauchy problem for the loop equation 
in 3D turbulence~\cite{turb}.

By definition adopted in the theory of random processes, a set of 
random quantities is called Gaussian, provided that all the cumulants 
of these quantities higher than the quadratic one vanish. According to 
the lattice data~\cite{di, digiac}, the stochastic ensemble of fields in 
QCD can with a good accuracy be considered as a Gaussian one, 
{\it i.e.}, the bilocal cumulant 

\begin{equation}
\label{cumul}
\left<\left<F_{\mu\nu}(x,x_0)F_{\lambda\rho}(y,x_0)
\right>\right>
\end{equation}
is much larger than all the higher cumulants, so 
that the cumulant expansion converges fastly. Due to this property 
of the QCD vacuum, one can disregard all the cumulants except the 
bilocal one, which leads to the bilocal or Gaussian 
approximation in the MFC in QCD. 
However one can immediately see that the 
neglection of all the cumulants higher than the bilocal one in 
the non-Abelian analogue of 
Eq.~(\ref{expans}) leads to the appearance of the artificial dependences 
of the R.H.S. of this equation on the reference point $x_0$ 
and on the shape of the surface $\Sigma$. 

The first problem can be 
solved by noting that the bilocal cumulant~(\ref{cumul}) decreases fastly 
when $\left|x-y\right|\simeq T_g$, where $T_g$ is the so-called 
correlation length of the vacuum, $T_g\simeq 0.13{\,}{\rm fm}$ 
in the $SU(2)$-case~\cite{zeitphys}, and $T_g\simeq 0.22{\,}{\rm fm}$ in 
the $SU(3)$-case~\cite{di}. On the 
other hand, as it has already been discussed in the previous Subsection, 
according to Ref.~\cite{ford}, the area law of the Wilson loop takes 
place when its size $R$ is of the order of $1.0{\,} {\rm fm}$. This 
means that in the confining regime, {\it i.e.}, 
for the loops of such size,  
one can in the general case ({\it i.e.}, 
for the dominant amount of cumulants) 
write down the following inequality 

$$\left|x-x_0\right|\simeq\left|y-x_0\right|\simeq R\gg T_g\simeq 
\left|x-y\right|.$$
According to it, we can with a good accuracy neglect the dependence 
on the point $x_0$ in the bilocal cumulant~(\ref{cumul}), {\it i.e.},  
approximate this quantity by the gauge- and translation-invariant 
cumulant as follows 

$$
{\rm tr}\left<\left<F_{\mu\nu}(x,x_0)F_{\lambda\rho}(y,x_0)
\right>\right>\simeq {\rm tr}\left<\left<F_{\mu\nu}(x)\Phi(x,y)
F_{\lambda\rho}(y)\Phi(y,x)\right>\right>.
$$
 
As far as the problem of the artificial dependence on the shape of 
the surface $\Sigma$ 
in the bilocal approximation is concerned, it 
cannot be solved on the basis of simple theoretical arguments, 
but one can choose 
$\Sigma$ to be the surface of the minimal area for a given 
contour $C$.
As we shall see in Subsection 2.1, this enables one to reproduce 
the area law behaviour of the Wilson loop. Since such a surface 
$\Sigma_{\rm min.}$ is uniquely defined by the contour $C$, 
we finally arrive 
at the following expression for the Wilson loop in the bilocal 
approximation 

$$
\left<W(C)\right>\equiv 
\left<W\left(\Sigma_{\rm 
min.}\right)\right>\simeq$$

\begin{equation}
\label{loop}
\simeq\frac{1}{N_c}{\rm tr}{\,}\exp
\left(-\frac{g^2}{2}\int\limits_{\Sigma_{\rm min.}}^{}d\sigma_{\mu\nu}(x)
\int\limits_{\Sigma_{\rm min.}}^{}d\sigma_{\lambda\rho}(x')
\left<\left<F_{\mu\nu}(x)\Phi(x,x')
F_{\lambda\rho}(x')\Phi(x',x)\right>\right>\right), 
\end{equation}
where we have denoted for brevity $x\equiv x(\xi)$ and $x'\equiv 
x(\xi')$. 
Due to the colour invariance of the Euclidean integration measure 
in QCD, the bilocal cumulant on the R.H.S. of Eq.~(\ref{loop}) is 
proportional to the unity matrix in the fundamental representation, 
$\hat 1_{N_c\times N_c}$, and therefore the $P$-ordering is not 
necessary any more.   

The bilocal cumulant standing in the exponent on the R.H.S. of 
Eq.~(\ref{loop}) can be now parametrized by two 
renormalization-group invariant coefficient functions $D$ and 
$D_1$ as follows

$$
\frac{g^2}{2}\left<\left<F_{\mu\nu}(x)\Phi(x,x')
F_{\lambda\rho}(x')\Phi(x',x)\right>\right>=\hat 1_{N_c\times 
N_c}\Biggl\{(\delta_{\mu\lambda}\delta_{\nu\rho}-\delta_{\mu\rho}
\delta_{\nu\lambda})D\left((x-x')^2\right)+$$

\begin{equation}
\label{dd1}
+\frac12\left[\frac{\partial}{\partial x_\mu}((x-x')_\lambda
\delta_{\nu\rho}-(x-x')_\rho\delta_{\nu\lambda})+\frac{\partial}
{\partial x_\nu}((x-x')_\rho\delta_{\mu\lambda}-(x-x')_\lambda
\delta_{\mu\rho})\right]D_1\left((x-x')^2\right)\Biggr\}.
\end{equation}
The 
parametrization~(\ref{dd1}) of the bilocal cumulant is chosen in such a 
way, that the term containing the function $D_1$ yields a perimeter type 
contribution to the Wilson loop~(\ref{loop}). Namely, by making use 
of the (ordinary) Stokes theorem, one can prove 
that~\footnote{From now on, 
for simplicity, we restore in Eq.~(\ref{loop}) the usual agreement 
on the summation over all $\mu$ and $\nu$ ($\lambda$ and 
$\rho$), not only $\mu<\nu$. This can always be done by the appropriate 
normalization of the functions $D$ and $D_1$.}  

$$\frac12\int
\limits_{\Sigma_{\rm min.}}^{}d\sigma_{\mu\nu}(x)
\int\limits_{\Sigma_{\rm min.}}^{}d\sigma_{\lambda\rho}(x')
\left[\frac{\partial}{\partial x_\mu}((x-x')_\lambda
\delta_{\nu\rho}-(x-x')_\rho\delta_{\nu\lambda})+\right.$$

\begin{equation}
\label{d1}
\left.+\frac{\partial}
{\partial x_\nu}((x-x')_\rho\delta_{\mu\lambda}-(x-x')_\lambda
\delta_{\mu\rho})\right]D_1\left((x-x')^2\right)=
\oint\limits_C^{}dx_\mu\oint\limits_C^{}dy_\mu G\left((x-y)^2\right),
\end{equation}
where 

\begin{equation}
\label{g}
G\left(x^2\right)\equiv\int\limits_{x^2}^{+\infty}d\lambda 
D_1(\lambda).
\end{equation}
In particular, we see that the 
one-gluon-exchange diagram contribution 
to the Wilson loop, equal to 

$$\exp\left(-C_2\frac{g^2}{4\pi^2}
\oint\limits_C^{}dx_\mu
\oint\limits_C^{}dy_\mu\frac{1}{(x-y)^2}\right)$$
(which for the contour $C$ without cusps yields the renormalization 
factor standing on the R.H.S. of Eq.~(\ref{regul})) 
is contained due to Eq.~(\ref{d1}) in the function $D_1$. 
According to Eq.~(\ref{g}), it is equal to 

\begin{equation}
\label{oge}
D_1^{\rm OGE}\left(x^2\right)=C_2\frac{g^2}{4\pi^2}\frac{1}
{|x|^4}.
\end{equation}
Eq.~(\ref{oge}) is the leading contribution to the function $D_1$. 
Therefore the effects of renormalization of the Wilson loop, discussed 
at the end of the previous Subsection, are 
taken into account in the MFC by virtue of the function $D_1$.  

Thus, the function $D_1$ is nonvanishing already in the order 
$g^2$, whereas the function $D$ in this order of perturbation theory 
vanishes. The nonvanishing contributions to the function $D$ emerging  
in the higher orders have been calculated 
in Ref.~\cite{jamin}. From now on, we shall not be interested any more 
in the perturbative contributions 
to these functions. 
The nonperturbative parts of the functions $D$ and 
$D_1$ have been calculated in the lattice experiments  
in Refs.~\cite{di, digiac}, where it has been 
shown that they 
are related to each other as  
$D_1\simeq\frac13 D$. 

Due to the Lorentz structure standing at the function $D$ in 
Eq.~(\ref{dd1}), its contribution to the Wilson loop~(\ref{loop}) 
cannot be reduced to that of a perimeter type. This function is 
responsible for the area law behaviour of the Wilson loop and 
gives rise to the QCD string effective action. The problems of a 
derivation of this action from Eq.~(\ref{loop}) and its further 
investigation will be the topic of the next Section.

\section{String Representation of QCD in the Framework of the 
Method of Field Correlators}

In the present Section, we shall demonstrate the usefulness of the MFC 
for the derivation and investigation of the local gluodynamics 
string effective action. Our interpretation of this topic in the 
next three Subsections 
will mainly follow the original papers~\cite{curv, inst}  
in Subsection 2.1,~\cite{pert} in Subsection 2.2, and~\cite{ham} in 
Subsection 2.3. A 
short review of all these papers can be found in Ref.~\cite{buckow}.   
\vspace{3mm}

\subsection{Gluodynamics String Effective Action from the Wilson 
Loop Expansion}

Let us start with the problem of a derivation of the effective 
action of the gluodynamics string, generated by the strong vacuum 
background 
fields in QCD, which ensure confinement and yield a dominant 
contribution to the bilocal cumulant. To this end we mention that, 
as it has already been discussed in the 
previous Section, the quantity $-\ln \left<W(C)\right>$ 
is nothing else, but 
a correction to the gluodynamics free energy due to the interaction 
of a test quark, moving along the contour $C$, with these background 
fields. This observation enables us to treat this quantity with 
$\left<W(C)\right>$ defined by Eq.~(\ref{loop}) as a nonlocal 
background-induced gluodynamics string effective action associated 
with the minimal area surface $\Sigma_{\rm min.}$. In what follows, 
by the terms ``local'' and ``nonlocal string effective action'', we 
shall imply the actions depending on a single string world-sheet 
coordinate $\xi$ 
or on two such coordinates, respectively.

Substituting  
Eq.~(\ref{dd1}) into Eq.~(\ref{loop}), one can see that the 
coefficient function $D$ plays the r\^ole 
of a propagator of the background 
field between the points $x(\xi)$ and $x(\xi')$, which lie on 
$\Sigma_{\rm min.}$. Though this propagator cannot be completely found 
analytically, its nonperturbative part can be parametrized with a 
good accuracy as 

\begin{equation}
\label{dnp}
D(x)=\alpha_s\left<\left(F_{\mu\nu}^a(0)
\right)^2\right>{\rm e}^{-|x|/T_g}, 
\end{equation} 
where the gluonic condensate 
$\alpha_s\left<\left(F_{\mu\nu}^a(0)\right)^2\right>$ is of the 
order of $0.038{\,}{\rm GeV}^2$~\cite{vain}~\footnote{The colour 
index ``a'' should not be confused with the index of the two-vector 
$\xi$.}. This will finally enable us 
to obtain the coupling constants of the few first terms of the 
resulting local string effective action expressed via the 
correlation length of the vacuum, $T_g$, and the gluonic condensate.

In order to derive the desired local effective action, let us proceed 
with the Taylor expansion of the nonlocal action

\begin{equation}
\label{nonloc}
S_{\rm eff.}\left(\Sigma_{\rm min.}\right)=2\int
\limits_{\Sigma_{\rm min.}}^{}d\sigma_{\mu\nu}(x)\int
\limits_{\Sigma_{\rm min.}}^{}d\sigma_{\mu\nu}(x')
D\left(\frac{(x-x')^2}{T_g^2}\right)
\end{equation}
in powers of the derivatives {\it w.r.t.} 
the string world-sheet coordinates 
$\xi^a$'s. Obviously, the nonlocality of the initial action will 
then display itself in the appearance of higher derivatives 
{\it w.r.t.} 
$\xi^a$'s in the final expression for the action. Notice also that 
in Eq.~(\ref{nonloc}), we have explicitly emphasized the 
form of the dependence on $T_g$, since, as we shall 
eventually see, the parameter of 
this expansion will be proportional to $T_g$. 

To perform an expansion, let us first 
rewrite infinitesimal surface elements 
on the R.H.S. of Eq.~(\ref{nonloc}) by making use of 
Eqs.~(\ref{element}), (\ref{metric}), and (\ref{curvature}). Next, we 
introduce instead of $\xi'^a$'s new integration variables $\zeta^a\equiv
\frac{(\xi'-\xi)^a}{T_g}$ and expand in power series of $\zeta^a$'s the 
quantities $\sqrt{g(\xi')}$, $t_{\mu\nu}(\xi')$, $x(\xi')-x(\xi)$, and 
finally $D\left(\frac{(x-x')^2}{T_g^2}\right)$. Such an expansion will 
automatically be a formal series in powers of $T_g$, {\it e.g.} 

$$
\left(x(\xi')-x(\xi)\right)^2=T_g^2\left[\zeta^a\zeta^bg_{ab}(\xi)+
T_g\zeta^a\zeta^b\zeta^c(\partial_ax_\mu)(\partial_b\partial_cx_\mu)+
{\cal O}\left(T_g^2\right)\right].
$$ 
In what follows, we shall be interested only in the terms in this 
series not higher than of the fourth order in $T_g$. 
Then, taking into account that for an 
arbitrary odd $n$, $\int d^2\zeta\zeta^{i_1}\cdots
\zeta^{i_n}D(y)=0$, where 
$y\equiv
\zeta^a\zeta^bg_{ab}(\xi)$, we obtain
 
$$S_{\rm eff.}=2T_g^2\int d^2\xi\sqrt{g(\xi)}\Biggl\{
2\sqrt{g(\xi)}\int d^2\zeta D(y)+T_g^2\Biggl\{\sqrt{g(\xi)}
\Biggl[-\frac12(\partial_at_{\mu\nu}(\xi))(\partial_bt_{\mu\nu}(\xi))
\int d^2\zeta\zeta^a\zeta^bD(y)+$$

$$+\Biggl(\frac12(\partial_a\partial_b
x_\mu)(\partial_c\partial_dx_\mu)+\frac23(\partial_ax_\mu)
(\partial_b\partial_c\partial_dx_\mu)\Biggr)\int d^2\zeta\zeta^a
\zeta^b\zeta^c\zeta^dD'(y)+$$

$$+(\partial_ax_\mu)(\partial_b\partial_cx_\mu)
(\partial_dx_\nu)(\partial_e\partial_fx_\nu)\int d^2\zeta\zeta^a
\zeta^b\zeta^c\zeta^d\zeta^e\zeta^fD''(y)\Biggr]+$$

$$+2\left(\partial_a
\sqrt{g(\xi)}\right)(\partial_bx_\mu)(\partial_c\partial_dx_\mu)
\int d^2\zeta\zeta^a\zeta^b\zeta^c\zeta^dD'(y)+\left(\partial_a\partial_b
\sqrt{g(\xi)}\right)\int d^2\zeta\zeta^a\zeta^bD(y)\Biggr\}\Biggr\},$$
where ``${\,}'{\,}$'' stands for the derivative {\it w.r.t.} 
the argument, and
we have arranged the terms in powers of the derivatives of 
$\sqrt{g(\xi)}$. In order to simplify the integrals over $\zeta^a$'s as 
much as possible, it is very useful to fix by reparametrization 
the conformal gauge for the induced metric, $g_{ab}(\xi)=\sqrt{g(\xi)}
\delta_{ab}$ (see {\it e.g.} Ref.~\cite{polbook}). After that, by making 
use of the equality $\zeta^a\frac{\partial}{\partial y}=
\frac{1}{2\sqrt{g(\xi)}}\frac{\partial}{\partial\zeta_a}$, it becomes 
possible to perform partial integrations over $\zeta^a$'s. It is also 
useful to replace the ordinary derivatives by the covariant ones 
according to the Gauss-Weingarten formulae 

$$D_aD_bx_\mu=\partial_a
\partial_bx_\mu-\Gamma_{ab}^c\partial_cx_\mu=K_{ab}^in_\mu^i.$$
Here, $n_\mu^i$'s stand for the unit normal vectors to the string 
world-sheet, $n_\mu^in_\mu^j=\delta^{ij}$, $n_\mu^i\partial_a
x_\mu=0$, $i,j=1,2$, $\Gamma_{ab}^c$ is a Christoffel symbol, and 
$K_{ab}^i$ is the second fundamental form of the world-sheet. 

Notice 
that in the conformal gauge adopted, all the quantities are greatly 
simplified, {\it e.g.} 
$\Gamma_{ab}^c=\frac12\left(\delta_a^c\partial_b+
\delta_b^c\partial_a-\delta_{ab}\partial^c\right)\ln\sqrt{g(\xi)}$. 
In particular, the expression for the scalar curvature ${\cal R}=\left(
K_a^{i{\,}a}\right)^2-K_b^{i{\,}a}K_a^{i{\,}b}$ of the 
world-sheet  
in this gauge takes the form 

\begin{equation}
\label{r}
{\cal R}=\frac{\partial^a\partial_a\ln
\sqrt{g(\xi)}}{\sqrt{g(\xi)}},
\end{equation} 
and one can prove the validity of 
the following equations 

$$T^{abcd}(\partial_a\partial_bx_\mu)(\partial_c\partial_dx_\mu)=
T^{abcd}(D_aD_bx_\mu)(D_cD_dx_\mu)+2\frac{
\left(\partial_a\ln\sqrt{g(\xi)}\right)^2}{\sqrt{g(\xi)}},$$

$$T^{abcd}(\partial_ax_\mu)(\partial_b\partial_c\partial_dx_\mu)=
T^{abcd}(D_ax_\mu)(D_bD_cD_dx_\mu)+2{\cal R},$$

$$T^{abcd}(D_aD_bx_\mu)(D_cD_dx_\mu)=3(D^aD_ax_\mu)(D^bD_bx_\mu)-2
{\cal R},$$

$$g^{ab}\frac{\partial_a\partial_b\sqrt{g(\xi)}}{\sqrt{g(\xi)}}={\cal R}+
\frac{\left(\partial_a\ln\sqrt{g(\xi)}\right)^2}{\sqrt{g(\xi)}},$$ 

$$T^{abcd}\left(\partial_a\sqrt{g(\xi)}\right)(\partial_bx_\mu)
(\partial_c\partial_dx_\mu)=2\left(
\partial_a\ln\sqrt{g(\xi)}\right)^2,$$
where $T^{abcd}\equiv g^{ab}g^{cd}+g^{ac}g^{bd}+g^{ad}g^{bc}$. Making 
use of all that, introducing a new integration variable $z^a\equiv
g^{1/4}\zeta^a$, and recovering the metric dependence by virtue 
of the equality 

\begin{equation}
\label{beltrami}
\left(D^aD_ax_\mu\right)\left(D^bD_bx_\mu\right)=g^{ab}
\left(\partial_at_{\mu\nu}\right)\left(\partial_bt_{\mu\nu}\right),
\end{equation}
we arrive after some straightforward algebra at the 
following expression for the string effective action up to the 
order $T_g^4$

\begin{equation}
\label{effact}
S_{\rm eff.}=\sigma\int d^2\xi\sqrt{g}+\kappa
\int d^2\xi\sqrt{g}{\cal R}+\frac{1}{\alpha_0}\int d^2\xi\sqrt{g}g^{ab}
(\partial_at_{\mu\nu})(\partial_bt_{\mu\nu})+
{\cal O}\left(\frac{T_g^6\alpha_s\left<\left(F_{\mu\nu}^a(0)\right)^2
\right>}{R^2}\right), 
\end{equation}
where $R$ is the size of the contour $C$ in the confining 
regime ({\it cf.} Eq.~(\ref{energy})).  

In Eq.~(\ref{effact}), the coupling constants are completely 
determined via the zeroth and the first moments of the coefficient 
function $D$ (as well as the omitted higher terms of the expansion 
are determined via the higher moments of the function $D$) 
as follows

\begin{equation}
\label{tens}
\sigma=4T_g^2\int d^2zD\left(z^2\right),
\end{equation}

$$
\kappa=\frac{T_g^4}{6}\int d^2zz^2D\left(z^2\right),
$$
and 

\begin{equation}
\label{ridconst}
\frac{1}{\alpha_0}=-\frac{T_g^4}{4}\int d^2zz^2D\left(z^2\right).
\end{equation}
Making use of Eq.~(\ref{dnp}), one can estimate them as 

$$\sigma=4\pi T_g^2\alpha_s\left<\left(F_{\mu\nu}^a(0)\right)^2
\right>,~\kappa=\frac{\pi}{6}T_g^4 \alpha_s\left<\left(
F_{\mu\nu}^a(0)\right)^2\right>,~\frac{1}{\alpha_0}=-\frac{\pi}{4}
T_g^4 \alpha_s\left<\left(
F_{\mu\nu}^a(0)\right)^2\right>.$$
Due to the lattice data quoted after Eq.~(\ref{cumul}), we 
thus get for the $SU(2)$-case the following values of the coupling 
constants $\sigma\simeq 0.2{\,}{\rm GeV}^2$ 
({\it cf.} the corresponding value before Eq.~(\ref{low})), 
$\kappa\simeq 0.003$, and $\frac{1}{\alpha_0}=
-0.005$. The obtained value of the Nambu-Goto term string tension 
demonstrates the agreement of the MFC with the present lattice results, 
while the two other obtained coupling constants are simply small 
numbers, which confirms the validity of the performed expansion. 

Before discussing various terms in Eq.~(\ref{effact}), let us 
comment on the parameter of the performed expansion of the initial 
Eq.~(\ref{nonloc}). First of all, as it has already been mentioned, 
this is an expansion in formal power series of $T_g$, which means 
that it is valid only when $T_g$ is sufficiently small. That was a 
reason for the author of Ref.~\cite{chiral} to call the limit when 
$T_g$ is small but $\sigma$ is kept fixed as a ``string limit of QCD''. 
An $n$-th term of the expansion has the order of magnitude 
$\alpha_s\left<\left(F_{\mu\nu}^a(0)\right)^2
\right>R^4\left(\frac{T_g}{R}\right)^{2n}$, which means that the 
parameter of the expansion is $\left(T_g/R\right)^2$. In the confining 
regime, this parameter 
is of the order of $0.04$ (see the lattice data after 
Eq.~(\ref{cumul})), {\it i.e.}, 
is indeed a small number. Therefore in the 
``string limit'', the operators of the lowest orders in the 
derivatives {\it w.r.t.} $\xi^a$'s dominate in the 
expansion of the full nonlocal 
action~(\ref{nonloc})~\footnote{Examples of the 
geometric structures, emerging in the 
expansion in the order $T_g^6$, are listed in the Appendix to the 
first paper of Ref.~\cite{curv}.}. Contrary, in the 
QCD sum rule limit~\cite{vain}, the effects brought about by the 
nonlocality of the functions $D$ and $D_1$ are disregarded, and these 
functions  
are simply replaced by the gluonic condensate. In the language of the 
parametrization~(\ref{dnp}), this means that $T_g\to\infty$, and our 
expansion diverges. This observation clarifies once more the 
relevance of the string picture of QCD to the description of confinement. 
We also see, that it is the MFC (where the correlation length of the 
vacuum can be considered as a variable parameter),  
which provides us with such a picture.    

Let us now proceed with the physical discussion of the obtained 
local string effective action~(\ref{effact}). The first term on the 
R.H.S. of this equation 
is the celebrated Nambu-Goto term with the positive 
string tension $\sigma$. This term ensures the area law behaviour of the 
Wilson loop~(\ref{area}), 
since $\int d^2\xi\sqrt{g(\xi)}$ is nothing else, 
but the area of the minimal surface, bounded by $C$. The obtained 
expression for the string tension coincides 
with the one obtained in Ref.~\cite{strten}.
Notice also that though the 
Nambu-Goto term alone is known to suffer from the 
problem of appearing of a tachyon in its spectrum, this problem is absent 
for the full nonlocal string effective action~(\ref{nonloc}).

The second term is known to be a full 
derivative in 2D, which can be most easily seen in the conformal 
gauge adopted, {\it i.e.}, from Eq.~(\ref{r}). This term is a topological 
invariant proportional to the Euler character of the world-sheet 

$$\chi=\frac{1}{4\pi}\int d^2\xi\sqrt{g(\xi)}{\cal R}=2-2\times 
(\mbox{number of handles})-(\mbox{number of boundaries}).$$
In particular, for the surface $\Sigma_{\rm min.}$ 
under study, $\chi=1$. 

The most interesting term in the obtained effective 
action~(\ref{effact}) is the third one, which is usually referred to 
as the rigidity term~\cite{crump, rigid}. 
This term is not a full derivative, since 
it can be rewritten (just modulo full derivative terms) as 

$$\frac{1}{\alpha_0}\int d^2\xi\sqrt{g(\xi)}K_b^{i{\,}a}K_a^{i{\,}b},$$
{\it i.e.}, the integrand does not contain the complete expression for 
$\sqrt{g}{\cal R}$.  The rigidity 
term has been introduced into string theory in the 
above mentioned papers from the general arguments as the only possible 
one, which is invariant under the scale transformation $x_\mu(\xi)\to
\lambda x_\mu(\xi)$. 

It is worth noting that 
during the derivation of the effective action~(\ref{effact}),
we have not used explicitly the fact 
that the surface $\Sigma_{\rm min.}=\Sigma_{\rm min.}[C]$, 
described by the induced metric $g_{ab}(\xi)$, is the one of the 
minimal area for a given contour $C$. However, this surface possesses 
an important property distinguishing it from all the other surfaces 
bounded by $C$. 
Namely, it is defined by the 
(nonlinear) equation $D^aD_ax_\mu(\xi)=0$
together with the boundary condition 
$\left.x_\mu(\xi)\right|_C=x_\mu(s)$.
Due to this exceptional property of $\Sigma_{\rm min.}[C]$
and Eq.~(\ref{beltrami}), we see that at this surface the rigidity 
term vanishes. However, in a complete string picture of QCD including 
quantum fluctuations above confining background, there should appear 
an average over all world-sheets $\Sigma$'s 
bounded by the contour $C$. 
It is natural to expect that the dominant contribution to this 
integral is brought about by $\Sigma_{\rm min.}[C]$, 
{\it i.e.}, Eq.~(\ref{loop}) can be represented as 

\begin{equation}
\label{complete}
\left<W\left(\Sigma_{\rm min.}\right)\right>=
\int Dx_\mu(\xi)\left<W(\Sigma)\right>.
\end{equation} 
Here, $x_\mu(\xi)$ parametrizes the surface $\Sigma$, and the 
measure of the average over world-sheets, 
$Dx_\mu(\xi)$, 
should just be determined by the above mentioned quantum fluctuations.
In writing down Eq.~(\ref{complete}), we have assumed that the 
statistical weight of the average over world-sheets is given by 
the MFC-inspired Eq.~(\ref{loop}) with the replacement 
$\Sigma_{\rm min.}\to\Sigma$. If this assumption really holds, 
which should be justified by further investigations, then 
Eq.~(\ref{effact}), with $g_{ab}$ being the induced metric 
describing $\Sigma$,
yields the first few terms of the local string effective action 
associated with the statistical weight $\left<W(\Sigma)\right>$.
Since $\Sigma$ here is no more the surface of the minimal area, 
the rigidity term in this string effective action survives.

As it has been for the first time 
mentioned in Ref.~\cite{rigid} and then confirmed by related 
calculations in Refs.~\cite{maeda} and~\cite{orl},   
the negative sign of the coupling constant 
$\alpha_0$ of the rigidity term is an important property relevant to 
the stability of the string world-sheet. A simple argument in favour 
of this observation can be obtained by considering the propagator 
corresponding to the action~(\ref{effact}), which for a certain 
Lorentz index $\lambda$ reads

\begin{equation}
\label{xx}
\left<x_\lambda(\xi)x_\lambda(0)\right>=\int\frac{d^2p}{(2\pi)^2}
\frac{{\rm e}^{ip^a\xi_a}}{\sigma p^2-\frac{1}{\alpha_0}p^4}.
\end{equation}
For negative $\alpha_0$, this integral yields 

$$\left<x_\lambda(\xi)x_\lambda(0)\right>=
-\frac{1}{2\pi\sigma}\left[\ln\left(\mu\left|\xi\right|\right)+K_0\left(
\sqrt{\left|\alpha_0\right|\sigma}\left|\xi\right|\right)\right]$$
with $\mu$ denoting the IR momentum cutoff and $K_0$ standing for the 
modified Bessel function,  
while for positive $\alpha_0$ an unphysical 
pole in the propagator occurs. 
Another arguments concerning the necessity of the negative sign of 
$\alpha_0$ have been presented in the above mentioned 
papers. 
This property (called there ``anti-rigidity'' or ``negative stiffness'') 
has been demonstrated 
to ensure vanishing of the imaginary part of the frequencies of small 
fluctuations of the world-sheet, which might lead to the 
instabilities of the latter one. Further, the requirement 
of the negative sign of $\alpha_0$ 
has been employed in Ref.~\cite{negstif}. There, a 
new nonlocal string action, which manifests negative stiffness, has been 
proposed as a good candidate for modelling the gluodynamics string. 
However contrary to our calculations, 
it remained unclear in that paper how the action introduced there
can be derived either from the Wilson loop expansion or from some 
quantity relevant to gluodynamics ({\it e.g.} 
another vacuum amplitude or the gluodynamics Lagrangian).

Once being integrated out, the small transversal fluctuations of the 
world-sheet, mentioned above, produce a renormalization of 
the string tension, which at the two-loop level has been calculated 
in Ref.~\cite{german} and reads 

$$
\sigma_{\rm ren.}=\sigma\left[1+\alpha_0\frac{D-2}{16\pi}\left(1+
\ln\frac{4\Lambda^2}{\alpha_0^2\sigma}\right)+\alpha_0^2
\frac{(D-2)(D-1)}{256\pi^2}\left(\ln\frac{4\Lambda^2}{\alpha_0^2\sigma}
\right)^2\right],
$$
where $\Lambda$ is an UV momentum cutoff, and $D$ is the dimension 
of the space-time. It is worth noting that 
the appearance 
of the rigidity term yields also some modifications in 
the quark-antiquark potential, string tension behaviour  
at finite temperature, and thermal deconfinement properties, all 
of which have been surveyed in Ref.~\cite{germ1}.

An important problem associated with the action~(\ref{effact}) 
is that the corresponding propagator~(\ref{xx}) enjoys the limit of 
small separations, $|\xi|\ll\frac{1}{\sqrt{\left|
\alpha_0\right|\sigma}}$. Indeed, in this limit it 
goes over into  
$\left<x_\lambda(\xi)x_\lambda(0)\right>\simeq 
-\frac{1}{2\pi\sigma}\ln\frac{\mu}{\sqrt{\left|
\alpha_0\right|\sigma}}$ and therefore becomes infinitely large 
due to the $\mu$-dependence. 
This means that 
normals to the string world-sheet 
tend to be very short-ranged, {\it i.e.}, the world-sheet becomes  
extremely crumpled. That is the reason, why this problem is 
usually referred to as the problem of crumpling of the string 
world-sheet.

Let us now consider two possibilities of curing this problem. First 
of them has been put forward in Ref.~\cite{trug} for the 
case of  
the effective string theory emerging from 
compact QED in $D$ space-time dimensions 
(the so-called confining string theory~\cite{confstr}, which 
will be considered in details in Section 4). There, 
it has been demonstrated that 
for the case $D\to\infty$, the 
correlation function of two transversal fluctuations of the string 
world-sheet in this theory 
has an oscillatory behaviour at large distances. Such a 
behaviour indicates that the world-sheet is smooth rather than crumpled. 
One might expect that the same mechanism works in all gauge theories, 
whose confining phases admit a representation in the form of some 
effective string theory with a non-local interaction between the 
world-sheet elements. However, it is not obvious whether this 
mechanism can be extended to the non-Abelian case of 
gluodynamics, where in the nonlocal string 
effective action~(\ref{nonloc}) 
instead of the propagator of the massive vector field, 
appearing in the Abelian case,  
stands the coefficient function $D(x)$. 
Though according to the lattice data~\cite{di, digiac}, 
the large distance 
asymptotic behaviour of the latter is indeed similar to the one 
of the massive vector field propagator, there nevertheless remain 
significant differences.

Let us therefore turn ourselves to the second possibility, proposed in 
Ref.~\cite{crump} and 
elaborated on  
in Ref.~\cite{inst}, which is more applicable to gluodynamics. 
It is based on the introduction of the 
so-called topological term, which is 
equal to the algebraic number of self-intersections of the string 
world-sheet, into the string effective action. 
Then, by adjusting the coupling constant of this term, one 
can eventually 
arrange the cancellation of contributions to the string 
partition function 
coming from highly crumpled surfaces, whose intersection numbers differ  
by one from each other. 

Thus it looks natural to address the problem 
of a derivation of the topological term 
from the gluodynamics Lagrangian. 
Such a term has been recently derived in Ref.~\cite{diamant}  
for $4D$ compact QED with an additional $\theta$-term. In the dual 
formulation of the Wilson loop in this theory (which is nothing else 
but the $4D$ confining string theory mentioned above), 
the latter one occurred to 
be crucial for the formation of the topological string term. 
However, such 
a mechanism of generation of a topological term is difficult to 
work out in gluodynamics due to 
our inability to construct the exact dual formulation of the 
Wilson loop in 
this theory. Therefore, it looks suggestive to seek for some 
model of the gluodynamics vacuum, which might lead to the 
appearance of the topological term in the string representation of the 
Wilson loop in this theory. 

In Ref.~\cite{inst}, this idea has been realized by making use 
of recent results concerning the evaluation of the         
field strength correlators in the dilute 
instanton gas model~\cite{mull}. In the latter paper, it 
has been demonstrated that for the case of an instanton gas with broken 
$CP$-invariance, the bilocal field strength correlator contains 
a term proportional to the tensor $\varepsilon_{\mu\nu\lambda
\rho}$. This term is absent in the case of a $CP$-symmetric vacuum, 
since it is proportional to the topological charge of the system, 
$V\left(n_4-\bar n_4\right)$, 
where $V$ is the four-volume of observation, and within the notations of 
Ref.~\cite{mull},  
$n_4$ and $\bar n_4$ stand for the densities 
of instantons and antiinstantons ($I$'s and $\bar I$'s for brevity),  
respectively. 
Similarly, the paper~\cite{inst}  also dealt with the approximation of a 
dilute $I-\bar I$ gas with  
fixed equal sizes $\rho$ of $I$'s and $\bar I$'s. In the remaining 
part of this Subsection, we shall briefly consider the main points of 
this paper.

The new structure arising in the bilocal correlator in the 
$I-\bar I$ gas reads~\cite{mull}

\begin{equation}
\label{delta}
\Delta{\,}{\rm tr}\left<\left<F_{\mu\nu}(x, x_0)F_{\lambda\rho}
(x', x_0)\right>\right>=
8\left(n_4-\bar n_4\right)I_r\left(\frac{(x-x')^2}{\rho^2}
\right)\varepsilon_{\mu\nu\lambda\rho}. 
\end{equation}
In Eq.~(\ref{delta}), 
the asymptotic behaviour of the function $I_r\left(z^2\right)$ 
at $z\ll 1$ and $z\gg 1$ has the following form  

\begin{equation}
\label{sm}
I_r\left(z^2\right)\longrightarrow \frac{\pi^2}{6}, 
\end{equation}
and 

\begin{equation}
\label{la}
I_r\left(z^2\right)\longrightarrow\frac{2\pi^2}{|z|^4}
\ln z^2, 
\end{equation}
respectively. 

In what follows, we are going to present the leading term in the 
derivative expansion of the correction 
to the nonlocal string effective action~(\ref{nonloc}), which has the 
form 

\begin{equation}
\label{dels} 
\Delta S_{\rm eff.}=-\ln \Delta\left<W\left(\Sigma_{\rm min.}
\right)\right>.
\end{equation} 
Here, 
$\Delta\left<W\left(\Sigma_{\rm min.}\right)
\right>$ is the corresponding correction to the 
expression~(\ref{loop}) for the Wilson loop, following from 
Eq.~(\ref{delta}) in the 
$CP$-broken vacuum. 

We shall not 
be interested in calculating corrections to the 
Nambu-Goto and rigidity terms 
arising due to 
additional contribution from the $I-\bar I$ gas 
to the function $D(x)$. 
This can be easily done by carrying out the 
corresponding integrals~(\ref{tens}) and (\ref{ridconst})  
of the function $D(x)$ 
in this gas. 
Notice only that, as it has already been mentioned in Ref.~\cite{inst}, 
due to the reasons discussed in details in Refs.~\cite{gas} and 
\cite{ufn}, 
a correction to the string tension~(\ref{tens}), obtained 
in such a way from the $I-\bar I$ gas contribution to the function 
$D(x)$, should be cancelled by the contributions coming 
from the higher cumulants in this gas. Clearly, this does not 
necessarily mean 
that the corresponding correction to the rigid string inverse bare 
coupling constant~(\ref{ridconst}) vanishes. Indeed, one can imagine 
himself a function $D(x)$, for which $\int\limits_0^\infty dt
D(t)=0$, whereas $\int\limits_0^\infty dttD(t)\ne 0$, where $t=z^2$. 
For example, this is true for the function $D(t)$ defined as 

$$D(t)=\frac{1}{2c^3},{\,}0<t<c;~ D(t)=-\frac{1}{t^3},{\,}t>c,$$
which can obviously be made continuous by 
smoothering   
the jump at $t=c$ with some function, odd {\it w.r.t.} the line $t=c$. 

One can now expand the 
correction~(\ref{dels}), emerging from the term (\ref{delta}), 
in powers of 
the derivatives {\it w.r.t.} $\xi^a$'s  
in the same manner, as it has been 
done above for the nonlocal string effective action~(\ref{nonloc}).    
Noting that an analogue of the 
Nambu-Goto term in this expansion vanishes 
since $\varepsilon_{\mu\nu\lambda\rho} t_{\mu\nu}t_{\lambda\rho}=0$, 
we obtain 

$$\Delta S_{\rm eff.}=
\beta\nu+{\cal O}\left(\frac{\rho^6\left(n_4-\bar n_4\right)}
{R^2}\right).$$
Here, 

$$\nu\equiv\frac{1}{4\pi}\varepsilon_{\mu\nu\lambda\rho}
\int d^2\xi\sqrt{g}g^{ab}(\partial_a t_{\mu\nu})(\partial_b 
t_{\lambda\rho})$$ 
is the algebraic number of self-intersections of the string 
world-sheet and 

\begin{equation}
\label{beta}
\beta=16\pi\rho^4\left(n_4-\bar n_4\right)\int d^2zz^2 I_r\left(
z^2\right) 
\end{equation}
is the corresponding coupling constant. 

Note that 
the averaged separation 
between the nearest neighbors in the $I-\bar I$ gas  
is given by $L=\left(n_4+\bar n_4\right)^{-\frac14}$. 
According to phenomenological considerations one obtains for the 
$SU(3)$-case, $\rho/L\simeq 1/3$~\cite{shur} 
(see also Ref.~\cite{chu}, where the ratio $\rho/L$ 
has been obtained from direct 
lattice measurements to be $0.37-0.40$). 
The parameter $L$ should then serve as a distance cutoff in the 
integral standing on the R.H.S. of Eq.~(\ref{beta}). 
Taking this into account, we get from Eqs.~(\ref{sm}),~(\ref{la}), 
and~(\ref{beta})   
the following approximate value of $\beta$ 

\begin{equation}
\label{beta1}
\beta\simeq(2\pi\rho)^4\left(n_4-\bar n_4\right)\left[\frac{1}{12}+
\left(\ln\frac{L^2}{\rho^2}\right)^2\right]. 
\end{equation}
Noting that the second term in square brackets on the R.H.S. of 
Eq.~(\ref{beta1}), 
emerging due to Eq.~(\ref{la}),  
is much larger than the first one, emerging due to Eq.~(\ref{sm}), and 
making use of the value $\rho^{-1}\simeq 0.6{\,}{\rm GeV}$~\cite{mitya}, 
one obtains $\beta\simeq 57680.4 \mbox{GeV}^{-4}\cdot 
\left(n_4-\bar n_4\right)$.

In conclusion of this Subsection, 
we have found that in the $I-\bar I$ gas with a 
nonzero topological charge, there appears a topological term in the 
string representation of the Wilson loop. The coupling 
constant of this term is given by Eq.~(\ref{beta}). 
Together with the Nambu-Goto and 
rigidity terms (see Eq.~(\ref{effact})), this term forms the effective 
Lagrangian of the gluodynamics string following from the MFC.

\subsection{Incorporation of Perturbative Corrections}

As it has been discussed in the beginning of the previous Subsection, 
the origin of the nonlocal string effective 
action~(\ref{nonloc}), which served 
as a starting point for the derivation of the local 
action~(\ref{effact}), is essentially nonperturbative. 
This is because the dominant contribution to the 
coefficient function $D(x)$ 
is brought about by the strong background fields, which ensure 
confinement. However, as it has been argued in Ref.~\cite{nuovo}, 
in order to get the exponential growth of the multiplicity of states 
in the spectrum of the open bosonic string, one must account for the 
perturbative gluons, which interact with the string world-sheet. 
In the present Subsection, 
we shall proceed with studying this interaction   
by making use 
of the so-called perturbation theory in the nonperturbative QCD 
vacuum~\cite{temp, spring, stoch4}. 

To this end, we shall split the total field $A_\mu^a$ as 
$A_\mu^a=B_\mu^a+a_\mu^a$, where $B_\mu^a\sim\frac{1}{g}$ 
is a strong nonperturbative 
background, and $a_\mu^a$'s 
are perturbative fluctuations around the latter, 
$a_\mu^a\sim gB_\mu^a$. Thus, our strategy should be to perform an 
integration  
over $a_\mu^a$'s in the expression~(\ref{sto}) for the Wilson loop, 
where $\Sigma$ is 
replaced by $\Sigma_{\rm min.}$. However due to the path-ordering, which 
remained after rewriting the contour 
integral as a surface one 
in the version of the non-Abelian Stokes 
theorem proposed in Refs.~\cite{stok, stokes1},
such an integration is 
difficult to carry out starting with Eq.~(\ref{sto}). That is why, we 
find it convenient to adopt another version of this theorem, proposed 
in Ref.~\cite{diak}, where the path-ordering is replaced by the 
integration over an auxiliary field from the $SU(N_c)/
\left[U(1)\right]^{N_c-1}$ coset space. In what follows, we shall 
consider the $SU(2)$-case, where this field is a unit three-vector 
${\bf n}$, which characterizes the instant orientation in colour 
space, and the non-Abelian Stokes theorem takes a remarkably simple form

$$
\left<W(C)\right>=\Biggl<\int D{\bf n}\exp\left\{\frac{iJ}{2}\left[-g
\int\limits_{\Sigma_{\rm min.}}^{}d\sigma_{\mu\nu}(x(\xi))
n^a(x(\xi)){\cal F}_{\mu\nu}^a(x(\xi))+\right.\right.$$

\begin{equation}
\label{diakon}
\left.\left.+\int\limits_{\Sigma_{\rm min.}}^{}
d\sigma_{\mu\nu}\varepsilon^{abc}n^a\left({\cal D}_\mu{\bf n}\right)^b
\left({\cal D}_\nu{\bf n}\right)^c\right]\right\}\Biggr>.
\end{equation}
Here, ${\cal F}_{\mu\nu}^a=\partial_\mu A_\nu^a-\partial_\nu A_\mu^a+
g\varepsilon^{abc}A_\mu^bA_\nu^c$ is a strength tensor of the gauge field, 
${\cal D}_\mu^{ab}=\partial_\mu\delta^{ab}-g\varepsilon^{abc}
A_\mu^c$ is the covariant derivative, and $J=\frac12, 1,\frac32,\ldots$ 
is the colour ``spin'' of representation of the $SU(2)$ group under 
consideration, defined via its generators $T^a$'s as $T^aT^a=
J(J+1)\hat 1$. 
The last term in the exponent on the R.H.S. of Eq.~(\ref{diakon}) 
is usually referred to as a gauged Wess-Zumino term.

As we shall see below, such a version of the non-Abelian Stokes theorem 
will indeed enable us to carry out the one-loop integration over 
perturbative fluctuations, which will then lead to a new 
({\it w.r.t.} Eq.~(\ref{nonloc})) type of   
interaction between the string world-sheet elements. 
Once being expanded 
in powers of the derivatives {\it w.r.t.} 
$\xi^a$'s, this interaction will 
finally yield 
a correction to the rigidity term, keeping the Nambu-Goto term untouched.  

Let us start with the above mentioned splitting of the total gauge field 
$A_\mu^a$ in Eq.~(\ref{diakon}), which owing to the background field 
formalism~\cite{abbot, temp, spring} yields 

$$\left<W\left(C\right)\right>={\cal N}
\int DB_\mu^a\eta\left(B_\alpha^b\right)
D{\bf n}\exp\Biggl\{\int d^4x\Biggl[-\frac{1}{4}\left(F_{\mu\nu}^a
\right)^2+
\frac{iJ}{2}\int\limits_{\Sigma_{\rm min.}}^{} 
d\sigma_{\mu\nu}\varepsilon^{abc}n^a\left(D_\mu{\bf n}
\right)^b\left(D_\nu{\bf n}\right)^c\Biggr]\Biggr\}\times$$

$$\times\exp\Biggl(-\frac{igJ}{2}\int\limits_{\Sigma_{\rm min.}}^{} 
d\sigma_{\mu\nu}n^a F_{\mu\nu}^a
\Biggr)\times$$

$$\times\int Da_\mu^a\exp\Biggl\{\int d^4x\Biggl[a_\nu^a D_\mu^{ab}
F_{\mu\nu}^b+\frac{1}{2}a_\mu^a\left(D_\rho^{ac}D_\rho^{cb}
\delta_{\mu\nu}-2ig\hat F_{\mu\nu}^{ab}\right)a_\nu^b
-g\varepsilon^{acd}\left(D_\mu^{ab}a_\nu^b\right)a_\mu^c a_\nu^d+$$

$$
+\frac{g^2}{4}\left(a_\mu^aa_\nu^aa_\mu^ba_\nu^b-\left(a_\mu^a\right)^2
\left(a_\nu^b\right)^2\right)\Biggr]-igJ\int\limits_{\Sigma_{\rm min.}}^{}
d\sigma_{\mu\nu}\Biggl(n^aD_\mu^{ab}a_\nu^b+a_\nu^a\left(D_\mu{\bf n}
\right)^a\Biggr)
\Biggr\}\times$$

\begin{equation}
\label{pertloop}
\times\int D\bar\theta^a D\theta^a\exp\Biggl(-\int d^4x\bar\theta^a
D_\mu^{ac}{\cal D}_\mu^{cb}\theta^b\Biggr).
\end{equation}
Here, $F_{\mu\nu}^a$ and $D_\mu^{ab}$ are the background field 
strength tensor and the corresponding 
covariant derivative, defined identically to ${\cal F}_{\mu\nu}^a$ and 
${\cal D}_\mu^{ab}$ with the replacement $A_\mu^a\to B_\mu^a$. Secondly, 
in order to avoid double counting of fields during the integration 
and perform the averages 
over the background fields and quantum fluctuations separately, 
we have used the so-called 
't Hooft identity~\cite{spring}

$$\int DA_\mu^af\left(A_\alpha^b\right)=\frac{\int DB_\mu^a \eta\left(
B_\alpha^b\right)\int Da_\mu^af\left(B_\alpha^b+a_\alpha^b
\right)}{\int DB_\mu^a \eta\left(
B_\alpha^b\right)},$$
valid for an arbitrary functional $f$. Here, an integration weight 
$\eta\left(B_\alpha^b\right)$ should be fixed by the demand that all 
the cumulants and the string tension of the Nambu-Goto term acquire 
their observed values.
Notice also that in a derivation of Eq.~(\ref{pertloop}) we 
have adopted the background Feynman gauge, 
${\cal L}_{\rm gauge{\,} fix.}=-\frac12\left(D_\mu^{ab}a_\mu^b\right)^2$, 
and denoted $\hat F_{\mu\nu}^{ab}\equiv F_{\mu\nu}^c\left(T^c
\right)^{ab}$ with $(T^a)^{bc}=-i\varepsilon^{abc}$.

It should be commented that the term 

\begin{equation}
\label{term}
-igJ\int\limits_{\Sigma_{\rm min.}}^{}d\sigma_{\mu\nu}
a_\nu^a\left(D_\mu{\bf n}\right)^a
\end{equation}
in Eq.~(\ref{pertloop}) emerged from the expansion of the Wess-Zumino 
term as a result of the following sequence of transformations 

$$-\frac{igJ}{2}\int\limits_{\Sigma_{\rm min.}}^{}
d\sigma_{\mu\nu}\varepsilon^{abc}n^an^d\left[\varepsilon^{bde}
a_\mu^e\left(D_\nu{\bf n}\right)^c+\varepsilon^{cde}a_\nu^e
\left(D_\mu{\bf n}\right)^b\right]=$$

$$
=igJ\int\limits_{\Sigma_{\rm min.}}^{}d\sigma_{\mu\nu}
a_\nu^a\left[n^an^b\left(D_\mu{\bf n}\right)^b-\left(D_\mu{\bf n}
\right)^a\right]=
-igJ\int\limits_{\Sigma_{\rm min.}}^{}d\sigma_{\mu\nu}
a_\nu^a\left(D_\mu{\bf n}\right)^a,$$
where in the last equality we have used the facts that 
${\bf n}^2=1$ and $\varepsilon^{bcd}n^bn^c=0$. It is also worth 
mentioning that the terms quadratic in quantum fluctuations, which emerge 
from the expansion of the field strength tensor ${\cal F}_{\mu\nu}^a$ 
and the Wess-Zumino term, cancel each other. 

In what follows, we shall 
work in the one-loop approximation, and thus disregard in 
Eq.~(\ref{pertloop}) the terms 
cubic and quartic in quantum fluctuations, as well as the ghost term.   
For simplicity, we shall also neglect the interaction of two 
perturbative gluons with the field strength tensor 
$F_{\mu\nu}^a$ (gluon spin term). Notice that  
within the Feynman-Schwinger 
proper time path-integral representation for the perturbative 
gluon propagator, which will be used immediately below, such a term 
leads to insertions of the colour magnetic moment into the contour 
of integration (see Ref.~\cite{temp}). 

Bringing now together the term~(\ref{term}) and the term  
$-igJ\int\limits_{\Sigma_{\rm min.}}^{}d\sigma_{\mu\nu}n^a
D_\mu^{ab}a_\nu^b$ 
from Eq.~(\ref{pertloop}) and performing Gaussian integration 
over perturbative fluctuations, we obtain

$$\left<W(C)\right>=$$

\begin{equation}
\label{pertl}
=\left<\left<\exp\left(-\frac{igJ}{2}\int
\limits_{\Sigma_{\rm min.}}^{}d\sigma_{\mu\nu}n^aF_{\mu\nu}^a\right)
\exp\left[-\frac{(gJ)^2}{2}\int
\limits_{\Sigma_{\rm min.}}^{}d\sigma_{\mu\nu}(x)\int
\limits_{\Sigma_{\rm min.}}^{}d\sigma_{\rho\nu}(x')K_{\mu\rho}(x,x')
\right]\right>_{\bf n}\right>_{B_\mu^a}.
\end{equation}
Here, 

$$\left<\ldots\right>_{\bf n}\equiv \int D{\bf n}\left(\ldots\right)
\exp\Biggl[
\frac{iJ}{2}\int\limits_{\Sigma_{\rm min.}}^{} 
d\sigma_{\mu\nu}\varepsilon^{abc}n^a\left(D_\mu{\bf n}
\right)^b\left(D_\nu{\bf n}\right)^c\Biggr],$$
 
$$\left<\ldots\right>_{B_\mu^a}\equiv {\cal N}\int DB_\mu^a\left(
\ldots\right)
\eta\left(B_\alpha^b\right)\exp\Biggl[-\frac14\int d^4x\left(
F_{\mu\nu}^a\right)^2\Biggr],$$
and the (generally speaking, non-translation-invariant) 
interaction kernel $K_{\mu\rho}(x,x')$ is expressed via the perturbative 
gluon propagator as follows

\begin{equation}
\label{kernel}
K_{\mu\rho}(x,x')=
\frac{\partial^2}{\partial x_\mu\partial x_\rho'}n^b(x)n^c(x')
\int\limits_0^{+\infty}ds\int Dz{\rm e}^{-\frac14\int
\limits_0^s\dot z^2 d\lambda}\left[P{\,}\exp\left(ig\int\limits_0^s
d\lambda\dot z_\alpha B_\alpha\right)\right]^{bc},
\end{equation}
where $z(0)=x'$, $z(s)=x$. In the derivation of Eq.~(\ref{pertl}), we 
have neglected the interaction of the string world-sheet with the 
background sources of the type $D_\mu^{ab}F_{\mu\nu}^b(y)$, where $y$ 
is an arbitrary space-time point outside the world-sheet, which should 
be finally integrated over. 

In order to derive a correction, emerging due to exchanges by 
perturbative gluons, to the background-induced gluodynamics 
string effective action~(\ref{nonloc}),  
let us apply to Eq.~(\ref{pertl}) the following 
formula~\cite{van}

$$
\left<{\rm e}^AB\right>=\left<{\rm e}^A\right>\left(\left<B\right>+
\sum\limits_{n=1}^{\infty}\frac{1}{n!}\left<\left<A^nB\right>\right>
\right),
$$
where $A$ and $B$ stand for two commuting operators, and $\left<\ldots 
\right>$ is an arbitrary average. Then, the leading correction, we are 
interested in, corresponds to the complete neglection of  
correlations between the arguments of the first and second exponential 
factors  
standing on the R.H.S. of Eq.~(\ref{pertl}). Secondly, it corresponds 
to putting the ${\bf n}$- and $B_\mu^a$-averages 
of the second exponential factor  
inside it. Taking all this into 
account and following our definition of the string effective action 
$S_{\rm eff.}$ as $-\ln\left<W(C)\right>$, we obtain

\begin{equation}
\label{newact}
S_{\rm eff.}=-\ln\left<\left<\exp\left(
-\frac{igJ}{2}\int
\limits_{\Sigma_{\rm min.}}^{}d\sigma_{\mu\nu}n^aF_{\mu\nu}^a\right)
\right>_{\bf n}\right>_{B_\mu^a}+\Delta S_{\rm eff.},
\end{equation}
where the desired leading correction to the string effective action 
reads 

\begin{equation}
\label{lead}
\Delta S_{\rm eff.}= 
\frac{(gJ)^2}{2}\int
\limits_{\Sigma_{\rm min.}}^{}d\sigma_{\mu\nu}(x)\int
\limits_{\Sigma_{\rm min.}}^{}d\sigma_{\rho\nu}(x')\left<\left<K_{\mu
\rho}(x,x')
\right>_{\bf n}\right>_{B_\mu^a}.
\end{equation}
Clearly, in the bilocal approximation, the first term on the R.H.S. 
of Eq.~(\ref{newact}) yields pure background part of the 
string effective action~(\ref{nonloc}) ({\it cf.} Eq.~(\ref{diakon}) 
with $A_\mu^a$ replaced by $B_\mu^a$).  

The obtained correction~(\ref{lead}) to the pure background string 
effective action~(\ref{nonloc}) corresponds to a new type 
of interaction between the string world-sheet elements. Namely, 
instead of the propagator of the background gluon, represented by 
the function $D(x)$, in Eq.~(\ref{lead}) stands a propagator of the 
perturbative gluon in the nonperturbative gluodynamics vacuum. Due to 
the statistical weight ${\rm e}^{-\frac14\int
\limits_0^s\dot z^2 d\lambda}P{\,}\exp\left(ig\int\limits_0^s
d\lambda\dot z_\alpha B_\alpha\right)$ of this gluon, it is the region 
where $s$ is small, which mainly contributes to the interaction 
kernel~(\ref{kernel}). This means that the dominant contribution 
to the obtained correction~(\ref{lead}) stems from those points 
$x_\mu(\xi)$ and $x_\mu(\xi')$ of the world-sheet, which are very 
close to each other. That is in line with the performed derivative 
expansion of the nonlocal string effective action~(\ref{nonloc}), 
where $|x-x'|\le T_g\ll R$. Taking this into account, 
we can adopt the simplest, local, approximation for the propagator 
$\left<n^b(x)n^c(x')\right>_{\bf n}$, {\it i.e.}, replace it by 

$$
\frac{\delta^{bc}}{3}
\int D{\bf n}
\exp\Biggl[
\frac{iJ}{2}\int\limits_{\Sigma_{\rm min.}}^{} 
d\sigma_{\mu\nu}\varepsilon^{def}n^d\left(D_\mu{\bf n}
\right)^e\left(D_\nu{\bf n}\right)^f\Biggr].
$$
This expression is a functional of the world-sheet as a whole 
({\it i.e.}, it is independent of $x_\mu(\xi)$) 
and therefore can be absorbed 
into the irrelevant normalization constant ${\cal N}$.  

Finally, in order to perform an expansion of the nonlocal 
correction~(\ref{lead}) in powers of the derivatives {\it w.r.t.} 
$\xi^a$'s 
and derive from it the first few local terms, it is convenient 
to pass to the integration over the trajectories $u_\mu(\lambda)=
z_\mu(\lambda)+\frac{\lambda}{s}(x'-x)_\mu-x_\mu'$. This enables one to 
extract explicitly the dependence on the points $x_\mu$ and $x_\mu'$ 
from the integral over trajectories (necessary for the differentiation 
{\it w.r.t.} these points), which yields

$$
K_{\mu\rho}(x,x')=\frac{\partial^2}{\partial x_\mu\partial x_\rho'}
\int\limits_0^{+\infty}ds{\rm e}^{-\frac{(x-x')^2}{4s}}\int 
Du
{\rm e}^{-\frac14\int
\limits_0^s\dot u^2 d\lambda}\times$$

$$\times {\rm tr}{\,}P{\,}\exp
\left[ig\int\limits_0^s
d\lambda\left(\frac{x-x'}{s}+\dot u\right)_\alpha B_\alpha
\left(u+x'+\frac{\lambda}{s}(x-x')\right)\right]$$
with $u(0)=u(s)=0$. Then, in the bilocal approximation, the dominant 
contribution to the Nambu-Goto and rigidity terms comes about  
from taking the derivatives of the free propagation factor 
${\rm e}^{-\frac{(x-x')^2}{4s}}$ 
and replacing the parallel transporter factor by the one over the 
closed path, which has the form 
${\rm tr}{\,}P{\,}\exp
\left[ig\int\limits_0^s d\lambda\dot u_\alpha B_\alpha (u)\right]$. 
Finally, substituting the so-obtained expression for $K_{\mu\rho}(x,x')$ 
into Eq.~(\ref{lead}) and performing an expansion of this nonlocal 
correction in powers of the derivatives {\it w.r.t.} 
$\xi^a$'s similarly to the previous Subsection, 
we arrive at the desired correction to the local effective 
action~(\ref{effact}). In this way, it turns out that the Nambu-Goto 
term string tension does not acquire any correction due to perturbative 
gluonic exchanges, whereas the correction to the inverse bare coupling 
constant of the rigidity term reads

\begin{equation}
\label{deltarid}
\Delta\frac{1}{\alpha_0}=-\frac{\pi (gJ)^2}{3}\int\limits_0^{+\infty}
dss\int Du {\rm e}^{-\frac14\int
\limits_0^s\dot u^2 d\lambda}\left<
{\rm tr}{\,}P{\,}\exp
\left[ig\int\limits_0^s d\lambda\dot u_\alpha B_\alpha (u)\right]
\right>_{B_\mu^a}.
\end{equation}
Note that since we have proved in the previous Subsection 
that the rigidity term 
vanishes at the surface of the minimal area, $\Sigma_{\rm min.}$, 
the obtained 
result means that accounting for perturbative gluons in the 
lowest order of perturbation theory does not change the 
background-induced string effective action~(\ref{effact}) associated 
with this surface. However, the derived correction~(\ref{deltarid}) 
is essential for the action~(\ref{effact}) corresponding to the 
statistical weight $\left<W(\Sigma)\right>$ in the integral over 
world-sheets standing on the R.H.S. of Eq.~(\ref{complete}). 

It is also worth noting that since $B_\mu^a\sim\frac{1}{g}$, the 
parallel transporter factor on the R.H.S. of Eq.~(\ref{deltarid}) 
cannot be expanded in powers of $g$ and should be considered as a whole. 
The path-integral on the R.H.S. of Eq.~(\ref{deltarid}) is not simply 
the perturbative gluon propagator, since the integral over the 
proper time contains an additional power of $s$, which makes the 
whole quantity dimensionless, as it should be. Notice also 
that the sign of  
the obtained correction~(\ref{deltarid}) to the inverse bare coupling 
constant~(\ref{ridconst}) depends on the form of 
background $B_\mu^a$ entering the Wilson loop (which is the most 
nontrivial content of this correction) on the R.H.S. of 
Eq.~(\ref{deltarid}). However, since the derived correction 
is a pure perturbative effect (due to the 
factor $g^2$ present in Eq.~(\ref{deltarid})), even in the case when 
it is positive, it cannot change the negative sign of the leading 
term~(\ref{ridconst}).

\subsection{A Hamiltonian of the Straight-Line QCD String with 
Spinless Quarks}

In the present Subsection, we shall derive a Hamiltonian 
corresponding to the quark-antiquark Green function~(\ref{fourpoint}) in 
the confining QCD vacuum. 
To this end, we shall first write it down in the Feynman-Schwinger 
proper time path-integral representation~(\ref{green}), after which 
substitute for $\left<W(C)\right>$ the above obtained expression 
$\exp\left(-S_{\rm eff.}\right)$ with $S_{\rm eff.}$ defined by 
Eq.~(\ref{effact}) (and inverse bare coupling constant of the 
rigidity term~(\ref{ridconst}) modified by Eq.~(\ref{deltarid}), 
{\it i.e.},  
$\frac{1}{\alpha_0}\to\frac{1}{\alpha_0}+\Delta\frac{1}{\alpha_0}$). 
It is worth noting that for the first time 
such a Hamiltonian has been derived in Ref.~\cite{dub}, where, however,  
only the Nambu-Goto term in the string effective action~(\ref{effact}) 
has been accounted for. Our aim below will be the derivation of a 
correction to this result due to the rigidity term, which we assume 
to be nonvanishing. That means that the QCD string 
in the problem under 
study is excited, since its world-sheet differs from the one of the 
minimal area. As a byproduct, 
we shall also rederive the leading terms in the Hamiltonian of the 
QCD string with quarks obtained in Ref.~\cite{dub}. 
However as we shall see, the rigid string theory, being the theory 
with higher derivatives, leads to the interesting and phenomenologically 
relevant modifications of the Nambu-Goto--induced 
part of the Hamiltonian.  
We shall also generalize the result of Ref.~\cite{dub} by considering 
the case of different masses of a quark and antiquark.  

The two main approximations, under which we shall consider 
the Green function~(\ref{green}), are the same as the ones used 
in Ref.~\cite{dub}. First, we shall neglect quark trajectories  
with backward motion in the proper time, which 
might lead to creation of additional quark-antiquark pairs. Secondly,  
we shall use the straight-line
approximation for the string world-sheet $\Sigma[C]$, 
which, as it has been argued in Ref.~\cite{dub},
corresponds to the valence quark approximation. Such a string
may rotate and oscillate longitudinally. This approximation is inspired by
two limiting cases, $l=0$ and $l\to\infty$, where $l$ is the orbital 
quantum number of the system. 
The first case
will be investigated below in more details, and 
the correction to the Hamiltonian of the relativistic quark 
model~\cite{cea}
due to the rigidity term in the limit of large masses of a quark 
and antiquark will be derived. 

Let us now proceed with a derivation of the desired Hamiltonian. 
To this end, we shall start with the expression for the 
Green function~(\ref{green}) with $\left<W(C)\right>$ 
defined via Eq.~(\ref{effact}) and

$$K\equiv m_1^2s+\frac14\int\limits_0^s \dot z^2d\lambda,~ 
\bar K\equiv m_2^2\bar s+\frac14\int\limits_0^{\bar s}
\dot{\bar z}^2d\lambda$$
standing for the kinematical factors of a quark and antiquark, 
respectively. Then, by making use of the auxiliary field 
formalism~\cite{polbook}, one can represent it 
in the following form 

$$G\left(x, \bar x; y, \bar y\right)=
\int\limits_0^{+\infty}dT
\int D{\bf z} D\bar{\bf z} D\mu_1D\mu_2Dh_{ab}
\exp\left(-K^\prime-\bar K^\prime\right)
\exp\biggl[\left(-\sigma+2\bar\alpha
\right)\int
 d^2\xi\sqrt{h}{\,}\biggr]
\times$$

\begin{equation}
\label{green1}
\times \exp\Biggl[-\bar\alpha\int d^2\xi 
\sqrt{h} h^{ab} (\partial_a w_\mu)
(\partial_b w_\mu)-\frac{1}{\alpha_0}\int d^2\xi \sqrt{h}h^{ab}
(\partial_a t_{\mu\nu})
(\partial_b t_{\mu\nu})\Biggr],
\end{equation}
where we have integrated over the Lagrange multiplier $\lambda^{ab}(\xi)=
\alpha(\xi)h^{ab}(\xi)+f^{ab}(\xi), f^{ab}h_{ab}=0$, and $\bar \alpha$
is the mean value of $\alpha(\xi)$.
Here $t_{\mu\nu}=\frac{1}{\sqrt{h}}\varepsilon^{ab}\left(\partial_a
w_\mu\right)\left(\partial_b w_\nu\right)$,

\begin{equation}
\label{kin}
K^\prime+\bar K^\prime=\frac{1}{2}\int\limits_0^T
d\tau\Biggl[\frac{m_1^2}{\mu_1(\tau)}+\mu_1(\tau)\left(1+
\dot{\bf z}^2(\tau)\right)+
\frac{m_2^2}{\mu_2(\tau)}+\mu_2(\tau)\left(1+\dot{\bar{\bf z}}^2(\tau)
\right)\Biggr], 
\end{equation}

$$T=\frac{1}{2}
\left(x_0+\bar x_0-y_0-\bar y_0\right),~  
\mu_1(\tau)=\frac{T}{2s}\dot z_0(\tau),~ 
\mu_2(\tau)=\frac{T}{2\bar s}\dot{\bar z}_0(\tau),$$ 
and the no-backtracking
time approximation~\cite{dub}

\begin{equation}
\label{mu1}
\mu_1(\tau)>0,~ \mu_2(\tau)>0 
\end{equation}
has been used. 
Similarly to Ref.~\cite{dub}, 
we employ in the valence quark sector (\ref{mu1}) the
approximation that the string world-sheet $\Sigma[C]$ 
can be parametrized by the
straight lines, connecting 
points $z_\mu(\tau)$ and $\bar z_\mu(\tau)$ with
the same $\tau$, {\it i.e.}, 
the trajectories of a quark and antiquark are
synchronized, 
$z_\mu=(\tau, {\bf z}), \bar z_\mu=(\tau, \bar{\bf z}),
w_\mu(\tau,\beta)=\beta z_\mu(\tau)+(1-\beta)\bar z_\mu(\tau), 0\le\beta
\le 1$.

Introducing auxiliary fields~\cite{dub} 
$\nu(\tau,\beta)=T\sigma\frac{h_{22}}
{\sqrt{h}}, \eta(\tau,\beta)=\frac{1}{T}\frac{h_{12}}{h_{22}}$ and making
a rescaling $z_\mu\to\sqrt{\frac{\sigma}{2\bar\alpha}}z_\mu, \bar z_\mu\to
\sqrt{\frac{\sigma}{2\bar\alpha}}\bar z_\mu$, one gets from the last
exponent on the R.H.S. of Eq.~(\ref{green1}) 
the following action of the string 
without quarks

$$S_{\rm eff.}=\int\limits_0^T d\tau\int\limits_0^1 d\beta\frac{\nu}{2}
\Biggl\{\dot w^2+\biggl(\biggl(\frac{\sigma}{\nu}\biggr)^2+\eta^2\biggr)
r^2-2\eta(\dot w r)+\frac{\sigma T^2}{\alpha_0\bar\alpha^2}
\frac{1}{h}\Biggl[\ddot w^2r^2-(\ddot w r)^2+\dot w^2\dot r^2-
(\dot w\dot r)^2+$$

$$+2((\ddot w\dot w)(\dot r r)-(\ddot w \dot r)(\dot w r))+
\biggl(\biggl(\frac{\sigma}{\nu}\biggr)^2+\eta^2\biggr)\left(\dot r^2 r^2-
(\dot r r)^2\right)-$$

\begin{equation}
\label{noq}
-2\eta\left(
(\ddot w \dot r)r^2-(\ddot w r)(\dot r r)+(\dot w
\dot r)(\dot r r)-(\dot w r)\dot r^2\right)\Biggr]\Biggr\},
\end{equation}
where a dot stands for $\frac{\partial}{\partial\tau}$, and $r_\mu(\tau)=
z_\mu(\tau)-\bar z_\mu(\tau)$ is the relative coordinate.

Let us now introduce the centre of masses coordinate $R_\mu(\tau)=
\zeta(\tau)z_\mu(\tau)+(1-\zeta(\tau))\bar z_\mu(\tau)$, where
$\zeta(\tau)\equiv\zeta_1(\tau)+\frac{1}{\alpha_0}\zeta_2(\tau), 
0\le \zeta(\tau)\le 1$, should be determined from the requirement 
that $\dot R_\mu$ decouples from $\dot r_\mu$. These extremal 
values of $\zeta_1$ and $\zeta_2$ can be obtained from the corresponding 
saddle-point equation. Referring the reader for the details to 
Appendix A, we shall present here only the final result for the 
path-integral Hamiltonian of the straight-line QCD string with quarks.
It has the form

\begin{equation}
\label{fullham}
H=H^{(0)}+\frac{1}{\alpha_0}H^{(1)}.
\end{equation}
Here

\begin{equation}
\label{hamnol}
H^{(0)}=\frac{1}{2}\left[\frac{\left({\bf p}_r{\,}^2+m_1^2\right)}{\mu_1}+
\frac{\left({\bf p}_r{\,}^2+m_2^2\right)}{\mu_2}+\mu_1+\mu_2+\sigma^2
{\bf r}^2\int\limits_0^1\frac{d\beta}{\nu}+\nu_0+
\frac{{\bf L}^2}{\rho {\bf r}^2}\right]
\end{equation}
with

$$\rho= \mu_1+\nu_2-\frac{(\mu_1+\nu_1)^2}{\mu_1+\mu_2+\nu_0},~ 
\nu_i\equiv\int\limits_0^1 d\beta\beta^i \nu,~ {\bf p}_r^2\equiv
\frac{({\bf p}{\bf r})^2}{{\bf r}^2},~ {\bf L}\equiv \left[{\bf r}, 
{\bf p}\right]$$
is the Hamiltonian of the Nambu-Goto string with quarks,
which for the case of equal masses of a quark and an antiquark 
has been derived and investigated in Ref.~\cite{dub}.  
The Hamiltonian
$H^{(1)}$, which is a new one, reads

\begin{equation}
\label{hamodin}
H^{(1)}=\frac{a_1}{\rho^2}{\bf L}^2+
\frac{a_2}{2\tilde\mu^3}\left|{\bf r}\right|\left(
{\bf p}_r^2\right)^{\frac32}+
\frac{a_3}{\tilde \mu^4}\left({\bf p}_r^2\right)^2+
\frac{a_4}{2\tilde\mu\rho^2}\frac{\sqrt{{\bf p}_r^2}{\bf L}^2}
{\left|{\bf r}\right|}+
\frac{a_5}{2\tilde \mu^2 \rho^2}\frac{{\bf p}_r^2{\bf L}^2}
{{\bf r}^2}, 
\end{equation}
where 
$\tilde \mu=\frac{\mu_1\mu_2}{\mu_1+\mu_2}$, and the
coefficients $a_k$'s, $k=1,\ldots,5$, are listed in Appendix A.

We see that the obtained Hamiltonian~(\ref{hamodin}) 
contains not only corrections to the 
orbital momentum of the system, but also several operators higher than 
of the second order in the momentum. The latter ones emerge as a 
consequence of the 
fact that the rigid string theory is a theory with higher derivatives.

The obtained (path-integral) Hamiltonian~(\ref{fullham})-(\ref{hamodin}) 
contains auxiliary fields $\mu_1$,
$\mu_2$ and $\nu$. In order to construct out of it 
the operator Hamiltonian, which acts 
upon the wave functions, one should integrate these fields out. This 
implies the substitution of their extremal values, which can be obtained 
from the corresponding saddle-point equations, into 
Eqs.~(\ref{fullham})-(\ref{hamodin}) and 
performing the Weil ordering~\cite{lee}.

In conclusion of this Subsection, let us 
apply Hamiltonian~(\ref{hamodin}) to a derivation of the rigid string
correction to the Hamiltonian of the so-called 
relativistic quark model~\cite{cea}, 
{\it i.e.}, consider the case when the orbital 
momentum is equal to zero. Let us also 
set for simplicity the mass of a quark equal to the mass 
of an antiquark $m_1=m_2\equiv m$. 
In order to get $H^{(1)}$, one should substitute the extremal values of 
the fields $\mu_1$, $\mu_2$, and $\nu$  
of the zeroth order in $\frac{1}{\alpha_0},~
\mu_1^{\rm extr.}=\mu_2^{\rm extr.}=\sqrt{{\bf p}^2+m^2}$ and 
$\nu_{\rm extr.}=\sigma\left|{\bf r}\right|$,  
into Eq.~(\ref{hamodin}). 
The limit of large masses of a quark and 
antiquark means that $m\gg\sqrt{\sigma}$. In this case,  
we obtain from Eq.~(\ref{hamodin}) 
the rigid string Hamiltonian 
$H^{(1)}=-\frac{4\sigma\left|{\bf r}\right|}
{m^4}\left({\bf p}_r^2\right)^2$ and then from Eqs.~(\ref{fullham}) 
and~(\ref{hamnol}) the following expression 
for the 
total Hamiltonian $H$ 

\begin{equation}
\label{chetyre}
H=2m+\sigma\left|{\bf r}\right|+\frac{{\bf p}^2}{m}-
\left(\frac{1}{4m^3}+\frac{4\sigma\left|{\bf r}\right|}
{\alpha_0 m^4}\right)\left({\bf p}^2\right)^2. 
\end{equation}
One can see that the new rigid string-inspired term, quartic in the 
relative momentum of the quark-antiquark pair, 
may cause a sufficient influence 
to the dynamics of the system.

\section{String Representation of Abelian-Projected Theories}

\subsection{The Method of Abelian Projections}

The MFC, exploited in the previous Section, enabled us to investigate 
the string effective action associated with a certain 
({\it e.g.} minimal) 
world-sheet, but did not yield a prescription of a derivation of the 
full string partition function in the form of an integral over all 
world-sheets~(\ref{complete}). 
The reason for that is that within the MFC one looses the meaning of 
the path-integral average over the gluodynamics vacuum, replacing 
this average by the phenomenological one. As a consequence, it looks 
difficult to extract singularities corresponding to QCD strings out 
of the resulting vacuum correlation functions. 

In the present Section, we 
shall proceed with a derivation of such an integral over world-sheets 
in the QCD-inspired effective Abelian-projected theories and find an exact 
field-theoretical analogue of the phenomenological background gluon 
coefficient function $D(x)$. 
We shall argue that the general 
features of the gluodynamics vacuum 
are in line with the ones predicted by the so-called dual 
Meissner picture of confinement, first put forward by 't Hooft and 
Mandelstam~\cite{mand}. According to the 
't Hooft-Mandelstam scenario, the properties of the QCD string should 
be similar to those of the electric vortex, which emerges 
between two electrically charged particles immersed into a 
superconducting medium filled with a condensate of Cooper pairs of 
magnetic monopoles. In the case of the usual Abelian Higgs Model (AHM), 
which is a relativistic version of the Ginzburg-Landau theory of 
superconductivity, such vortices are referred to as 
Abrikosov-Nielsen-Olesen strings~\cite{ano}. 

Up to now, there does not exist any analytical proof of the existence 
of the condensate of Abelian monopoles in QCD, though in lattice 
QCD there are a lot of numerical data in favour of this 
conjecture (see {\it e.g.} Refs.~\cite{sch, bali, 
over, newover}). However in all theories 
allowing for an analytical description of confinement, the latter one 
is due to monopoles, which either form a gas or condense. 
These examples include 
compact QED and 3D Georgi-Glashow model~\cite{mon, polbook}, 
and supersymmetric Yang-Mills theories~\cite{seib}. 
As far as the origin of magnetic monopoles in QCD is concerned, they 
appear in the 
so-called Abelian projection method~\cite{th, maedan}, which we shall 
briefly describe below. 

The essence of this method is based on a certain 
partial gauge fixing procedure, which reduces the original  
gauge group 
$SU(N_c)$ to the maximal Abelian (or Cartan) 
subgroup $\left[U(1)
\right]^{N_c-1}$, {\it i.e.}, leaves Abelian degrees of freedom unfixed. 
Then, since the original $SU(N_c)$ group is compact, the 
resulting Abelian subgroup is compact as well, which is just the origin 
of Abelian magnetic monopoles in the original non-Abelian theory. 
To perform such a gauge fixing, one chooses a certain 
composite operator $X$, transforming by the adjoint representation 
of $SU(N_c)$, $X\to X'=UXU^\dag$, and diagonalizes it, 
{\it i.e.}, finds such a gauge (unitary matrix $U$), that 
$X'={\rm diag}{\,}
\left(\lambda_1,\ldots,\lambda_{N_c}\right)$. 
As an example of such 
an operator may serve {\it e.g.} $F_{12}(x)$ or, at finite temperatures, 
the so-called Polyakov line~\cite{rein}

$$
P{\,}\exp\left[ig\int\limits_{0}^{\beta}dx_4 A_4 
({\bf x}, x_4)\right],
$$
where $\beta=1/T$ is an inverse temperature. 
Notice that the most interesting 
numerical results 
(see {\it e.g.}~\cite{lat}) 
have been obtained for the $SU(2)$-case 
in the so-called Maximal Abelian gauge, 
defined by the condition of maximization of the functional 

$$
R\left[A_\mu\right]=-\int d^4x\left(\left(A_\mu^1\right)^2+
\left(A_\mu^2\right)^2\right),
$$
which reads $\max\limits_{U}^{}R\left[A_\mu'\right]$. Here, $A_\mu'$ 
is the transformed vector potential following from the original one 
after the gauge transformation with the matrix $U$. The condition 
of the local extremum for this functional, 
$\left(\partial_\mu\pm igA_\mu^{'3}\right)A_\mu^{'\pm}=0$, 
where $A_\mu^{'\pm}=A_\mu^{'1}\pm iA_\mu^{'2}$, means that in this gauge 
we make the field $A_\mu$ as diagonal as possible.

The matrix $U$, which diagonalizes the operator $X$, is obviously 
defined up to a left multiplication by the diagonal $SU(N_c)$
matrix, which belongs to the subgroup $\left[U(1)
\right]^{N_c-1}$. This is just the way how the original non-Abelian 
group reduces to the maximal Abelian subgroup. If we now perform a 
certain Abelian projection, {\it i.e.}, consider the transformed 
vector potential

\begin{equation}
\label{atransf}
A_\mu'=U\left(A_\mu+\frac{i}{g}\partial_\mu\right)U^\dag, 
\end{equation}
then it is straightforward to see 
that under the remained maximal Abelian subgroup, the 
diagonal field components of the matrix-valued vector potential 
$a_\mu^i\equiv\left(A_\mu'\right)^{ii}$, 
$i=1,\ldots,N_c$, transform as Abelian gauge fields, 
$a_\mu^i\to a_\mu^i+\frac{1}{g}\partial_\mu\alpha_i$,
whereas the off-diagonal components $c_\mu^{ij}=\left(A_\mu'\right)^{ij}$ 
transform as charged matter fields,  
$c_\mu^{ij}\to\exp\left[i\left(\alpha_i-\alpha_j\right)
\right]c_\mu^{ij}$. The latter ones can be further disregarded 
in a derivation of an effective IR Abelian-projected theory owing 
to the so-called Abelian dominance hypothesis~\cite{abdom}. 
As far as the monopoles ({\it w.r.t.} the ``photon'' fields of 
diagonal elements of the vector potential) are concerned, their 
r\^ole is played by 
the singularities of the matrix $U$, 
which might occur if some of the eigenvalues 
$\lambda_1,\ldots,\lambda_{N_c}$ of 
the operator $X$ coincide. This may happen 
in some 3D point ${\bf x}_0$, which becomes 
a world-line of magnetic monopole in 4D.

Let us start with illustrating 
the above described ideas at the simplest example 
of the $SU(2)$-gluodynamics. The initial Yang-Mills action reads 

\begin{equation}
\label{ym}
S_{\rm YM}\left[A_\mu^i\right]=\frac12{\rm tr}\int d^4x F_{\mu\nu}^2
\end{equation}
with $F_{\mu\nu}=F_{\mu\nu}^iT^i$. Here,  
$F_{\mu\nu}^i=\partial_\mu A_\nu^i-\partial_\nu A_\mu^i+
g\varepsilon^{ijk}A_\mu^j A_\nu^k$, 
$T^i=\frac{\tau^i}{2}$, $i=1,2,3$, and $\tau^i$'s stand for 
Pauli matrices.

One can perform the gauge transformation~(\ref{atransf}) 
so that the gauge-transformed field $A_\mu'$ obeys {\it e.g.} 
the above-mentioned 
maximal Abelian gauge fixing condition 
$\left(\partial_\mu\pm iga'_\mu\right)A_\mu^{'\pm}=0$, 
where $a'_\mu\equiv A^{'3}_\mu$~\footnote{ 
Notice that the maximal Abelian gauge fixing condition 
can be written as follows 
${\cal D}_\mu^{'ab}A^{'b}_\mu=0$, where 
${\cal D}_\mu^{'ab}=\partial_\mu\delta^{ab}-g\varepsilon^{ab3}
a'_\mu$. Once being rewritten in this form, 
this gauge can be easily recognized as the standard 
background gauge~\cite{abbot}  
with the field $a'_\mu$ playing the r\^ole of the background.}.   
The gauge transformed field strength tensor then reads
$F'_{\mu\nu}=U\left(F_{\mu\nu}+F_{\mu\nu}^{\rm sing.}\right)U^\dag$,
where the singular contribution has the form 
$F_{\mu\nu}^{\rm sing.}=\frac{i}{g}\left(\left[\partial_\mu, 
\partial_\nu\right]U^\dag\right)U$.
This contribution comes about from the generally singular character 
of the matrix $U$ of the gauge transformation, mentioned above, 
and describes world-sheets of the Dirac strings~\cite{kon}. 
Clearly, integration over all possible singular gauge transformations 
results to an integration over $F_{\mu\nu}^{\rm sing.}$'s. 

Let us next single out the diagonal (neutral) component 
$a_\mu\equiv A_\mu^3$ of the field 
$A_\mu$ by making use of the decomposition

$$A_\mu=a_\mu T^3+A_\mu^aT^a\equiv {\cal A}_\mu+{\cal C}_\mu,$$ 
where $a=1,2$. Consequently, 
one has for the field strength tensor 

\begin{equation}
\label{cart}
F_{\mu\nu}\equiv F_{\mu\nu}\left[{\cal A}+{\cal C}\right]=
F_{\mu\nu}\left[{\cal A}\right]+\left(D\left[{\cal A}\right]\wedge{\cal C}
\right)_{\mu\nu}-ig\left[{\cal C}_\mu,{\cal C}_\nu\right], 
\end{equation}
where 

$$\left({\cal O}\wedge{\cal G}\right)_{\mu\nu}\equiv {\cal O}_\mu 
{\cal G}_\nu-{\cal O}_\nu{\cal G}_\mu,~ \mbox{and}~ 
D_\mu\left[{\cal A}\right]
=\partial_\mu-ig\left[{\cal A}_\mu,\cdot\right].$$ 
Eq.~(\ref{cart}) can be straightforwardly rewritten as follows 

$$
F_{\mu\nu}=\left(f_{\mu\nu}+C_{\mu\nu}\right)T^3+S_{\mu\nu}^aT^a.$$
Here, 
$f_{\mu\nu}=\left(\partial\wedge a\right)_{\mu\nu}$ and  
$C_{\mu\nu}=g\varepsilon^{ab3}A_\mu^a A_\nu^b$ stand for the 
contributions of diagonal and off-diagonal components of the gluon field 
to the diagonal part of the field strength tensor, respectively, and 
$S_{\mu\nu}^a=\left({\cal D}^{ab}\wedge A^b\right)_{\mu\nu}$ is the 
off-diagonal part of the field strength tensor with 
${\cal D}_\mu^{ab}=\partial_\mu\delta^{ab}-g\varepsilon^{ab3}
a_\mu$.
This yields the following decomposition of the 
action~(\ref{ym}) 
taken now on the gauge transformed fields, 

$$
S_{\rm YM}\left[A^{'i}_\mu\right]=
\frac14\int d^4x\left(f_{\mu\nu}+C_{\mu\nu}+\left(F_{\mu\nu}^{\rm 
sing.}\right)^3\right)^2 
+\frac14\int d^4x \left(S_{\mu\nu}^a+\left(F_{\mu\nu}^{\rm sing.}
\right)^a\right)^2,
$$
where $\left(F_{\mu\nu}^{\rm sing.}
\right)^i=2{\,}{\rm tr}{\,}\left(T^iF_{\mu\nu}^{\rm sing.}\right)$.

Next, since our aim 
will be the 
investigation of the confining ({\it i.e.}, infrared) 
properties of the Abelian-projected 
$SU(2)$-gluodynamics (rather than the problems of its renormalization, 
related 
to the region of asymptotic freedom), we 
shall disregard the $A_\mu^a$-dependent terms~\cite{abdom}. 
Within this approximation, the resulting 
effective action takes the form

\begin{equation}
\label{et}
S_{\rm eff.}\left[a_\mu, f_{\mu\nu}^{\rm sing.}\right]=
\frac14\int d^4x\left(f_{\mu\nu}+
f_{\mu\nu}^{\rm sing.}
\right)^2, 
\end{equation}
where we have denoted for brevity $f^{\rm sing.}_{\mu\nu}\equiv 
\left(F_{\mu\nu}^{\rm sing.}\right)^3$. The monopole current is defined 
via the modified Bianchi identities as 

$$
j_\nu^M=\partial_\mu\left(\tilde f_{\mu\nu}+\tilde f^{\rm sing.}_{\mu\nu}
\right)=\frac12\varepsilon_{\mu\nu\lambda\rho}\partial_\mu
f^{\rm sing.}_{\lambda\rho}$$
with $\tilde f_{\mu\nu}=\frac12\varepsilon_{\mu\nu\lambda\rho}
f_{\lambda\rho}$. Thus the obtained effective theory~(\ref{et}) 
can be regarded as a $U(1)$ gauge theory with magnetic monopoles. 
To proceed with the investigation of the monopole ensemble, 
it is reasonable to cast 
the partition function under study, ${\cal Z}=\int 
Df_{\mu\nu}^{\rm sing.}
Da_\mu{\rm e}^{-S_{\rm eff.}}$, to the dual form~\footnote{Notice that 
the gauge fixing term for the Abelian field is assumed to be included 
into the integration measure $Da_\mu$.}. This can be done by making use 
of the first-order formalism, {\it i.e.}, linearizing the square 
$f_{\mu\nu}^2$ in Eq.~(\ref{et}) by introducing an integration 
over an auxiliary antisymmetric tensor field $B_{\mu\nu}$ as follows

\begin{equation}
\label{et1}
{\cal Z}=\int Df_{\mu\nu}^{\rm sing.}Da_\mu DB_{\mu\nu}\exp\left\{
-\int d^4x\left[\frac14B_{\mu\nu}^2+\frac{i}{2}\tilde B_{\mu\nu}
f_{\mu\nu}+\frac12 f_{\mu\nu}f_{\mu\nu}^{\rm sing.}+\frac14
\left(f_{\mu\nu}^{\rm sing.}\right)^2\right]\right\}.
\end{equation}
Integration over the $a_\mu$-field leads to the constraint 
$\partial_\mu\left(\tilde B_{\mu\nu}-if_{\mu\nu}^{\rm sing.}\right)=0$,
whose resolution yields $B_{\mu\nu}=i\tilde f_{\mu\nu}^{\rm sing.}
+\left(\partial\wedge B\right)_{\mu\nu}$, where $B_\mu$ is now the 
``magnetic'' potential dual to the ``electric'' potential $a_\mu$.
Substituting this representation for $B_{\mu\nu}$ into Eq.~(\ref{et1}),
we obtain 

\begin{equation}
\label{et2}
{\cal Z}=\left<\int DB_\mu\exp\left\{-\int d^4x\left[\frac14F_{\mu\nu}^2
-iB_\mu j_\mu^M\right]\right\}\right>_{j_\mu^M},
\end{equation}
where from now on $F_{\mu\nu}$ denotes $\left(\partial\wedge
B\right)_{\mu\nu}$. In Eq.~(\ref{et2}), the integration over 
$f_{\mu\nu}^{\rm sing.}$'s has transformed to a certain average over 
the monopole currents, $\left<\ldots\right>_{j_\mu^M}$, whose concrete 
form will be specified immediately below. 

Clearly, the next crucial step necessary to proceed with the summation 
over the ensemble of monopoles is to postulate the properties 
of their ensemble. In a derivation of effective Abelian-projected 
theories, it is usually assumed that those form a condensate of 
their Cooper pairs of the charge $2g_m$, where $g_m$ stands for the 
magnetic coupling constant~\footnote{
Within this assumption, we restrict ourselves to the sector of 
the full yet unknown Abelian-projected theory where there do not 
exist antimonopoles. Possible interference between these two 
sectors requires further investigations.}. 
This assumption is realized firstly by  
setting for the collective current of $N$ magnetic 
Cooper pairs the expression 

$$j_\mu^{M{\,}(N)}(x)=2g_m\sum\limits_{n=1}^{N}\oint dx_\mu^n(s)
\delta\left(x-x^n(s)\right),$$
where the world-line of the $n$-th Cooper pair is parametrized 
by the vector $x_\mu^n(s)$, and secondly by specifying the measure 
$\left<\ldots\right>_{j_\mu^M}$ to the following form~\cite{bard, 
for, zinn} 

$$
\left<\exp\left(i\int d^4xB_\mu j_\mu^M\right)\right>_{j_\mu^M}=
1+\sum\limits_{N=1}^{\infty}\frac{1}{N!}\left[\prod\limits_{n=1}^{N}
\int\limits_{0}^{+\infty}\frac{ds_n}{s_n}{\rm e}^{4\lambda\eta^2s_n}
\int\limits_{u(0)=u(s_n)}^{}Du(s_n')\right]\times$$

\begin{equation}
\label{et3}
\times\exp\left[\sum\limits_{l=1}^{N}\int\limits_{0}^{s_l}ds_l'\left(
-\frac14\dot u^2(s_l')+2ig_m\dot u_\mu(s_l')B_\mu(u(s_l'))\right)-
4\lambda\sum\limits_{l,k=1}^{N}\int\limits_{0}^{s_l}ds_l'
\int\limits_{0}^{s_k}ds_k''\delta\left[u(s_l')-u(s_k'')\right]\right].
\end{equation}
Here, the vector $u_\mu(s_n')$ parametrizes the same contour as the 
vector $x_\mu^n(s)$. Clearly, the world-line action standing in the 
exponent on the R.H.S. of Eq.~(\ref{et3}) contains besides the usual 
free part also a short-range interaction term, which after carrying out 
the path-integral yields the Higgs potential of magnetic Cooper pairs. 
Indeed, Eq.~(\ref{et3}) can be rewritten as an integral over the 
effective magnetic Higgs field describing Cooper pairs as follows

\begin{equation}
\label{et4} 
\left<\exp\left(i\int d^4xB_\mu j_\mu^M\right)\right>_{j_\mu^M}=
\int D\Phi D\Phi^{*}\exp\left\{-\int d^4x\left[\frac12\left|D_\mu
\Phi\right|^2+\lambda\left(|\Phi|^2-\eta^2\right)^2\right]\right\},
\end{equation}
where $D_\mu=\partial_\mu-2ig_mB_\mu$ is the covariant derivative, and 
an inessential constant factor has been referred to the integration 
measure~\footnote{A seeming divergency at large proper times produced 
in Eq.~(\ref{et3}) by the factor ${\rm e}^{4\lambda\eta^2s_n}$ is 
actually apparent, since the last term in the exponent on the R.H.S. 
of this equation yields the desired damping.}. Finally, substituting 
Eq.~(\ref{et4}) into Eq.~(\ref{et2}), we arrive at the following 
effective Abelian-projected theory of the $SU(2)$-gluodynamics

\begin{equation}
\label{et5}
{\cal Z}=\int \left|\Phi\right| D\left|\Phi\right|
D\theta DB_\mu\exp\left\{-\int d^4x\left[
\frac14 F_{\mu\nu}+
\frac12\left|D_\mu
\Phi\right|^2+\lambda\left(|\Phi|^2-\eta^2\right)^2\right]\right\},
\end{equation}
where $\Phi(x)=\left|\Phi(x)\right|{\rm e}^{i\theta(x)}$. Clearly, 
in Eq.~(\ref{et5}) one can recognize the partition function of the 
dual AHM (DAHM).

Next, analogous considerations can be applied to the $SU(3)$-gluodynamics.
Demanding the condensation of magnetic monopoles, one can expect 
that in this case the resulting Abelian-projected theory should be also 
of the DAHM type, but with the $[U(1)]^2$ gauge invariance. To derive it, 
let us start with the Cartan decomposition of the action~(\ref{ym}) 
in the $SU(3)$-case. Then, 
$F_{\mu\nu}^i=\partial_\mu A_\nu^i-\partial_\nu A_\mu^i+gf^{ijk}A_\mu^j
A_\nu^k$, and the $SU(3)$-generators 
$T^i=\frac{\lambda^i}{2}$, $i=1,\ldots,8$, obey 
the commutation relations $\left[T^i,T^j\right]=if^{ijk}T^k$, where 
$\lambda^i$'s denote the Gell-Mann matrices. 
The Cartan decomposition 
of the field $A_\mu$, which singles out its diagonal part, has the form 

$$A_\mu={\bf a}_\mu{\bf H}+\left(C_\mu^{*a}E_a+C_\mu^aE_{-a}\right)
\equiv{\cal A}_\mu+{\cal C}_\mu,$$
$a=1,2,3$, where ${\bf a}_\mu\equiv\left(A_\mu^3, A_\mu^8\right)$, and 
the diagonal $SU(3)$-generators, 
which generate the Cartan subalgebra, read
${\bf H}\equiv\left(H_1,H_2\right)=\left(T^3,T^8\right)$. We have also 
introduced the so-called step operators $E_{\pm a}$'s (else called 
raising operators for positive $a$'s and lowering operators otherwise) 
by redefining the rest (non-diagonal) $SU(3)$-generators as follows 

$$E_{\pm 1}=\frac{1}{\sqrt{2}}\left(T^1\pm iT^2\right),~ 
E_{\pm 2}=\frac{1}{\sqrt{2}}\left(T^4\mp iT^5\right),~ 
E_{\pm 3}=\frac{1}{\sqrt{2}}\left(T^6\pm iT^7\right).$$
Clearly, these operators are non-Hermitean in the sense that 
$(E_a)^\dag=E_{-a}$. The complete Lie algebra of the so-redefined 
$SU(3)$-generators reads

$$\left[{\bf H},E_{\pm a}\right]=\pm{\bf e}_aE_{\pm a},~ 
\left[E_{\pm a},E_{\pm b}\right]=\mp\frac{1}{\sqrt{2}}
\varepsilon_{abc}E_{\mp c},~ 
\left[E_a,E_{-b}\right]=\delta_{ab}{\bf e}_a{\bf H}.$$
Here, we have introduced the so-called root vectors 

$$
{\bf e}_1=\left(1,0\right),~ 
{\bf e}_2=\left(-\frac12,-\frac{\sqrt{3}}{2}\right),~ 
{\bf e}_3=\left(-\frac12,\frac{\sqrt{3}}{2}\right),$$ 
which thus play the r\^ole of the structural constants in the first of the 
above commutation relations. Eq.~(\ref{cart}) then still holds, and we 
obtain for various terms on its R.H.S. the following expressions 

$$F_{\mu\nu}[{\cal A}]={\bf f}_{\mu\nu}{\bf H},~ 
{\bf f}_{\mu\nu}=\left(\partial\wedge {\bf a}\right)_{\mu\nu},~ 
\left(D[{\cal A}]\wedge{\cal C}\right)_{\mu\nu}=E_a
\left({\cal D}^{*a}\wedge{\cal C}^{*a}\right)_{\mu\nu}+
E_{-a}\left({\cal D}^a\wedge{\cal C}^a\right)_{\mu\nu},$$

$$
-ig\left[{\cal C}_\mu,{\cal C}_\nu\right]=ig\left\{{\bf e}_a{\bf H}
\left(C^a\wedge C^{*a}\right)_{\mu\nu}+\frac{1}{\sqrt{2}}
\varepsilon_{abc}\left(C_\mu^{*a}C_\nu^{*b}E_{-c}-C_\mu^aC_\nu^bE_c
\right)\right\},$$
where ${\cal D}_\mu^a=\partial_\mu+ig{\bf e}_a{\bf a}_\mu$. Bringing 
all these expressions together, we arrive at the following Cartan 
decomposition of the field strength tensor 

$$F_{\mu\nu}=\left({\bf f}_{\mu\nu}+{\bf C}_{\mu\nu}\right){\bf H}+
S_{\mu\nu}^{*a}E_a+S_{\mu\nu}^aE_{-a},$$
where ${\bf C}_{\mu\nu}=ig{\bf e}_a\left(C^a\wedge 
C^{*a}\right)_{\mu\nu}$ and $S_{\mu\nu}^a=\left({\cal D}^a\wedge
C^a\right)_{\mu\nu}+\frac{ig}{\sqrt{2}}\varepsilon_{abc}C_\mu^{*b}
C_\nu^{*c}$. Next, it is worth employing the $SU(3)$-version of the 
maximal Abelian gauge, which for every index $a$ reads ${\cal D}_\mu^a
C_\mu^a=0$. Analogously to the $SU(2)$-case, 
this leads to the appearance of the singular contributions to 
the field strength tensor.
Accounting for them as well as for the equations 
${\rm tr}{\,}{\bf H}E_{\pm a}=
{\rm tr}{\,}E_{\pm a}E_{\pm b}=0$ and ${\rm tr}{\,}E_aE_{-b}=\frac12
\delta_{ab}$, we finally 
obtain the following Cartan decomposition of the $SU(3)$ 
Yang-Mills action 

$$S_{\rm YM}=\frac14\int d^4x\left[\left({\bf f}_{\mu\nu}+{\bf C}_{\mu\nu}
+{\bf f}_{\mu\nu}^{\rm sing.}\right)^2+
2\left(S_{\mu\nu}^{*a}+\left(F_{\mu\nu}^{\rm sing.}\right)^{*a}\right)
\left(S_{\mu\nu}^a+\left(F_{\mu\nu}^{\rm sing.}\right)^a\right)\right],$$
where ${\bf f}_{\mu\nu}^{\rm sing.}=2{\,}{\rm tr}{\,}\left({\bf H}
F_{\mu\nu}^{\rm sing.}\right)$ and $\left(F_{\mu\nu}^{\rm sing.}
\right)^a=2{\,}{\rm tr}{\,}\left(E_a F_{\mu\nu}^{\rm sing.}\right)$. 
Again, disregarding the off-diagonal part of the 
action owing to the Abelian dominance hypothesis, 
we are left with the $SU(3)$-analogue of the effective 
action~(\ref{et}), which reads 

\begin{equation}
\label{pureglue}
S_{\rm eff.}\left[{\bf a}_\mu, {\bf f}_{\mu\nu}^{\rm sing.}\right]=
\frac14\int d^4x\left({\bf f}_{\mu\nu}+
{\bf f}_{\mu\nu}^{\rm sing.}\right)^2.
\end{equation}
Next, by the dualization of the obtained theory we arrive at the 
following partition function ({\it cf.} Eq.~(\ref{et2}))

\begin{equation}
\label{puredual}
{\cal Z}=\left<
\int D{\bf B}_\mu
\exp\left\{-\int d^4x\left[\frac14{\bf F}_{\mu\nu}^2
-i{\bf B}_\mu {\bf j}_\mu^M
\right]\right\}\right>_{{\bf j}_\mu^M},
\end{equation}
where ${\bf F}_{\mu\nu}=\left(\partial\wedge{\bf B}\right)_{\mu\nu}$ 
is the field strength tensor of the field ${\bf B}_\mu$, which is 
dual to the field ${\bf a}_\mu$, and ${\bf j}_\nu^M=\partial_\mu
\tilde{\bf f}_{\mu\nu}^{\rm sing.}$.

At this point, it is worth mentioning that the root vectors 
define the lattice at which monopole charges 
are distributed.
Taking this into account, it is straightforward 
to write down the expression for the collective current of 
$N$ magnetic Cooper pairs 

$${\bf j}_\mu^{M{\,}(N)}(x)=2g_m
\sum\limits_{n=1}^{N}\sum\limits_{a=1}^{3}{\bf e}_a\oint 
dx_\mu^{(a){\,}n}(s)
\delta\left(x-x^{(a){\,}n}(s)\right).$$
The average over magnetic currents then has the form  

$$
\left<\exp\left(i\int d^4x{\bf B}_\mu {\bf j}_\mu^M\right)
\right>_{{\bf j}_\mu^M}=\prod\limits_{a=1}^{3}\left\{
1+\sum\limits_{N=1}^{\infty}\frac{1}{N!}\left[\prod\limits_{n=1}^{N}
\int\limits_{0}^{+\infty}\frac{ds_n}{s_n}{\rm e}^{4\lambda\eta^2s_n}
\int\limits_{u^{(a)}(0)=u^{(a)}(s_n)}^{}Du^{(a)}
(s_n')\right]\times\right.$$

$$
\times\exp\left[\sum\limits_{l=1}^{N}
\int\limits_{0}^{s_l}ds_l'\left(
-\frac14\left(\dot u^{(a)}(s_l')\right)^2
+2ig_m\dot u_\mu^{(a)}(s_l'){\bf e}_a
{\bf B}_\mu\left(u^{(a)}(s_l')\right)\right)-\right.
$$

$$
\left.\left.-4\lambda\sum\limits_{l,k=1}^{N}\int\limits_{0}^{s_l}ds_l'
\int\limits_{0}^{s_k}ds_k''\delta\left[u^{(a)}(s_l')-u^{(a)}(s_k'')
\right]\right]\right\}=$$

\begin{equation}
\label{glue1}
=\int D\Phi_a D\Phi_a^{*}\exp\left\{-\int d^4x
\sum\limits_{a=1}^{3}\left[\frac12\left|\left(\partial_\mu-
2ig_m{\bf e}_a{\bf B}_\mu\right)\Phi_a\right|^2+
\lambda\left(|\Phi_a|^2-\eta^2\right)^2\right]\right\},
\end{equation}
where the vector $u_\mu^{(a)}(s_n')$ on the L.H.S. parametrizes the 
same contour as the vector $x_\mu^{(a){\,}n}(s)$. Finally, it is 
worth noting that since monopoles are distributed over the root lattice,
whose vectors are related to each other by the condition 
$\sum\limits_{a=1}^{3}{\bf e}_a=0$, the dual Higgs fields 
$\Phi_a$'s are also not completely independent of each other.
In Ref.~\cite{maedan}, it was argued that the relevant constraint
for these fields reads $\sum\limits_{a=1}^{3}\theta_a=0$. 
Taking this into account 
we arrive at the following partition function describing an effective 
$[U(1)]^2$ gauge invariant Abelian-projected theory of the 
$SU(3)$-gluodynamics~\cite{maedan}

$$
{\cal Z}=\int \left|\Phi_a\right| D\left|\Phi_a\right|
D\theta_a D{\bf B}_\mu
\delta\left(\sum\limits_{a=1}^{3}
\theta_a\right)\times$$

\begin{equation}
\label{et6}
\times\exp\left\{-\int d^4x\left[\frac14{\bf F}_{\mu\nu}^2+
\sum\limits_{a=1}^{3}\left[\frac12\left|\left(\partial_\mu-
2ig_m{\bf e}_a{\bf B}_\mu\right)\Phi_a\right|^2+
\lambda\left(|\Phi_a|^2-\eta^2\right)^2\right]\right]\right\},
\end{equation}
where $\Phi_a=\left|\Phi_a\right|{\rm e}^{i\theta_a}$.
 
Notice that in general for the gauge group 
$SU\left(N_c\right)$
the number of independent fields of distinct 
monopoles in the corresponding 
effective Abelian-projected theory is equal to $N_c-1$.
This number is just equal to the number of independent 
closed Abrikosov-Nielsen-Olesen type strings in this theory. 

In the present Section, we shall investigate string representations 
of the models~(\ref{et5}) and~(\ref{et6}) and their extensions 
due to the introduction of external electrically charged particles, 
which we shall call ``quarks'', as well as various correlators 
in these models. Various topological properties of Abelian-projected 
theories will also be considered. Besides that, we shall briefly discuss 
the phenomenon of chiral symmetry breaking from the point of view 
of these theories.

\subsection{Nonperturbative Field Correlators and String Representation 
of the Dual Abelian Higgs Model in the London Limit}

\subsubsection{String Representation for the Partition Function of 
Extended DAHM}

Our first aim in the present Subsection will be the derivation 
of a string representation for the partition function 
of DAHM, extended by introduction of external electrically charged 
({\it w.r.t.} the maximal Abelian $U(1)$ subgroup of the 
original $SU(2)$ group) particles, which we shall refer to as 
``quarks''. Such an extension can be performed by adding to the 
action~(\ref{et}) the term $i\int d^4xa_\mu j^E_\mu$ with  
$j_\mu^E(x)\equiv e\oint
\limits_{C}^{}dx_\mu(s)\delta(x-x(s))$ standing for the conserved 
electric current of a quark, which moves along the closed contour $C$
({\it cf.} the notations to Eq.~(\ref{perim})). 
The electric coupling 
constant $e$ is related to the magnetic one according to the  
topological quantization 
condition (which replaces the standard Dirac quantization 
condition in the case of QCD-inspired theories) $eg_m=4\pi n$,
where $n$ is an integer.
In what follows, we shall 
for concreteness restrict ourselves to the case of monopoles 
possessing the minimal charge, {\it i.e.}, set $n=1$.

Then, performing the dualization~(\ref{et})-(\ref{et2}) with the 
new term included and summing up over 
monopole currents according to Eq.~(\ref{et3}), 
we arrive at Eq.~(\ref{et5}) with $F_{\mu\nu}$ replaced by
$F_{\mu\nu}+F_{\mu\nu}^E$. Here,  
$F_{\mu\nu}^E$ stands for the field strength tensor 
generated by quarks according to the equation $\partial_\mu\tilde 
F_{\mu\nu}^E=j_\nu^E$. A solution to this equation reads $F_{\mu\nu}^E=-e
\tilde\Sigma_{\mu\nu}^E$, where   
$\Sigma_{\mu\nu}^E(x)\equiv\int\limits_{\Sigma^E}^{}
d\sigma_{\mu\nu}(\bar x(\xi))\delta(x-\bar x(\xi))$ is the so-called vorticity 
tensor current~\cite{sato} defined on an arbitrary surface $\Sigma^E$
(which is just the world-sheet of an open dual 
Abrikosov-Nielsen-Olesen string), bounded by the contour $C$.
Due to the Stokes theorem, 
the vorticity tensor current is related to the quark current according 
to the equation 
$e\partial_\nu\Sigma_{\mu\nu}=j_\mu^E$. 
In particular, 
this equation means that in the case when there are no external 
quarks, the vorticity tensor current is conserved, {\it i.e.}, due to 
the conservation of electric flux all the strings in that case are 
closed. Notice that 
when external quarks are introduced into the system, 
some amount of closed strings might nevertheless survive. 

Let us now proceed with the string representation for the
partition function of the theory~(\ref{et5}), 
extended by external quarks,
in the so-called London limit, $\lambda\to\infty$. 
In this limit, the radial part of the 
magnetic Higgs field can be integrated out, becoming fixed to 
its {\it v.e.v.}, $\left|\Phi\right|\to\eta$, and 
the partition function under study 
takes the form 

\begin{equation}
\label{vosem}
{\cal Z}=\int DB_\mu D\theta^{{\rm sing.}} 
D\theta^{{\rm reg.}}\exp\left\{-\int d^4x\left[\frac14
\left(F_{\mu\nu}+F_{\mu\nu}^E\right)^2+\frac{\eta^2}{2}
\left(\partial_\mu\theta-2g_mB_\mu\right)^2\right]\right\}, 
\end{equation}
where from now on constant normalization factors will be omitted. 
In our further interpretation of the topic of the string representation 
for the partition function and field correlators of 
the extended DAHM, we shall mainly follow 
Refs.~\cite{mpla3} and~\cite{euro} 
(for related investigations 
see Refs.~\cite{orl, for, kleinert, sato, 
wiese, zubkov}). 

In Eq.~(\ref{vosem}), we have performed a 
decomposition of the phase of the magnetic ({\it i.e.}, dual) 
Higgs field $\theta=
\theta^{{\rm sing.}}+\theta^{{\rm reg.}}$. Here, $\theta^{{\rm sing.}}(x)$ 
describes a certain configuration of electric strings and 
obeys the equation (see {\it e.g.}~\cite{lee1})  
 
\begin{equation}
\label{devyat}
\varepsilon_{\mu\nu\lambda\rho}\partial_\lambda
\partial_\rho\theta^{{\rm sing.}}(x)=2\pi\Sigma_{\mu\nu}(x). 
\end{equation}
This equation is just the covariant formulation of the 4D analogue of the 
Stokes theorem for the field $\partial_\rho\theta^{\rm sing.}(x)$, 
written in the local form. Here, $\Sigma_{\mu\nu}$ stands for the 
vorticity tensor current, defined at the world-sheet $\Sigma$ of a 
closed electric string, parametrized by the vector $x_\mu(\xi)$.
On the other hand, the field 
$\theta^{\rm reg.}(x)$ 
describes simply a single-valued fluctuation around the above-mentioned
string configuration. 
Note that as it has been argued in Ref.~\cite{zubkov}, the 
integration measure over the field $\theta$ factorizes into the 
product of measures over the fields $\theta^{\rm sing.}$ and 
$\theta^{\rm reg.}$.

Performing   
the path-integral duality transformation of Eq.~(\ref{vosem}) along 
the lines described in Ref.~\cite{lee1}, we get 

\begin{equation}
\label{odinnad}
{\cal Z}=
\int Dx_\mu(\xi) Dh_{\mu\nu} 
\exp\Biggl\{-\int d^4x\left[\frac1{12\eta^2}H_{\mu\nu
\lambda}^2+g_m^2h_{\mu\nu}^2+i\pi h_{\mu\nu}\hat\Sigma_{\mu\nu}
\right]\Biggr\},
\end{equation}
where $\hat\Sigma_{\mu\nu}\equiv4\Sigma_{\mu\nu}^E-\Sigma_{\mu\nu}$ and   
$H_{\mu\nu\lambda}\equiv\partial_\mu h_{\nu\lambda}+
\partial_\lambda h_{\mu\nu}+\partial_\nu h_{\lambda\mu}$ is the field 
strength 
tensor of a massive antisymmetric tensor field $h_{\mu\nu}$ (the so-called 
Kalb-Ramond field~\cite{kalb}). This antisymmetric spin-1 tensor field 
emerged via some constraints from the integration over 
$\theta^{\rm reg.}$ and describes a 
massive dual gauge boson. As far as the integration over the 
world-sheets of closed strings, 
$Dx_\mu(\xi)$, is concerned, it appeared from the 
integration over $\theta^{\rm sing.}$ by virtue of  
Eq.~(\ref{devyat}), owing to which there exists a one-to-one 
correspondence between $\theta^{\rm sing.}$ and $x_\mu(\xi)$. 
Physially this correspondence stems from the fact that the singularity 
of the phase of the Higgs field just takes place at the string 
world-sheets (Notice that since in what follows we shall be 
interested in effective actions rather than the integration measures, 
the Jacobian emerging during the change of the integration variables 
$\theta^{\rm sing.}\to x_\mu(\xi)$, which has been evaluated 
in Ref.~\cite{zubkov}, will not be discussed below and is assumed 
to be included into the measure $Dx_\mu(\xi)$.).
The details of a 
derivation of Eq.~(\ref{odinnad}) 
are outlined in Appendix B. 
Thus, the path-integral duality transformation 
is just a way of getting a coupling of the dual gauge 
boson, described now by the field $h_{\mu\nu}$, to a string world-sheet, 
rather than to a world-line (as it takes place in the usual 
case of the Wilson loop). 

Finally, the Gaussian 
integration over the field $h_{\mu\nu}$ in Eq.~(\ref{odinnad}) (see 
Appendix C) 
leads to the following expression for the  
partition function~(\ref{vosem})

$$
{\cal Z}=\exp\left[-\frac{e^2}{2}\oint\limits_C^{}dx_\mu
\oint\limits_C^{}dy_\mu D_m^{(4)}(x-y)\right]\times$$

\begin{equation}
\label{pyatnad}
\times\int Dx_\mu(\xi)\exp\left[-(\pi\eta)^2\int d^4x\int d^4y
\hat\Sigma_{\mu\nu}(x)D_m^{(4)}(x-y)\hat\Sigma_{\mu\nu}(y)
\right]. 
\end{equation}
Here, $D_m^{(4)}(x)\equiv\frac{m}{4\pi^2|x|}K_1(m|x|)$ is the 
propagator of the dual vector boson, whose mass $m$ is equal to
$2g_m\eta$, and $K_i$'s, $i=0,1,2,3$, 
henceforth stand for the modified Bessel functions.

Note that since quarks and antiquarks were from the very beginning 
considered as classical particles, the external contour $C$  
explicitly enters the final result. 
Would one consider them on the quantum level, Eq.~(\ref{pyatnad}) 
must be supplied by a certain prescription of the summation over the 
contours. This can be done by linearizing the 
current~$\times$~current 
interaction standing in the first exponential factor 
on the R.H.S. of Eq.~(\ref{pyatnad}) by integration over an 
auxiliary massive electric 
vector field interacting with the current $j_\mu^E$
and further averaging over $j_\mu^E$'s by virtue of some formula 
{\it \'a la} Eq.~(\ref{et3}).

Clearly, the above mentioned 
exponential factor 
is the standard result, which can be obtained without accounting for the 
Abrikosov-Nielsen-Olesen type electric strings. Contrary to that, 
the integral over string world-sheets on the R.H.S. of 
this equation stems just from the contribution of strings to the 
partition function and is therefore 
the essence of the desired string representation ({\it cf.}
Eq.~(\ref{complete})). The respective string effective action describes
both the interaction of the closed world-sheets $\Sigma$'s with the  
open world-sheets $\Sigma^E$'s and self-interactions of these
objects.

In particular, 
comparing the obtained 
nonlocal string effective action, standing in the 
second exponential factor on the R.H.S. of 
Eq.~(\ref{pyatnad}), with Eq.~(\ref{nonloc})
we obtain by virtue of Eqs.~(\ref{tens}) and 
(\ref{ridconst}) the following values of the string tension of the 
Nambu-Goto term and the 
inverse bare coupling constant of the rigidity term for the surface 
$\Sigma$:

\begin{equation}
\label{sigahm}
\sigma=2\pi\eta^2 K_0(c)\simeq
2\pi\eta^2\ln\frac{1}{c}
\end{equation}
and $\frac{1}{\alpha_0}=-\frac{\pi}{16g_m^2}$. 
Obviously, these two quantities, corresponding to the world-sheet 
$\Sigma^E$, are in the factor 16 larger.  
Here, $c$ 
stands for a characteristic small dimensionless parameter. 
In the London limit, this parameter is of the order of the ratio 
of $m$ to the mass of magnetic Cooper pair (which
plays the r\^ole of the 
UV momentum cutoff somehow analogous to the inverse lattice 
spacing $1/a$ ({\it cf.} Eq.~(\ref{low}))), {\it i.e.}, 
$c\sim \frac{g_m}{\sqrt{\lambda}}$. Moreover, in a derivation 
of Eq.~(\ref{sigahm}) we have assumed that not only $\frac{1}{c}\gg 1$,
but also $\ln\frac{1}{c}\gg 1$, {\it i.e.}, the obtained 
expression for the string tension is valid with the logarithmic 
accuracy. Note that 
the logarithmic divergency of the string tension 
in the 3D version of AHM, the Ginzburg-Landau model, 
is a well known result, which 
can be obtained directly from the definition of this quantity as a free 
energy per unit length of the Abrikosov vortex     
(see {\it e.g.}~\cite{pit}). The physical origin of this result is that 
in the centre of the vortex, the condensate 
is destroyed, and dual gauge bosons remain massless. 
Possible phenomenological consequences of this effect 
were recently discussed in~\cite{bramb}. 

Clearly, both the string tension 
and the inverse bare coupling constant of the rigidity term are 
nonanalytic in  
$g_m$, which means that these quantities are essentially 
nonperturbative similarly to the QCD case ({\it cf.} 
Eq.~(\ref{sigqcd})). Notice also that the finite temperature 
behaviour of the magnetic Higgs field {\it v.e.v.}~\cite{pit}, 
$\eta(T)\propto 
\sqrt{1-\frac{T}{T_c}}$ with $T_c$ standing for the critical 
temperature, obviously governs the corresponding 
behaviour of the string tension~(\ref{sigahm}) and, in particular, 
the deconfinement phase transition. 

Finally, it is worth noting that since contrary to the previous 
Section, the obtained nonlocal 
string effective action is no more associated with the world-sheet 
of the minimal area, $\Sigma_{\rm min.}[C]$, the rigidity term 
generally does not vanish.

\subsubsection{String Representation  
of the Bilocal Field Strength Correlator}

The partition function of extended DAHM, investigated in the 
previous Subsection, can obviously be applied to the derivation 
of the string representation for the bilocal correlator of the 
field strength tensors. Indeed, owing to the Stokes theorem,
this partition function 
can be written as $\left<\exp\left(-\frac{ie}{2}\int d^4x
\Sigma^E_{\mu\nu}f_{\mu\nu}\right)\right>_{a_\mu, j_\mu^M}$, 
where 
$\left<\ldots\right>_{a_\mu, j_\mu^M}\equiv\left<\int Da_\mu\exp
\left(-S_{\rm eff.}\left[a_\mu, f_{\mu\nu}^{\rm sing.}\right]\right) 
\left(\ldots\right)\right>_{j_\mu^M}$ with 
$S_{\rm eff.}$ and $\left<\ldots\right>_{j_\mu^M}$
given by Eqs.~(\ref{et}) and (\ref{et3}), respectively. Applying to this 
expression the cumulant expansion, we have in the bilocal approximation:

\begin{equation}
\label{Zonehand}
{\cal Z}\simeq\exp\left[-\frac{e^2}{8}\int d^4x\int d^4y
\Sigma_{\mu\nu}^E(x)\Sigma_{\lambda\rho}^E(y)\left<\left<
f_{\mu\nu}(x)f_{\lambda\rho}(y)\right>\right>_{a_\mu, j_\mu^M}\right].
\end{equation}
Following the MFC, let us parametrize the bilocal cumulant  
by the two Lorentz structures similarly to Eq.~(\ref{dd1}): 

$$\left<\left<f_{\mu\nu}(x)f_{\lambda\rho}(0)\right>
\right>_{a_\mu, j_\mu^M}=
\Biggl(\delta_{\mu\lambda}\delta_{\nu\rho}-\delta_{\mu\rho}
\delta_{\nu\lambda}\Biggr){\cal D}\left(x^2\right)+$$

\begin{equation}
\label{dvaddva}
+\frac12\Biggl[\partial_\mu
\Biggl(x_\lambda\delta_{\nu\rho}-x_\rho\delta_{\nu\lambda}\Biggr)
+\partial_\nu\Biggl(x_\rho\delta_{\mu\lambda}-x_\lambda\delta_{\mu\rho}
\Biggr)\Biggr]{\cal D}_1\left(x^2\right). 
\end{equation}
Owing to Eqs.~(\ref{d1}) and (\ref{g}), Eq.~(\ref{dvaddva}) 
eventually yields

\begin{equation}
\label{eventual}
{\cal Z}\simeq\exp\left\{-\int d^4x\int d^4y\left[\frac{e^2}{4}
\Sigma_{\mu\nu}^E(x)\Sigma_{\mu\nu}^E(y){\cal D}\left((x-y)^2\right)+
\frac18j_\mu^E(x)j_\mu^E(y)\int\limits_{(x-y)^2}^{+\infty}d\lambda
{\cal D}_1(\lambda)\right]\right\}.
\end{equation} 

On the other hand, this expression should coincide with 
Eq.~(\ref{pyatnad}) divided by ${\cal Z}\left[\Sigma_{\mu\nu}^E=0
\right]$~\footnote{This is just the standard normalization 
condition, encoded in the measure $Dx_\mu(\xi)$.}, 
{\it i.e.}, it reads

$${\cal Z}=\exp\left\{-\int d^4x\int d^4y D_m^{(4)}(x-y)\left[
(4\pi\eta)^2\Sigma_{\mu\nu}^E(x)\Sigma_{\mu\nu}^E(y)+\frac12
j_\mu^E(x)j_\mu^E(y)\right]\right\}\times$$

\begin{equation}
\label{otherhand}
\times\left<\exp\left[8(\pi\eta)^2\int d^4x\int d^4yD_m^{(4)}
(x-y)\Sigma_{\mu\nu}^E(x)\Sigma_{\mu\nu}(y)\right]\right>_{x_\mu(\xi)},
\end{equation}
where 

$$\left<\ldots\right>_{x_\mu(\xi)}\equiv\frac{\int Dx_\mu(\xi)
\left(\ldots\right)\exp\left[-(\pi\eta)^2\int d^4x\int d^4y
\Sigma_{\mu\nu}(x)D_m^{(4)}(x-y)\Sigma_{\mu\nu}(y)\right]}{
\int Dx_\mu(\xi)
\exp\left[-(\pi\eta)^2\int d^4x\int d^4y
\Sigma_{\mu\nu}(x)D_m^{(4)}(x-y)\Sigma_{\mu\nu}(y)\right]}.$$
In the confining regime of the Wilson loop describing external
quarks, we are interested with, the area of the surface $\Sigma^E$ 
is much larger than the typical area of the surface $\Sigma$, swept
out by closed electric strings. Owing to this, the average over 
world-sheets standing on the R.H.S. of Eq.~(\ref{otherhand}) can be 
disregarded {\it w.r.t.} the first exponential factor. Then, the comparison 
of the latter one with Eq.~(\ref{eventual}) straightforwardly yields 
for the function ${\cal D}$ the following expression

\begin{equation}
\label{dvadtri}
{\cal D}\left(x^2\right)=\frac{m^3}{4\pi^2}
\frac{K_1(m|x|)}{\left|x\right|},   
\end{equation}
whereas for the function ${\cal D}_1$ we get the equation
$\int\limits_{x^2}^{+\infty}d\lambda{\cal D}_1(\lambda)=4D_m^{(4)}(x)$,
which leads to:

\begin{equation}
\label{dvadchetyr}
{\cal D}_1\left(x^2\right)=
\frac{m}{2\pi^2x^2}\Biggl[\frac{K_1(m|x|)}{\left|x\right|}
+\frac{m}{2}\Biggl(K_0(m|x|)+K_2(m|x|)\Biggr)\Biggr]. 
\end{equation}
We see that in the limit $\left|x\right|\gg\frac1m$, 
the asymptotic behaviours of the coefficient functions~(\ref{dvadtri}) 
and~(\ref{dvadchetyr}) 
are given by  

\begin{equation}
\label{dvadpyat}
{\cal D}\longrightarrow\frac{m^4}{4\sqrt{2}\pi^{\frac32}}
\frac{{\rm e}^{-m\left|x\right|}}{\left(m\left|x\right|\right)^
{\frac32}}  
\end{equation}
and 

\begin{equation}
\label{dvadshest}
{\cal D}_1\longrightarrow\frac{m^4}{2\sqrt{2}\pi^{\frac32}}
\frac{{\rm e}^{-m\left|x\right|}}{\left(m\left|x\right|\right)^
{\frac52}}. 
\end{equation}
For bookkeeping purposes, let us also list here the asymptotic 
behaviours of these functions 
in the opposite case, $\left|x\right|
\ll\frac1m$. Those read 
 
\begin{equation}
\label{dvadsem} 
{\cal D}\longrightarrow\frac{m^2}{4\pi^2x^2}  
\end{equation}
and

\begin{equation}
\label{dvadvosem}
{\cal D}_1\longrightarrow\frac{1}{\pi^2\left|x\right|^4}. 
\end{equation}

One can now see that according to the lattice data~\cite{di, digiac},  
the asymptotic behaviours~(\ref{dvadpyat}) and~(\ref{dvadshest}) 
are very similar 
to the large distance ones of the nonperturbative parts of the 
functions $D$ and $D_1$ ({\it cf.} Eq.~(\ref{dnp})). 
In particular, both 
functions decrease exponentially, and the function ${\cal D}$ 
is much larger 
than the function ${\cal D}_1$ 
due to the preexponential power-like behaviour.
We also see that the dual gauge boson mass $m$ corresponds to the 
inverse correlation length of the vacuum $T_g^{-1}$. In particular, 
in the string limit of QCD, when $T_g\to 0$ 
while the value of the string tension is kept fixed, $m$ corresponds to 
$\sqrt{\frac{D(0)}{\sigma}}$. 
Clearly, 
the found similarity in the large-distance 
asymptotic behaviours of 
the bilocal cumulant of the field strength tensors 
in DAHM and the gauge-invariant cumulant in QCD  
supports the original conjecture 
by 't Hooft and Mandelstam concerning the dual Meissner nature of 
confinement.  

Moreover, the short distance asymptotic behaviours~(\ref{dvadsem}) 
and~(\ref{dvadvosem}) also 
parallel the results obtained within the MFC of QCD in 
the lowest order 
of perturbation theory. Namely, at such distances the 
function $D_1$ to the lowest order 
also behaves as $\frac{1}{\left|x\right|^4}$ (see Eq.~(\ref{oge})) 
and is 
much larger than the function $D$ to the same order. 
However, it should be realized that the effects of asymptotic 
freedom (which is the most important UV feature of 
QCD, distinguishing it from all the Abelian gauge theories) obviously 
cannot be obtained from DAHM.

Hence we see that within the approximation 
$|\Sigma|\ll\left|\Sigma^E\right|$, relevant to the confinement of 
external quarks, the bilocal approximation to the MFC is an exact
result in the London limit of DAHM, {\it i.e.}, all the cumulants
of the orders higher than the second one vanish. Higher cumulants
are naturally calculable upon performing the average 
over world-sheets on the R.H.S. of 
Eq.~(\ref{otherhand}) (In particular, this average yields also  
the important modification of the bilocal cumulant, which will be discussed  
below, in Subsection 4.3.). Clearly, higher cumulants appeared in 
this way are dependent on the properties of the ensemble 
of closed strings, one averages over. In particular, this means 
that the relations between cumulants of various orders do depend on 
these properties as well. However, it turns out that such relations may be 
established even without performing the average over world-sheets.
This occurs to be possible 
in the London limit of DAHM 
with an additional term, which describes 
an interaction of an axion with two dual gauge bosons~\cite{axion}. 
The reader is referred to that paper for a detailed discussion 
of the relation between the bilocal and threelocal cumulants in such an 
extended model.

\subsubsection{
Dynamical Chiral Symmetry Breaking within the Method of Field 
Correlators}

The QCD Lagrangian with $N_f$ massless flavours is known to be invariant  
under the global symmetry transformations, which are the $U(N_f)\times 
U(N_f)$ independent rotations of left- and right-handed quark fields. 
This symmetry is referred to as chiral symmetry. The above mentioned 
rotations of the two-component Weyl spinors are equivalent to the 
independent vector and axial $U(N_f)$ rotations of the full four-component 
Dirac spinors, under which the QCD Lagrangian remains invariant as well. 
At the same time, the axial transformations mix states with different 
$P$-parities. Therefore, we conclude that if the chiral symmetry remains 
unbroken, one would observe parity degeneracy of all the states, whose 
other quantum numbers are the same. The observed splittings between 
such states occur, however, 
to be too large to be explained by the small bare or 
current quark masses. Namely, this splitting is of the order of hundreds 
MeV, whereas the current masses of light $u$- and $d$-quarks are of the 
order of a few MeV~\footnote{The current mass of the $s$-quark, which  
is around 
150 MeV, is still smaller than the typical splitting 
values at least in a factor of three.}. This observation tells us that 
the chiral symmetry of the QCD Lagrangian is broken down spontaneously. 
This phenomenon of the {\it spontaneous chiral symmetry breaking (SCSB)} 
naturally leads to the appearance of light pseudoscalar Goldstone 
bosons, whose r\^ole is played by pions, which are indeed the lightest 
of all the hadrons. Besides confinement, the explanation of SCSB is 
known to be the other 
most fundamental problem of the modern theory of strong 
interactions. 

The order parameter of SCSB is the so-called {\it quark 
condensate} (else called chiral condensate) 
$\left<\bar\psi\psi\right>\simeq -(250{\,}{\rm MeV})^3$.
This is nothing else, 
but the quark Green function taken at the origin or a closed quark 
loop in the momentum representation. Clearly, if the quarks from the 
beginning are treated as massless 
({\it i.e.}, only the kinetic term is present in their propagators), the 
phenomenon of SCSB leads to the appearance 
of a non-zero 
{\it dynamical} quark mass, which in general is momentum-dependent. 
At zero momentum, the value of the dynamical quark mass is of the order of 
350-400 MeV, which is just the value of the so-called constituent 
quark mass. 

Let us now proceed to the quantitative description of SCSB. Integrating 
the quarks out of the QCD Lagrangian obviously yields  

$${\cal Z}_{\rm QCD}=\prod\limits_{f=1}^{N_f}\left<
\det\left(i\hat D^{\rm fund.}+im_f\right)\right>_{A_\mu^a},$$
where $\hat D^{\rm fund.}\equiv \gamma_\mu D_\mu^{\rm fund.}$ 
with $D_\mu^{\rm fund.}=\partial_\mu-igA_\mu$ standing for 
the covariant derivative in the fundamental representation.  
As it follows from its definition, the chiral condensate for a given 
flavour $f$ then has the form 

\begin{equation}
\label{condens} 
\left<\bar\psi_f\psi_f\right>=
-\frac{1}{V}\left(\frac{\partial}{\partial m_f}
\ln {\cal Z}_{\rm QCD}\right)_{m_f\to 0},
\end{equation}
where $V$ stands for the four-volume of observation. 
In what follows, let us for simplicity restrict ourselves to the case 
of one flavour only. 
Next, let $\Psi_n$ be an eigenfunction of the Dirac operator, 
corresponding 
to a nonvanishing eigenvalue $\lambda_n$,  
$i\hat D^{\rm fund.}\Psi_n=\lambda_n\Psi_n$. Then, since $\gamma_5$ 
anticommutes with $i\hat D^{\rm fund.}$, the function $\Psi_{n'}=
\gamma_5\Psi_n$ is also an eigenfunction of the Dirac operator, 
corresponding to the eigenvalue $\lambda_{n'}=-\lambda_n$. Owing to this  
observation, the determinant can be rewritten as follows

$$ 
\det\left(i\hat D^{\rm fund.}+im\right)=\prod\limits_{n}^{}\left(
\lambda_n+im\right)\sim\sqrt{\prod\limits_{n}^{}\left(\lambda_n^2+m^2
\right)}=\exp\left[\frac12\sum\limits_{n}^{}\ln 
\left(\lambda_n^2+m^2
\right)\right]=
$$

$$
=\exp\left[\frac12\int\limits_{-\infty}^{+\infty}d\lambda\nu(\lambda) 
\ln 
\left(\lambda^2+m^2\right)\right],
$$
where $\nu(\lambda)\equiv\sum\limits_{n}^{}\delta(\lambda-\lambda_n)$ 
stands for the so-called spectral density of the Dirac operator. 
By virtue of Eq.~(\ref{condens}), we thus obtain for the chiral condensate 

\begin{equation}
\label{avdens}
\left.
\left<\bar\psi\psi\right>=-\frac{1}{V}\int\limits_{-\infty}^{+\infty}
d\lambda\left<\nu(\lambda)\right>\frac{m}{\lambda^2+m^2}\right|_{m\to 0},
\end{equation}
where $\left<\nu(\lambda)\right>$ denotes the spectral density averaged 
over the full QCD partition function, including the weight given by 
the determinant itself. The latter can be disregarded in the quenched 
approximation (justified at large-$N_c$), in which one neglects 
the backward influence of quarks to the dynamics.

For a finite-volume system, the R.H.S. of Eq.~(\ref{avdens}) obviously 
vanishes. However, for the case when the volume increases, the 
spectrum becomes continuous, and we should use the formula 

$$
\lim_{m\to 0}^{}\frac{m}{\lambda^2+m^2}={\rm sign}{\,}(m)\pi
\delta(\lambda).
$$
Taking this into account, one finally gets the celebrated 
Banks-Casher relation~\cite{banks}

\begin{equation}
\label{caser}
\left<\bar\psi\psi\right>=-\frac{1}{V}{\rm sign}{\,}(m)\pi\left<\nu(0)
\right>.
\end{equation}
Thus, we see that the chiral condensate is proportional to the averaged 
spectral density of the Dirac operator at the origin. Notice that 
the sign function in Eq.~(\ref{caser}) 
means that at small $m$, the QCD partition function 
depends on $m$ nonanalytically, which is typical for the situation 
when the symmetry is spontaneously broken. 

Up to now, there exist several microscopic models of 
SCSB in QCD~\cite{ufn}. The most elaborated and therefore 
popular of them is the one due to 
instantons~\cite{mitya0}~\footnote{
Recently in Ref.~\cite{gon}, there has been 
proposed a model based on the so-called dyonic gas.
The advantage 
of this model {\it w.r.t.} the instanton one 
is that, as it has been argued there, it yields not only SCSB, but 
also the 
confinement property.}. 
It is based on the observation that in the background field of one 
(anti)instanton, the Dirac operator has an exact zero mode~\cite{thinst}. 
The estimate of $\left<\nu(0)\right>$ for the case of a gas of $I$'s 
and $\bar I$'s has been performed in 
Ref.~\cite{mitya0} directly by calculating the overlap of the 
zero modes and has the form $\left<\nu(0)\right>
\sim\frac{V}{L^2\rho}$~\footnote{
Here, we adopt the same notations as the ones at 
the end of Subsection 2.1. However, now $\rho$ and $L$ stand not for 
fixed, but for the averaged size and separation in the $I-\bar I$ 
ensemble, respectively.}.
This result then has been rederived in Ref.~\cite{pobyl} by solving 
a closed equation for the averaged quark propagator as an expansion 
in powers of the so-called packing fraction parameter 
$\frac{N\rho^4}{VN_c}$, where $N$ stands for the total 
number of $I$'s and $\bar I$'s. 
By virtue of Eq.~(\ref{caser}), the chiral condensate reads   
$\left<\bar\psi\psi\right>\sim -\frac{1}{L^2\rho}$.

Notice that this result can be also derived if we first average 
over the $I-\bar I$ gas and only after that calculate the chiral 
condensate by making use of the so-obtained effective 
theory~\cite{mitya}. In that case, quark interactions are due to the 
scattering of two or more (anti)quarks over the same $I$ or $\bar I$, 
{\it i.e.}, there arise 
four- (or more) fermion interaction terms. The range of such 
interaction, which is 
usually referred to as 't Hooft interaction~\cite{thinst}, 
is obviously of the order of $\rho$. In the case of two flavours, 
this interaction is the four-fermion one and yields an effective theory 
similar to the well known Nambu-Jona-Lasinio model~\cite{vlnjl, dietmar}. 

There have been also derived other types of nonlocal NJL models
in an approximate way 
from QCD by making use of path-integral techniques in bilocal fields.  
In particular, exact bosonization of 2D QCD has been 
performed in Ref.~\cite{dietmar2d}, and 
several attempts to bosonization of 4D QCD have been 
proposed~\cite{dietmar4d, fieldstr}. The 
general strategy of these approaches includes several steps. 
Firstly, one casts the effective four-fermion interaction (which can 
be either local or nonlocal with a certain interaction kernel) 
into the form of a Yukawa interaction by 
introducing a set of collective bosonic fields, which is just the 
essence of the bosonization procedure. Secondly, one can integrate 
over the fermions and derive an effective action in 
terms of these collective fields. Next, for large-$N_c$ the saddle-point
of this effective action yields the 
so-called Schwinger-Dyson (or gap) 
equation, which determines    
the dynamical mass, responsible for SCSB. After that, an expansion
of the resulting meson action 
in small field fluctuations around this stationary solution 
yields the so-called Bethe-Salpeter equation, which determines the 
spectrum of meson excitations. 
This finally leads to a description 
of the chiral sector of QCD in terms of Effective Chiral Hadron 
Lagrangians~\cite{njl3, dietmar, jurke} 
containing higher order derivative terms with fixed 
structure constants. The form of these Lagrangians agrees 
with that of the phenomenological Hadron Lagrangians postulated 
at the end of the Sixties on the basis of pure group-theoretical 
arguments when considering nonlinear realizations of chiral 
symmetry~\cite{efflagr}. However, contrary to those Lagrangians, 
the new ones are not obtained by symmetry principles alone, but rather 
derived from an underlying microscopic quark (diquark) picture, which 
enables one to estimate masses and coupling constants of composite 
hadrons.  

However, it is worth noting that though the QCD-motivated NJL type 
models model well 
the SCSB phenomenon in QCD and describe with a good accuracy 
the hadron spectrum and coupling constants, they do not reproduce 
the confinement property.
Recently, an attempt of a derivation  
of an effective 
Lagrangian for a light quark propagating in the confining QCD 
vacuum, which could account simultaneously for both phenomena, 
has been done~\cite{chiral}. Within this approach, one starts with 
an expression for the Green function of a system consisting of an 
infinitely heavy quark and light antiquark. The propagation of a 
light (anti)quark alone in the QCD vacuum then leads to SCSB, whereas 
the presence of a heavy 
quark enables one to take into account the QCD string, joining both 
objects, and thus to incorporate confinement. Averaging over gluonic 
fields by making use of the cumulant expansion in the 
bilocal approximation, it is then straightforward to derive an effective 
NJL-type Lagrangian for the light quark with a certain nonlocal kernel. 
However, the form of this kernel is now determined via the coefficient 
function $D(x)$ ({\it cf.} Eq.~(\ref{dd1})), which, as it has been 
demonstrated above, yields the string picture ({\it cf.} 
Eqs.~(\ref{effact}), (\ref{tens}), (\ref{ridconst})), 
{\it i.e.}, is responsible for 
confinement. Such an interpolation between confinement and SCSB within 
this approach leads  
to a relation between chiral and gluonic condensates, 
which reads

\begin{equation}
\label{relation}
\left<\bar\psi\psi\right>\propto -D(0)T_g.  
\end{equation}
Following Ref.~\cite{chiral}, let us  
briefly demonstrate how this relation can be obtained in the 
Abelian-projected theories. To this end, we for simplicity consider 
the case of Abelian-projected $SU(2)$-gluodynamics, studied above. 
(Our analysis can be straightforwardly extended to the Abelian-projected 
$SU(3)$-gluodynamics, which will be investigated in the next Subsection.)  
To proceed with, let us assume for a while that a monopole has 
an effective infinitesimal size $b$ (which is known to become 
finite for the 't Hooft-Polyakov monopole). Then, every monopole 
with the world-line 
of the length $T$ produces $T/b$ quasizero modes~\cite{modes}. The 
total number of quasizero modes produced by all the monopoles from the  
3D volume $V^{(3)}$ in the interval of modes $\Delta\lambda$ thus 
has the form $\left<\nu(\lambda)\right>
\Delta\lambda=\frac{T}{b}V^{(3)}n_3$, where $n_3$ is 
the 3D density of monopoles. Since $V=TV^{(3)}$, one gets an estimate

\begin{equation}
\label{rel1}
\frac{\left<\nu(0)\right>}{V}\sim\frac{n_3}{\Delta\lambda b}.
\end{equation}
Since the quasizero modes of all the monopoles are mixed due to the 
interaction, the denominator of the R.H.S. 
of this relation is of the order of unity. 
On the other hand, 
the 3D density of monopoles can be estimated via 
the following equation~\cite{ufn} 

\begin{equation}
\label{tridodin}
\left<j_\mu(x)j_\nu(y)\right>=\Biggl(\frac{\partial^2}
{\partial x_\lambda\partial y_\lambda}\delta_{\mu\nu}-
\frac{\partial^2}{\partial x_\mu\partial y_\nu}\Biggr)
{\cal D}\left((x-y)^2\right), 
\end{equation}
which follows from Eq.~(\ref{dvaddva}) due to equations of motion.
Indeed, owing to Eq.~(\ref{tridodin}), one has 

\begin{equation}
\label{rel2}
n_3\sim\int d^3x\left<j_\mu\left({\bf x}, x_4\right)j_\mu
\left({\bf 0}, x_4\right)\right>\sim {\cal D}(0)T_g.
\end{equation}
Here, as it has been argued after Eq.~(\ref{dvadvosem}), the correlation 
length of the vacuum, $T_g$, for the Abelian-projected 
$SU(2)$-gluodynamics is equal to the inverse mass of the dual gauge 
boson, $m^{-1}$.
Notice also that in a derivation of the first estimate in 
Eq.~(\ref{rel2}), one uses the following observation~\cite{chiral}. 
The correlator $\left<j_\mu(x)j_\mu(0)\right>$ estimates the 
probability of finding a monopole at the 4D point $x$, if there is one at 
the origin. Integrating over $d^3x$, one finds the probability of 
having a monopole at the origin, while another one is anywhere. 
Fixing $x_4$ means that the probability refers to a given moment. 
It has been also assumed that one magnetic monopole yields one 
quasizero fermion mode per unit length of its world-line. This is 
true for an isolated monopole, and this result has been extrapolated 
to the QCD vacuum as a whole.

Finally, substituting Eq.~(\ref{rel2}) into Eq.~(\ref{rel1}) and 
making use of Eq.~(\ref{caser}), we arrive at Eq.~(\ref{relation}).

\subsection{Nonperturbative Field Correlators and String Representation 
of the $SU(3)$-QCD within the Abelian Projection Method}

In the present Subsection, we shall extend the results of Subsection 3.2
to the case of the $SU(3)$-QCD. Our interpretation of this 
subject will mainly follow Ref.~\cite{suzuki, bohmplb}. We shall start with  
the string representation for the partition function~(\ref{et6}) in the 
London limit.

\subsubsection{String Representation for the Partition Function of the 
Abelian-Projected $SU(3)$-Gluodynamics}

Similarly to the $SU(2)$-case, in the London limit, the 
monopole fields become infinitely heavy, and their
radial parts can be integrated out. After that, we are left with  
the following partition function

$$
{\cal Z}=\int D{\bf B}_\mu D\theta_a^{\rm sing.}
D\theta_a^{\rm reg.} Dk\delta\left(\sum\limits_{a=1}^{3}
\theta_a^{\rm sing.}\right)\times
$$

\begin{equation}
\label{suz2}
\times\exp\Biggl\{\int d^4x\Biggl[
-\frac14{\bf F}_{\mu\nu}^2-\frac{\eta^2}{2}\sum\limits_{a=1}^{3}
\left(\partial_\mu\theta_a-2g_m{\bf e}_a{\bf B}_\mu\right)^2+
ik\sum\limits_{a=1}^{3}\theta_a^{\rm reg.}\Biggr]\Biggr\}. 
\end{equation}
Similarly to DAHM, 
in the model~(\ref{et6}), there exist   
string-like singularities of the 
Abrikosov-Nielsen-Olesen 
type. That is why, in Eq.~(\ref{suz2}) 
we have again decomposed the total phases 
of the monopole 
fields into singular and regular parts, $\theta_a=
\theta_a^{\rm sing.}+
\theta_a^{\rm reg.}$, and imposed the constraint of vanishing of the 
sum of regular parts by introducing the integration over the 
Lagrange multiplier $k(x)$. Analogously to the DAHM, in the 
model~(\ref{suz2}), 
$\theta_a^{\rm sing.}$'s 
describe a given electric string configuration and 
are related to the world-sheets $\Sigma_a$'s of strings of three types 
via the equations 

\begin{equation}
\label{suz3}
\varepsilon_{\mu\nu\lambda\rho}\partial_\lambda\partial_\rho
\theta_a^{\rm sing.}(x)=2\pi\Sigma_{\mu\nu}^a(x)\equiv
2\pi\int\limits_{\Sigma_a}^{}d\sigma_{\mu\nu}(x_a(\xi))
\delta(x-x_a(\xi)),
\end{equation}
where $x_a\equiv x_\mu^a(\xi)$ is a four-vector, which parametrizes 
the world-sheet $\Sigma_a$. 

The path-integral duality transformation of the partition 
function~(\ref{suz2}) parallels
that of Subsection 3.2.1. The only seeming problem brought 
about by the additional integration over the Lagrange multiplier occurs 
to be trivial due to the explicit form of the root vectors. Indeed, let 
us first cast Eq.~(\ref{suz2}) into the following form 

$$
{\cal Z}=   
\int D{\bf B}_\mu 
{\rm e}^{-\frac14\int d^4x {\bf F}_{\mu\nu}^2} 
D\theta_a^{\rm sing.}
\delta\left(\sum\limits_{a=1}^{3}
\theta_a^{\rm sing.}\right)\times
$$

\begin{equation}
\label{suz4} 
\times\int Dk D\theta_a^{\rm reg.} DC_\mu^a\exp\Biggl\{\int d^4x
\Biggl[-\frac{1}{2\eta^2}\left(C_\mu^a\right)^2+iC_\mu^a\left(
\partial_\mu\theta_a-2g_m{\bf e}_a{\bf B}_\mu\right)
+ik\sum\limits_{a=1}^{3}\theta_a^{\rm reg.}\Biggr]\Biggr\} 
\end{equation}
and carry out the integration over $\theta_a^{\rm reg.}$'s. 
In this way, one needs to solve the equation $\partial_\mu C_\mu^a=k$, 
which should hold for an arbitrary index $a$. 
The solution to this equation reads

$$
C_\mu^a(x)=\partial_\nu\tilde
h_{\mu\nu}^a(x)-\frac{1}{4\pi^2}\frac{\partial}{\partial x_\mu}
\int d^4y\frac{k(y)}{(x-y)^2},$$
where $h_{\lambda\rho}^a$ stands for the Kalb-Ramond field of the 
$a$-th type. Next, making use of the constraint $\sum\limits_{a=1}^{3}
\theta_a^{\rm sing.}=0$, 
replacing then the integrals over $\theta_a^{\rm sing.}$'s 
with the integrals over $x_\mu^a(\xi)$'s by virtue of Eq.~(\ref{suz3}),  
and discarding again for simplicity the 
Jacobians~\cite{zubkov} emerging during such 
changes of the integration variables, we arrive 
at the following representation for the partition function

$${\cal Z}=
\int D{\bf B}_\mu{\rm e}^{-\frac14\int d^4x
{\bf F}_{\mu\nu}^2}\times$$

$$\times\int Dk\exp\Biggl\{\frac{1}{4\pi^2}\int d^4x\int d^4y\Biggl[
-\frac{3}{2\eta^2}
\frac{k(x)k(y)}{(x-y)^2}+2ig_m
\left(\frac{\partial}{\partial x_\mu}
\frac{k(y)}{(x-y)^2}\right)\sum\limits_{a=1}^{3}
{\bf e}_a{\bf B}_\mu(x) 
\Biggr]\Biggr\}\times$$

$$\times\int Dx_\mu^a(\xi)\delta\left(\sum\limits_{a=1}^{3}
\Sigma_{\mu\nu}^a\right) Dh_{\mu\nu}^a
\exp\Biggl\{\int d^4x\Biggl[-\frac{1}{12\eta^2}
\left(H_{\mu\nu\lambda}^a\right)^2+i\pi h_{\mu\nu}^a\Sigma_{\mu\nu}^a-
ig_m\varepsilon_{\mu\nu\lambda\rho}{\bf e}_a
{\bf B}_\mu\partial_\nu h_{\lambda\rho}^a\Biggr]\Biggr\}.$$ 
Here, $H_{\mu\nu\lambda}^a=\partial_\mu h_{\nu\lambda}^a+
\partial_\lambda h_{\mu\nu}^a+\partial_\nu h_{\lambda\mu}^a$ stands 
for the field strength tensor of the Kalb-Ramond field $h_{\mu\nu}^a$. 
Clearly, due to the explicit form of the root vectors, 
the sum $\sum\limits_{a=1}^{3}{\bf e}_a{\bf B}_\mu$ vanishes, and 
the integration over the Lagrange multiplier thus yields an 
inessential determinant factor. Notice also that due to 
Eq.~(\ref{suz3}),  
the constraint 
$\sum\limits_{a=1}^{3}\theta_a^{\rm sing.}=0$ resulted into a  
constraint for the world-sheets of strings of three types
$\sum\limits_{a=1}^{3}\Sigma_{\mu\nu}^a=0$. This means that actually 
only the world-sheets of two types are independent of each other, 
whereas the third one is unambiguously fixed by the demand that the above 
constraint holds.

Straightforward integrations over the dual gauge 
field ${\bf B}_\mu$ as well as over the Kalb-Ramond fields 
then yield the following desired 
string representation for the partition function

$$
{\cal Z}=\int Dx_\mu^a(\xi) 
\delta\left(\sum\limits_{a=1}^{3}
\Sigma_{\mu\nu}^a\right)\times
$$

\begin{equation}
\label{suz5}
\times
\exp\left[
-\frac{g_m\eta^3}{2}\sqrt{\frac32}\int\limits_{\Sigma_a}^{}
d\sigma_{\mu\nu}(x_a(\xi))
\int\limits_{\Sigma_a}^{}
d\sigma_{\mu\nu}(x_a(\xi'))\frac{K_1\left(m_B
\left|x_a(\xi)-x_a(\xi')\right|
\right)}{\left|x_a(\xi)-x_a(\xi')\right|}\right].
\end{equation}
Here, $m_B=\sqrt{6}g_m\eta$ is the mass of the fields ${\bf B}_\mu$, 
which they acquire due to the Higgs mechanism. 
Finally, it is 
possible to integrate out one of the three world-sheets, for 
concreteness $x_\mu^3(\xi)$. This yields the expression for the 
partition function in terms of the integral over two independent 
string world-sheets

$$
{\cal Z}=\int Dx_\mu^1(\xi) Dx_\mu^2(\xi)\times
$$

$$
\times\exp\left\{-g_m\eta^3\sqrt{\frac32}\left[
\int\limits_{\Sigma_1}^{}
d\sigma_{\mu\nu}(x_1(\xi))
\int\limits_{\Sigma_1}^{}
d\sigma_{\mu\nu}(x_1(\xi'))\frac{K_1\left(m_B
\left|x_1(\xi)-x_1(\xi')\right|
\right)}{\left|x_1(\xi)-x_1(\xi')\right|}+\right.\right.$$

$$+\int\limits_{\Sigma_1}^{}
d\sigma_{\mu\nu}(x_1(\xi))
\int\limits_{\Sigma_2}^{}
d\sigma_{\mu\nu}(x_2(\xi'))\frac{K_1\left(m_B
\left|x_1(\xi)-x_2(\xi')\right|
\right)}{\left|x_1(\xi)-x_2(\xi')\right|}+$$

\begin{equation}
\label{suz6}
\left.\left.
+\int\limits_{\Sigma_2}^{}
d\sigma_{\mu\nu}(x_2(\xi))
\int\limits_{\Sigma_2}^{}
d\sigma_{\mu\nu}(x_2(\xi'))\frac{K_1\left(m_B
\left|x_2(\xi)-x_2(\xi')\right|
\right)}{\left|x_2(\xi)-x_2(\xi')\right|}
\right]\right\}. 
\end{equation}
According to Eq.~(\ref{suz6}), 
in the language of the effective string theory, 
the partition function~(\ref{suz2}) 
has the form of two independent string 
world-sheets, which (self-)interact by the exchanges of 
massive dual gauge bosons. 
Notice also that as it follows from Eq.~(\ref{suz6}), 
the energy density 
corresponding to the obtained effective nonlocal string Lagrangian  
increases not only 
with the distance between two points lying on the same world-sheet, but 
also with the distance between two different world-sheets, which means 
that also the ensemble of strings as a whole displays confining 
properties. This observation will be elaborated on in the next 
Section, where the collective effects in the string ensemble will 
be studied.

\subsubsection{String Representation of Abelian-Projected $SU(3)$-QCD
and the Bilocal Field Strength Correlator}

An external quark of a certain colour $c=R,B,G$ (red, blue, green, 
respectively) can be introduced
into Abelian-projected $SU(3)$-gluodynamics by adding to the 
initial action~(\ref{pureglue}) the interaction term $i{\bf Q}^{(c)}
\int d^4x{\bf a}_\mu j_\mu$. Here, 
$j_\mu(x)=\oint\limits_{C}^{}dx_\mu(s)\delta(x-x(s))$, 
and the vectors of colour charges read 

$${\bf Q}^{(R)}=\left(\frac{g}{2}, \frac{g}{2\sqrt{3}}\right),~ 
{\bf Q}^{(B)}=\left(-\frac{g}{2}, \frac{g}{2\sqrt{3}}\right),~ 
{\bf Q}^{(G)}=\left(0, -\frac{g}{\sqrt{3}}\right).
$$
The electric coupling constant $g$ here is again related to the 
magnetic one, $g_m$, via the topological quantization condition 
$gg_m=4\pi k$ with $k$ standing for an integer, henceforth set to 
unity (the generalization to an arbitrary $k$ is straightforward).
Then, applying the Stokes theorem and the cumulant expansion in the bilocal 
approximation, we get for the partition function of the Abelian-projected
$SU(3)$-QCD with an external quark of the colour $c$ the following 
expression:

\begin{equation}
\label{colour}
{\cal Z}_c\simeq\exp\left[-\frac18Q^{(c)i}Q^{(c)j}\int d^4x\int d^4y
\Sigma_{\mu\nu}(x)\Sigma_{\lambda\rho}(y)\left<\left<f_{\mu\nu}^i(x)
f_{\lambda\rho}^j(y)\right>\right>_{{\bf a}_\mu, {\bf j}_\mu^M}\right].
\end{equation}
Here, $i,j=1,2$ are the $[U(1)]^2$-indices, referring to the 
Cartan generators ${\bf H}$, and $\Sigma_{\mu\nu}$ is the vorticity tensor 
current defined at a certain world-sheet $\Sigma$, bounded by the 
contour $C$. The average 
on the R.H.S. of Eq.~(\ref{colour}) is defined as follows:
$\left<\ldots\right>_{{\bf a}_\mu, {\bf j}_\mu^M}
\equiv\left<\int Da_\mu\exp
\left(-S_{\rm eff.}\left[{\bf a}_\mu, {\bf f}_{\mu\nu}^{\rm sing.}
\right]\right) 
\left(\ldots\right)\right>_{{\bf j}_\mu^M}$ with 
$S_{\rm eff.}$ and $\left<\ldots\right>_{{\bf j}_\mu^M}$
given by Eqs.~(\ref{pureglue}) and (\ref{glue1}), respectively.
Upon the parametrization of the bilocal cumulant as 

$$\left<\left<f_{\mu\nu}^i(x)f_{\lambda\rho}^j(0)\right>
\right>_{{\bf a}_\mu, {\bf j}_\mu^M}=\delta^{ij}\Biggl\{
\Biggl(\delta_{\mu\lambda}\delta_{\nu\rho}-\delta_{\mu\rho}
\delta_{\nu\lambda}\Biggr)\hat D\left(x^2\right)+$$

\begin{equation}
\label{colorcorrel}
+\frac12\Biggl[\partial_\mu
\Biggl(x_\lambda\delta_{\nu\rho}-x_\rho\delta_{\nu\lambda}\Biggr)
+\partial_\nu\Biggl(x_\rho\delta_{\mu\lambda}-x_\lambda\delta_{\mu\rho}
\Biggr)\Biggr]\hat D_1\left(x^2\right)\Biggr\}, 
\end{equation}
we can write for Eq.~(\ref{colour}) the following expression

\begin{equation}
\label{Zapr}
{\cal Z}_c\simeq\exp\left\{-\frac{g^2}{24}\int d^4x\int d^4y\left[
2\Sigma_{\mu\nu}(x)\Sigma_{\mu\nu}(y)\hat D\left((x-y)^2\right)+
j_\mu(x)j_\mu(y)\int\limits_{(x-y)^2}^{+\infty}d\lambda\hat D_1
(\lambda)\right]\right\},
\end{equation}
where it has been used that for every $c$, 
$({\bf Q}^{(c)})^2=\frac{g^2}{3}$.

On the other hand, the partition function ${\cal Z}_c$ can be 
calculated exactly. Indeed, the dualization of the action~(\ref{pureglue})
with the term $i{\bf Q}^{(c)}
\int d^4x{\bf a}_\mu j_\mu$ added, leads to Eq.~(\ref{puredual}) 
with ${\bf F}_{\mu\nu}$ replaced by ${\bf F}_{\mu\nu}+
{\bf F}_{\mu\nu}^{(c)}$. Here, ${\bf F}_{\mu\nu}^{(c)}$ stands for the 
field strength tensor of a test quark of the colour $c$, which obeys the 
equation $\partial_\mu\tilde{\bf F}_{\mu\nu}^{(c)}={\bf Q}^{(c)}j_\nu$
and thus can be written as ${\bf F}_{\mu\nu}^{(c)}=-{\bf Q}^{(c)}
\tilde\Sigma_{\mu\nu}$. Then, the summation over magnetic currents 
in the sense of Eq.~(\ref{glue1}) yields Eq.~(\ref{et6}) with the 
same extension of ${\bf F}_{\mu\nu}$. In the London limit under study, 
the path-integral duality transformation of this action 
and further integration over the Kalb-Ramond fields (see Ref.~\cite{bohmplb}
for details) yield 

$${\cal Z}_c=\int Dx_\mu^a(\xi)\delta\left(\sum\limits_{a=1}^{3}
\Sigma_{\mu\nu}^a\right)\times$$

\begin{equation}
\label{exactZ}
\times\exp\left\{-\pi^2\int d^4x\int d^4y
D_{m_B}^{(4)}(x-y)\left[\eta^2\bar\Sigma_{\mu\nu}^a(x)
\bar\Sigma_{\mu\nu}^a(y)+\frac{8}{3g_m^2}j_\mu(x)j_\mu(y)\right]\right\}.
\end{equation}
Here, $\bar\Sigma_{\mu\nu}^a\equiv\Sigma_{\mu\nu}^a-
2s_a^{(c)}\Sigma_{\mu\nu}$ with the following numbers $s_a^{(c)}$'s:
$s_a^{(c)}$'s: $s_3^{(R)}=s_2^{(B)}=
s_1^{(G)}=0$, $s_1^{(R)}=s_3^{(B)}=s_2^{(G)}=
-s_2^{(R)}=-s_1^{(B)}=-s_3^{(G)}=1$, which obey the relation
${\bf Q}^{(c)}=\frac{g}{3}{\bf e}_as_a^{(c)}$. Taking into account 
that for every $c$, $\left(s_a^{(c)}\right)^2=2$, we eventually arrive 
at the following expression for the partition function ({\it cf.}
Eq.~(\ref{otherhand})):

$${\cal Z}_c=\exp\left\{-8\pi^2\int d^4x\int d^4yD_{m_B}^{(4)}(x-y)\left[
\eta^2\Sigma_{\mu\nu}(x)\Sigma_{\mu\nu}(y)+\frac{1}{3g_m^2}
j_\mu(x)j_\mu(y)\right]\right\}\times$$

\begin{equation}
\label{Zexact}
\times\left<\exp\left[(2\pi\eta)^2s_a^{(c)}\int d^4x\int d^4y
\Sigma_{\mu\nu}^a(x)D_{m_B}^{(4)}(x-y)\Sigma_{\mu\nu}(y)\right]
\right>_{x_\mu^a(\xi)}
\end{equation}
with the average over world-sheets defined as 

$$\left<\ldots\right>_{x_\mu^a(\xi)}\equiv\frac{
\int Dx_\mu^a(\xi)\delta\left(\sum\limits_{a=1}^{3}
\Sigma_{\mu\nu}^a\right)\left(\ldots\right)\exp\left[
-(\pi\eta)^2\int d^4x\int d^4y\Sigma_{\mu\nu}^a(x)
D_{m_B}^{(4)}(x-y)\Sigma_{\mu\nu}^a(y)\right]}{
\int Dx_\mu^a(\xi)\delta\left(\sum\limits_{a=1}^{3}
\Sigma_{\mu\nu}^a\right)\exp\left[
-(\pi\eta)^2\int d^4x\int d^4y\Sigma_{\mu\nu}^a(x)
D_{m_B}^{(4)}(x-y)\Sigma_{\mu\nu}^a(y)\right]}.$$

Comparing now Eq.~(\ref{Zapr}) with Eq.~(\ref{Zexact}), we see that 
in the approximation $|\Sigma^a|\ll|\Sigma|$, valid in the
confining regime for an external quark, the fuctions $\hat D$ and 
$\hat D_1$ are given by Eqs.~(\ref{dvadtri}) 
and~(\ref{dvadchetyr}) with the replacement 
$m\to m_B$. 
Besides that, it is obvious that 
the bilocal cumulant~(\ref{colorcorrel}) is 
nonvanishing only for the gluonic field strength tensors of the same 
kind, {\it i.e.}, for $i=j=1$ or $i=j=2$. 
Hence, for these diagonal cumulants (whose large-distance 
asymptotic behaviours match those of the $SU(3)$-gluodynamics as 
well as Eqs.~(\ref{dvadpyat}) and~(\ref{dvadshest}) correspond to the 
$SU(2)$-case), 
the vacuum of the Abelian-projected $SU(3)$-QCD in the London limit 
does exhibit a nontrivial correlation length  
$T_g=\frac{1}{m_B}$. 

Finally, it is also worth noting that there has also been derived 
a string representation of the Abelian-projected $SU(3)$-QCD, extended 
by the introduction of the $\Theta$-term. The reader is referred 
to Ref.~\cite{bohmplb} for details of this investigation.

\subsection{Representation of Abelian-Projected Theories in Terms 
of the Monopole Currents}

In the present Subsection, we shall 
derive a representation 
for the partition functions of the Abelian-projected theories directly in 
terms of the monopole currents. In our interpretation we 
shall follow Ref.~\cite{quitenew}.

Let us again start our analysis with the $SU(2)$-case, {\it i.e.}, 
DAHM and for simplicity consider the unextended case, 
where intuitively the 
resulting monopole effective action should contain besides the 
free part, quadratic in the monopole currents, also the interaction 
of these currents with the closed string world-sheets $\Sigma$'s.   
As our starting point will serve Eqs.~(B.2) and (B.3) of 
Appendix B, where, 
however, we shall not perform any hypergauge 
fixing for the field $h_{\mu\nu}$. 
Then, the partition function has the form

\begin{equation}
\label{mon1}
{\cal Z}=
\int DA_\mu Dx_\mu(\xi) Dh_{\mu\nu}\exp\Biggl\{
-\int d^4x\Biggl[\frac{1}{12\eta^2}H_{\mu\nu\lambda}^2-i\pi
h_{\mu\nu}\Sigma_{\mu\nu}+\left(g_mh_{\mu\nu}+\partial_\mu A_\nu-
\partial_\nu A_\mu\right)^2\Biggr]\Biggr\}.
\end{equation}

Notice that according to the equation of motion for the field $A_\mu$,  
the absence of external electric currents is expressed by the 
equation $\partial_\mu {\cal F}_{\mu\nu}=0$, where 
${\cal F}_{\mu\nu}\equiv \partial_\mu A_\nu-\partial_\nu A_\mu+
g_mh_{\mu\nu}$. 
Regarding ${\cal F}_{\mu\nu}$ 
as a full electromagnetic field strength tensor, one can write for it 
the corresponding Bianchi identity modified by the monopoles, 
$\partial_\mu\tilde {\cal F}_{\mu\nu}=
g_m\partial_\mu\tilde h_{\mu\nu}$. This identity means that the 
monopole 
current can be written in terms of the Kalb-Ramond field $h_{\mu\nu}$ 
as 

\begin{equation}
\label{jmon}
j_\mu=g_m\partial_\nu\tilde h_{\nu\mu},
\end{equation}
which manifests its 
conservation. (Notice that this current is related to the field 
$C_\mu$ from Eq.~(B.1) of Appendix B  
as $j_\mu=-g_mC_\mu$. This means that 
the $\delta$-function in the last equality on the R.H.S. of this equation  
just imposes once more the conservation of the current $j_\mu$.)

It is also instructive to write down the equation of motion for the 
Kalb-Ramond field in terms of the introduced full electromagnetic 
field strength tensor. This equation has the form 
${\cal F}_{\nu\lambda}=\frac{g_m}{m^2}\partial_\mu H_{\mu\nu\lambda}+
\frac{i\pi}{2g_m} \Sigma_{\nu\lambda}$.
By virtue of conservation of the vorticity tensor current for the 
closed string world-sheets, $\partial_\mu
\Sigma_{\mu\nu}=0$, this equation again yields the condition of 
absence of external electric currents, $\partial_\mu   
{\cal F}_{\mu\nu}=0$.

Let us now turn ourselves to a derivation of the 
monopole current representation for the partition function of DAHM. 
To this end, we shall first resolve the equation $\frac{g_m}{2}
\varepsilon_{\mu\nu\lambda\rho}\partial_\nu h_{\lambda\rho}=-j_\mu$ 
{\it w.r.t.} $h_{\mu\nu}$, which yields 

$$h_{\mu\nu}(x)=\frac{1}{2\pi^2g_m}\varepsilon_{\mu\nu\lambda\rho}
\int d^4y\frac{(x-y)_\lambda}{|x-y|^4}j_\rho (y).$$
Next, we get the following expressions for various terms 
on the R.H.S. of Eq.~(\ref{mon1})

$$H_{\mu\nu\lambda}^2=\frac{6}{g_m^2}j_\mu^2,~~  
\int d^4xh_{\mu\nu}^2=\frac{1}{2\pi^2g_m^2}\int d^4x\int d^4yj_\mu(x)
\frac{1}{(x-y)^2}j_\mu(y).$$
Bringing all this together 
and performing in Eq.~(\ref{mon1}) again  
the hypergauge transformation 
$h_{\mu\nu}\to h_{\mu\nu}+\partial_\mu\lambda_\nu-
\partial_\nu\lambda_\mu$ with the gauge function $\lambda_\mu=
-\frac{1}{g_m}A_\mu$, which eliminates 
the field $A_\mu$, 
we finally arrive at the desired 
monopole current representation, which has the form 

$$
{\cal Z}=
\int Dx_\mu(\xi) Dh_{\mu\nu}\exp\Biggl\{-\Biggl[\frac{1}{2\pi^2}
\int d^4xd^4yj_\mu(x)\frac{1}{(x-y)^2}j_\mu(y)+\frac{2}{m^2}
\int d^4xj_\mu^2+
$$

\begin{equation}
\label{mon2}
+\frac{2\pi i}{g_m}S_{\rm int.}(\Sigma, j_\mu)\Biggr]
\Biggr\}.
\end{equation}
The first term in the exponent on the 
R.H.S. of Eq.~(\ref{mon2}) 
has the form of the Biot-Savart energy of the electric field generated 
by monopole currents~\cite{for}, the second term corresponds to 
the (gauged) kinetic energy of Cooper pairs,   
and the term 

\begin{equation}
\label{sint}
S_{\rm int.}(\Sigma, j_\mu)=
\frac{1}{4\pi^2}\varepsilon_{\mu\nu\lambda\rho}
\int d^4xd^4y 
j_\mu(x)\frac{(y-x)_\nu}{|y-x|^4}\Sigma_{\lambda\rho}(y)
\end{equation}
describes the 
interaction of the string world-sheet with the monopole 
current $j_\mu$. Obviously, this interaction can be rewritten 
in the form $S_{\rm int.}=\int d^4x j_\mu H_\mu^{\rm str.}$, where 
$H_\mu^{\rm str.}$ is the four-dimensional analogue of the 
magnetic induction, produced by the 
electric string according to the equation

\begin{equation}
\label{induct}
\varepsilon_{\mu\nu\lambda\rho}\partial_\lambda H_\rho^{\rm str.}=
\Sigma_{\mu\nu}.
\end{equation}  
Notice that if we approximate the current $j_\mu$ by the classical 
expression, {\it i.e.}, set

\begin{equation}
\label{part}
j_\mu(x)=2g_m\oint\limits_{\Gamma}^{}dy_\mu(\tau)
\delta(x-y(\tau)),
\end{equation} 
the interaction term~(\ref{sint}) takes the form 
$S_{\rm int.}=2g_m\hat L(\Sigma,\Gamma)$, 
where $\hat L(\Sigma,\Gamma)$ is simply the 
Gauss linking number of the world-sheet $\Sigma$ 
with the trajectory of the monopole Cooper pair, 
$\Gamma$~\footnote{Topological 
interactions of this kind are sometimes interpreted as a 4D analogue 
of the Aharonov-Bohm effect. In particular, this interaction, albeit 
for the current of an external electrically charged 
particle with the 
world-sheet of magnetic string, 
emerges in the string representation for the 
Wilson loop of this particle in AHM~\cite{zubkov}.}.

It is straightforward to extend the above analysis to the case 
of the Abelian-projected $SU(3)$-gluodynamics, where the string 
representation 
for the partition function~(\ref{suz2}) 
has the form

$$
{\cal Z}=\int Dx_\mu^a(\xi)\delta\left(\sum\limits_{a=1}^{3}
\Sigma_{\mu\nu}^a\right)Da_\mu^aDh_{\mu\nu}^a\exp\left\{
-\int d^4x\left[\frac{1}{12\eta^2}\left(H_{\mu\nu\lambda}^a
\right)^2-\right.\right.
$$

\begin{equation}
\label{su3}
\left.\left.
-i\pi h_{\mu\nu}^a\Sigma_{\mu\nu}^a+\left(g_m\sqrt{\frac32}
h_{\mu\nu}^a+\partial_\mu a_\nu^a-\partial_\nu a_\mu^a\right)^2
\right]\right\}
\end{equation}
with $a_\mu^a\equiv{\bf e}_a{\bf a}_\mu$. Analogously to the 
argumentation following after Eq.~(\ref{mon1}), 
Eq.~(\ref{su3}) 
means that the arising monopole currents 
can be expressed in terms of 
three Kalb-Ramond fields as $j_\mu^a=
g_m\sqrt{\frac32}
\partial_\nu \tilde h_{\nu\mu}^a$~\footnote{Clearly, these currents 
are related to the current ${\bf j}_\mu^M$, which enters 
Eq.~(\ref{puredual}), as $j_\mu^a={\bf e}_a{\bf j}_\mu^M$.}.  
Finally, rewriting Eq.~(\ref{su3}) via these currents and 
resolving the constraint $\sum\limits_{a=1}^{3}\Sigma_{\mu\nu}^a=0$ 
by integrating over one of the world-sheets (for concreteness, again 
$x_\mu^3(\xi)$), we obtain 

$$
{\cal Z}=\int Dx_\mu^1(\xi)Dx_\mu^2(\xi)
Dh_{\mu\nu}^a
\exp\Biggl\{-\Biggl[\frac{1}{2\pi^2}
\int d^4xd^4yj_\mu^a(x)\frac{1}{(x-y)^2}j_\mu^a(y)+\frac{2}{m_B^2}
\int d^4x\left(j_\mu^a\right)^2+
$$

\begin{equation}
\label{su31}
+2\pi i\sqrt{
\frac23}\frac{1}{g_m}\Biggl[S_{\rm int.}\left(\Sigma^1, j_\mu^1\right)+
S_{\rm int.}\left(\Sigma^2, j_\mu^2\right)-S_{\rm int.}
\left(\Sigma^1, j_\mu^3\right)-S_{\rm int.}
\left(\Sigma^2, j_\mu^3\right)\Biggr]\Biggr]\Biggr\}.
\end{equation}
We see that the last four 
terms on the R.H.S. of Eq.~(\ref{su31}) describe an interference between 
various possibilities of the interaction between string 
world-sheets and monopole currents in the Abelian-projected 
$SU(3)$-gluodynamics to occur. 

For illustrations, 
let us establish a correspondence 
of the above 
results to the 3D ones. Namely, let us derive 
a 3D analogue of 
Eq.~(\ref{mon2}), {\it i.e.}, find a representation in terms of the 
monopole currents of the dual Ginzburg-Landau model. There, 
Eq.~(\ref{devyat}) is replaced by~\cite{for} 

\begin{equation}
\label{vortex}
\varepsilon_{\mu\nu\lambda}\partial_\nu\partial_\lambda
\theta^{\rm sing.}({\bf x})=2\pi\delta_\mu({\bf x}).
\end{equation}
Here, on the R.H.S. stands the so-called vortex density with 
$\delta_\mu({\bf x})\equiv\int\limits_{L}^{}
dx_\mu(\tau)\delta({\bf x}-{\bf x}(\tau))$ being the 
$\delta$-function defined {\it w.r.t.} the electric 
vortex line $L$, parametrized by the vector 
${\bf x}(\tau)$. This line is closed in the case under study, 
{\it i.e.}, 
in the absence of external quarks, which means that $\partial_\mu 
\delta_\mu=0$. 
Performing  by virtue of Eq.~(\ref{vortex}) 
the path-integral duality 
transformation of the partition function of 3D DAHM in the London limit, 

\begin{equation}
\label{ginland}
{\cal Z}=\int DB_\mu D\theta^{\rm sing.}
D\theta^{\rm reg.}\exp\Biggl\{-\int d^3x\Biggl[\frac{1}{4}
F_{\mu\nu}^2+\frac{\eta^2}{2}(\partial_\mu\theta-2g_mB_\mu)^2
\Biggr]\Biggr\},
\end{equation}
we get for it the 
following representation

$$
{\cal Z}=\int D{\bf x}(\tau)D{\bf h}D\varphi
\exp\Biggl\{-\int d^3x\Biggl[\frac{1}{4\eta^2}\left(\partial_\mu
h_\nu-\partial_\nu h_\mu\right)^2-
$$

\begin{equation}
\label{threed}
-2\pi ih_\mu\delta_\mu
+\left(g_m\sqrt{2}h_\mu+\partial_\mu\varphi\right)^2\Biggr]
\Biggr\}.
\end{equation}
Notice that the Kalb-Ramond field has now reduced to a massive one-form 
field ${\bf h}$ with the mass $m=2g_m\eta$, 
as well as the $A_\mu$-field has reduced to a 
scalar $\varphi$. Analogously to the 4D case, the field 
${\bf E}\equiv g_m\sqrt{2}{\bf h}+\nabla\varphi$ 
can be regarded as a full electric field, defined via the 
full dual electromagnetic field strength tensor as $E_\mu=
\frac12\varepsilon_{\mu\nu\lambda}{\cal F}_{\nu\lambda}$. 
The absence of external quarks is now expressed by the equation 
$\nabla {\bf E}=0$, following from the equation of motion 
for the field $\varphi$. 
Correspondingly, the monopole currents are defined as 
$j_\nu=\partial_\mu {\cal F}_{\mu\nu}=g_m\sqrt{2} 
\varepsilon_{\mu\nu\lambda}\partial_\mu h_\lambda$ and are 
manifestly conserved. Notice also that the condition 
of closeness of the vortex lines,  
$\partial_\mu\delta_\mu=0$, unambiguously exhibits itself as a condition 
of absence of external quarks, $\nabla {\bf E}=0$, by virtue 
of the equation of motion for the field ${\bf h}$, which can be 
written in the form 
$E_\mu=\frac{1}{g_m\sqrt{2}}\left[\frac{1}{2\eta^2}
\partial_\nu\left(\partial_\nu h_\mu-\partial_\mu h_\nu\right)+
i\pi\delta_\mu\right]$.
 
Next, after performing 
the gauge transformation ${\bf h}\to {\bf h}+\nabla\gamma$ 
with the gauge function $\gamma=-\frac{1}{g_m\sqrt{2}}\varphi$, 
the field $\varphi$ drops out. 
Expressing $h_\mu$ via $j_\mu$ as follows 

$$
h_\mu({\bf x})=-\frac{1}{4\sqrt{2}\pi g_m}
\varepsilon_{\mu\nu\lambda}\frac{\partial}{\partial x_\nu}
\int d^3y\frac{j_\lambda({\bf y})}{|{\bf x}-{\bf y}|}
$$
and substituting this expression into the R.H.S. of Eq.~(\ref{threed}), 
we finally arrive at the desired representation for the partition 
function of 3D DAHM in terms of the monopole currents

$$
{\cal Z}=\int D{\bf x}(\tau)D{\bf h}\exp\Biggl\{
-\Biggl[\frac{1}{4\pi}\int d^3xd^3yj_\mu({\bf x})
\frac{1}{|{\bf x}-{\bf y}|}j_\mu({\bf y})+
$$

\begin{equation}
\label{threed1}
+\frac{1}{m^2}\int d^3xj_\mu^2+\frac{\sqrt{2}\pi i}{g_m}
S_{\rm int.}(L, j_\mu)\Biggr]\Biggr\}.
\end{equation}
The 
interaction term of the electric vortex line with the 
monopole current now takes the form

$$
S_{\rm int.}(L, j_\mu)=\frac{1}{4\pi}\varepsilon_{\mu\nu\lambda}
\int d^3xd^3yj_\mu({\bf x})
\frac{({\bf y}-{\bf x})_\nu}{|{\bf y}-{\bf x}|^3}\delta_\lambda
({\bf y}).
$$
This interaction 
term can be again rewritten as $S_{\rm int.}=\int d^3x {\bf j}
{\bf H}^{\rm vor.}$, where the magnetic induction, generated by 
the electric vortex  
line, obeys the equation $\varepsilon_{\mu\nu\lambda}
\partial_\nu H_\lambda^{\rm vor.}=\delta_\mu$.  
Note again that if we set for the monopole current the classical 
expression~(\ref{part}), the interaction term takes the form 
$S_{\rm int.}=2g_m
\hat L(L, \Gamma)$ with $\hat L(L, \Gamma)$ 
standing for the Gauss linking number 
of the contours $L$ and $\Gamma$. 

In conclusion of this Section notice that our results demonstrate the 
usefulness of the Abelian projection method and path-integral 
duality transformation for the solution of the problem of string 
representation of non-Abelian gauge theories and provide us with 
some new insights concerning the vacuum structure of these theories. 
They give a new field-theoretical status to the MFC as well as to 
the 't Hooft-Mandelstam scenario of confinement.

\section{Ensembles 
of Topological Defects in the Abelian-Projected Theories and 
String Representation of Compact QED}

In the previous Section, we have demonstrated the relevance of the 
massive Kalb-Ramond field coupled to the string world-sheet to the 
construction of the string representation of Abelian-projected 
$SU(2)$-gluodynamics. 
We have also argued that the large-distance 
asymptotic behaviour of the propagator of this field matches the one 
of the bilocal field strength correlator in MFC of QCD. 
All this tells us 
that the Kalb-Ramond field 
coupled to the string world-sheet is indeed quite adequate for modelling 
the QCD string effective action. 

In the present Section, we shall find the dual formulation of one more 
model allowing for the analytical description of confinement, which 
is compact QED~\cite{mon, polbook} 
in three and four space-time dimensions. 
There, confinement is also due to monopoles, which however 
form not a condensate of their Cooper pairs 
(as it is argued to take place 
in the Abelian-projected theories), but a dilute gas.  
The resulting dual action also turns out to be some 
kind of a Kalb-Ramond field action, albeit quite nonlinear one. 
As we shall see, this nonlinearity will eventually realize 
the independence of the Wilson loop, describing an external 
electrically charged test particle, of the form of a certain 
surface bounded by the contour of this Wilson loop. Namely, 
it will be demonstrated that this independence is achieved by 
the summation over the branches of a certain (highly nonlinear) 
multivalued 
effective potential of the monopole densities in 3D or monopole 
currents in 4D, both of which are unambiguously related to the 
Kalb-Ramond field via the modified Bianchi identities. This is 
the essence of the string representation of compact QED, which 
is thus alternative to the DAHM one. Note that the inapplicability 
of the method of a derivation of the string representation of DAHM 
to compact QED is clearly not surprising
due to the absence of magnetic Higgs field in the latter case 
(no condensation of 
monopole Cooper pairs). Correspondingly, there is no way of getting the 
integration over string world-sheets from the integration 
over the singular part of the phase of this field.

However in the low-energy limit, the real branch of the above mentioned 
effective monopole potential goes over into a simple quadratic functional, 
and the resulting action of the Kalb-Ramond field becomes 
analogous to the one of  
Eq.~(\ref{odinnad}). This will enable us to 
derive the string tension and 
the bilocal field strength correlator in 3D- and 4D compact QED 
in this limit. 
Next, we shall apply the 
partition function of the type~(\ref{odinnad}) to 
the description of the string world-sheet excitations in DAHM, 
4D compact QED, and QCD and derive the effective action quadratic 
in these excitations. 
In our interpretation of the above mentioned topics, which 
will be covered in the next 
two Subsections,
we shall mainly follow the papers~\cite{confstr, mpla2} 
and~\cite{quitenew}. 

In the third Subsection, we shall demonstrate that the developed 
techniques of investigation of the grand canonical ensembles of 
monopoles are quite appropriate to the description of similar 
ensembles of topological defects in the Abelian-projected 
theories. 
This is a natural extension of the results 
of the previous Section, where these defects have been treated 
as individual ({\it i.e.}, noninteracting) ones. Here, contrary to that, 
collective effects in the dilute gases of such defects will be studied. 
In our interpretation of this topic, we shall follow Refs.~\cite{ijnew, 
mplanew}.

\subsection{Vacuum Correlators and String 
Representation of Compact QED}

\subsubsection{3D-case}

In the present Subsection, we shall derive a representation 
of 3D compact QED in terms of the monopole densities and then 
employ it for the construction of the string representation of 
the Wilson loop in this theory. 
Besides that, 
vacuum correlators 
in the low-energy limit of 3D compact QED will be also investigated.

The most important feature of 3D compact QED, which 
distinguishes it from the noncompact case, is 
the existence of magnetic 
monopoles. Their general configuration is the Coulomb gas with the 
action~\cite{mon}

\begin{equation}
\label{smon}
S_{\rm mon.}=g^2\sum\limits_{a<b}^{}q_aq_b\left(\Delta^{-1}
\right)\left({\bf z}_a, {\bf z}_b\right)+S_0\sum\limits_{a}^{}
q_a^2,
\end{equation}
where $\Delta$ is the 3D Laplace operator, and 
$S_0$ is the action of a single monopole, $S_0=\frac{{\rm const.}}{e^2}$.
Here, similarly to Ref.~\cite{mon}, we have adopted the  
standard Dirac notations, where $eg=2\pi n$, 
restricting ourselves to the monopoles of the minimal charge, {\it i.e.},  
setting $n=1$. Then, the partition function of the grand canonical 
ensemble of monopoles corresponding to the 
action~(\ref{smon}) reads 

\begin{equation}
\label{zmon}
{\cal Z}_{\rm mon.}=1+\sum\limits_{N=1}^{\infty}\sum\limits_{q_a=\pm 1}^{}
\frac{\zeta^N}{N!}\prod\limits_{i=1}^{N}\int d^3z_i
\exp\left[-\frac{\pi}{2e^2}\int d^3xd^3y \rho_{\rm gas}({\bf x})
\frac{1}{|{\bf x}-{\bf y}|}\rho_{\rm gas}({\bf y})\right],
\end{equation}
where $\rho_{\rm gas}({\bf x})=\sum\limits_{a=1}^{N}q_a
\delta\left({\bf x}-{\bf z}_a\right)$ is the monopole density, 
corresponding to the gas configuration. 
Here, a single monopole weight $\zeta\propto\exp\left(-S_0
\right)$ has the dimension of $({\rm mass})^3$ 
and is usually referred to as fugacity. Notice also that 
we have 
restricted ourselves to the values $q_a=\pm 1$, since 
at large 
values of the magnetic coupling constant $g$, monopoles 
with $|q|>1$ turn out to be unstable and 
tend to dissociate into those with $|q|=1$.

Next, Coulomb interaction can be made local, albeit nonlinear one, 
by introduction an 
auxiliary scalar field~\cite{mon}

\begin{equation}
\label{zcos}
{\cal Z}_{\rm mon.}=\int D\chi\exp\left\{-\int d^3x
\left[\frac12\left(\partial_\mu\chi\right)^2-2\zeta
\cos (g\chi)\right]\right\}.
\end{equation}
The magnetic mass $m=g\sqrt{2\zeta}$ of the field $\chi$, 
following from the quadratic term in the expansion of the 
cosine on the R.H.S. of Eq.~(\ref{zcos}), is due to the Debye 
screening in the monopole gas. The next, quartic, term of the 
expansion determines the coupling constant of the 
diagrammatic expansion for this gas, which is therefore 
exponentially small and proportional to $g^4\exp\left(-{\rm const.}
g^2\right)$. 

Before proceeding with the investigation of 3D compact QED, 
let us present the derivation of its partition  
function~(\ref{zmon}) (as well as the related Eq.~(\ref{zcos})) 
from the partition function of the usual 3D QED 
extended by the external monopoles. The statistical weight 
of such a model ({\it i.e.}, its partition function 
for the given monopole charges $q_a$'s and the 
number of monopoles $N$, where the summation 
over the grand canonical ensemble is not yet performed)
is the same as the one of the Ginzburg-Landau theory extended by 
external monopoles, but  
without the Higgs field ({\it cf.} the previous Section) and reads 

\begin{equation}
\label{zofrho}
{\cal Z}\left[\rho_{\rm gas}\right]=\int DA_\mu\left[
-\frac{1}{4e^2}\int d^3x\left(F_{\mu\nu}+F_{\mu\nu}^{\rm gas}\right)^2
\right].
\end{equation}
Here, $F_{\mu\nu}=\partial_\mu A_\nu-\partial_\nu A_\mu$, and 
$F_{\mu\nu}^{\rm gas}$ is the monopole field strength tensor, 
responsible for the violation of the Bianchi identities, 

\begin{equation}
\label{biamod}
\frac12\varepsilon_{\mu\nu\lambda}\partial_\mu 
F_{\nu\lambda}^{\rm gas}=2\pi\rho_{\rm gas}.
\end{equation}
Next, the term 
$\frac{1}{4e^2}F_{\mu\nu}^2$ in the Lagrangian on the R.H.S. 
of Eq.~(\ref{zofrho}) can be linearized by introducing the integration 
over an auxiliary vector field $B_\mu$ as follows 

$$\exp\left(-\frac{1}{4e^2}\int d^3x F_{\mu\nu}^2\right)=
\int DB_\mu\exp\left[-\int d^3x\left(\frac12 B_\mu^2+\frac{i}{2e}
\varepsilon_{\mu\nu\lambda}B_\mu F_{\nu\lambda}\right)\right].$$
After that, the integration over the $A_\mu$-field leads to a 
constraint $\partial_\nu\left(ie\varepsilon_{\mu\nu\lambda}B_\mu+
F_{\nu\lambda}^{\rm gas}\right)=0$, whose resolution yields 
$B_\mu=\frac{i}{2e}\varepsilon_{\mu\nu\lambda}F_{\nu\lambda}^{\rm gas}+
\partial_\mu\chi$. Substituting this representation for the field 
$B_\mu$ into the remained action $\int d^3x\left[\frac12 B_\mu^2+
\frac{1}{4e^2}\left(F_{\mu\nu}^{\rm gas}\right)^2\right]$ 
and making use of Eq.~(\ref{biamod}),
we finally 
arrive at the following expression for the 
partition function~(\ref{zofrho})  

\begin{equation}
\label{zofchi}
{\cal Z}\left[\rho_{\rm gas}\right]=\int D\chi\exp\left\{
-\int d^3x\left[\frac12(\partial_\mu\chi)^2-ig\chi\rho_{\rm gas}
\right]\right\}.
\end{equation}
Clearly, the summation over the grand canonical ensemble of monopoles 
leads now to Eq.~(\ref{zcos}). On the other hand, integration 
over the field $\chi$ in Eq.~(\ref{zofchi}) yields the argument of 
the exponent standing on the R.H.S. of Eq.~(\ref{zmon}). This conclusion  
completes the derivation of Eqs.~(\ref{zmon}) and~(\ref{zcos}) 
from the partition function~(\ref{zofrho}).

Let us now cast the partition function~(\ref{zmon}) into the form 
of an integral over the monopole densities. This can be done by 
introducing into Eq.~(\ref{zmon}) a unity of the form 

\begin{equation}
\label{legendre}
1=\int D\rho\delta\left(\rho({\bf x})-\rho_{\rm gas}({\bf x})
\right)=\int D\rho D\lambda\exp\left\{i\left[
\sum\limits_{a=1}^{N}q_a\lambda({\bf z}_a)-\int d^3x\lambda\rho
\right]\right\}.
\end{equation}
Then, performing the summation over the monopole ensemble in the 
same way as it has been done in a derivation of 
the representation~(\ref{zcos}), we get

\begin{equation}
\label{zrhophi}
{\cal Z}_{\rm mon.}=
\int D\rho D\lambda\exp\left\{-\frac{\pi}{2e^2}\int d^3xd^3y
\rho({\bf x})\frac{1}{|{\bf x}-{\bf y}|}
\rho({\bf y})+\int d^3x\left(2\zeta\cos\lambda-i\lambda\rho
\right)\right\}.
\end{equation}
Finally, integrating over the Lagrange multiplier 
$\lambda$ by resolving the corresponding 
saddle-point equation, 

\begin{equation}
\label{speq}
\sin\lambda=-\frac{i\rho}{2\zeta},
\end{equation}
we arrive at the following expression for 
the partition function 

\begin{equation}
\label{zdens}
{\cal Z}_{\rm mon.}=
\int D\rho \exp\left\{-\left[\frac{\pi}{2e^2}\int d^3xd^3y
\rho({\bf x})\frac{1}{|{\bf x}-{\bf y}|}
\rho({\bf y})+V_{\rm real}[\rho]\right]\right\}. 
\end{equation}
Here,  

\begin{equation}
\label{pot}
V_{\rm real}[\rho]
\equiv \int d^3x\left\{\rho\ln\left[\frac{\rho}{2\zeta}+
\sqrt{1+\left(\frac{\rho}{2\zeta}\right)^2}\right]-2\zeta
\sqrt{1+\left(\frac{\rho}{2\zeta}\right)^2}\right\}
\end{equation}
is the parabolic-type effective monopole potential, whose asymptotic 
behaviour at $\rho\ll\zeta$ reads 

\begin{equation}
\label{lowdens}
V_{\rm real}[\rho]
\longrightarrow\int d^3x\left(-2\zeta+\frac{\rho^2}{4\zeta}
\right).
\end{equation}
Owing to Eq.~(\ref{legendre}), the obtained 
representation~(\ref{zdens}) is natural to be called as a 
representation for the 
partition function in terms of the 
monopole densities.

Note that the reason for the notation ``$V_{\rm real}[\rho]$'' is 
that during the integration over the field $\lambda$ 
in Eq.~(\ref{zrhophi})   
we have chosen only the real branch of the solution to the 
saddle-point equation~(\ref{speq}) and 
disregarded the complex ones. Owing to this, Eq.~(\ref{zdens}) 
describes actually not the full expression for the partition 
function, but only its real part. The corresponding full expression 
will be discussed below.

The obtained representation for the partition function 
can be straightforwardly applied to the calculation 
of the coefficient function ${\cal D}^{\rm mon.}\left(x^2\right)$, 
related to the bilocal correlator of the field strength tensors 
as follows 

$$\left<{\cal F}_{\lambda\nu}({\bf x}) 
{\cal F}_{\mu\rho}({\bf 0})
\right>_{A_\mu,\rho}=
\Biggl(\delta_{\lambda\mu}\delta_{\nu\rho}-\delta_{\lambda\rho}
\delta_{\nu\mu}\Biggr){\cal D}^{\rm mon.}\left(x^2\right)+$$

\begin{equation}
\label{cora}
+\frac12\Biggl[\partial_\lambda
\Biggl(x_\mu\delta_{\nu\rho}-x_\rho\delta_{\nu\mu}\Biggr)
+\partial_\nu\Biggl(x_\rho\delta_{\lambda\mu}-x_\mu\delta_{\lambda\rho}
\Biggr)\Biggr]{\cal D}_1^{\rm full}\left(x^2\right), 
\end{equation}
where the average over the monopole densities is defined by the 
partition function~(\ref{zdens}), whereas the $A_\mu$-average 
is defined as 

$$
\left<\ldots\right>_{A_\mu}\equiv\frac{\int DA_\mu\left(\ldots\right)
\exp\left(-\frac{1}{4e^2}\int d^3x F_{\mu\nu}^2
\right)}{\int DA_\mu 
\exp\left(-\frac{1}{4e^2}\int d^3x F_{\mu\nu}^2
\right)}.
$$
In Eq.~(\ref{cora}), 
${\cal F}_{\mu\nu}=F_{\mu\nu}+F_{\mu\nu}^M$ 
stands for the full electromagnetic field strength tensor, which  
includes also the monopole part

$$
F_{\mu\nu}^M({\bf x})=-\frac12\varepsilon_{\mu\nu\lambda}
\frac{\partial}{\partial x_\lambda}\int d^3y
\frac{\rho({\bf y})}{|{\bf x}-{\bf y}|}.
$$
This monopole part yields the R.H.S. of 
the Bianchi identities modified by the monopoles, 

\begin{equation}
\label{fullbia}
\partial_\mu {\cal H}_\mu=2\pi\rho,
\end{equation}
where ${\cal H}_\mu=
\frac12
\varepsilon_{\mu\nu\lambda}{\cal F}_{\nu\lambda}$ 
stands for the full magnetic 
induction. 
Eqs.~(\ref{cora}) and~(\ref{fullbia}) then lead to the 
following equation for the function ${\cal D}^{\rm mon.}$

\begin{equation}
\label{lapld}
\Delta{\cal D}^{\rm mon.}\left(x^2\right)=-4\pi^2\left<\rho
({\bf x})\rho({\bf 0})\right>_\rho,
\end{equation}
which is a 
3D analogue of the 4D equation~(\ref{tridodin}). 
The correlator standing on the R.H.S. of 
Eq.~(\ref{lapld}) can be found in the limit of small monopole 
densities (low-energy limit), 
$\rho\ll\zeta$. By making use of Eqs.~(\ref{zdens}) 
and~(\ref{lowdens}), we obtain 

$$
\left<\rho({\bf x})\rho({\bf 0})\right>_\rho=-\frac{\zeta}{2\pi}
\Delta
\frac{{\rm e}^{-m|{\bf x}|}}{|{\bf x}|}.
$$   
Then, demanding that ${\cal D}^{\rm mon.}\left(
x^2\to\infty\right)\to 0$, we get by the maximum 
principle for the harmonic functions the desired expression for the 
function ${\cal D}^{\rm mon.}$ in the low-energy limit 

\begin{equation}
\label{dmon} 
{\cal D}^{\rm mon.}\left(x^2\right)=2\pi\zeta 
\frac{{\rm e}^{-m|{\bf x}|}}{|{\bf x}|}.
\end{equation}
We see that in the model under study, the correlation length of the 
vacuum $T_g$, {\it i.e.}, the distance at which the function 
${\cal D}^{\rm mon.}$ decreases, 
corresponds to the inverse Debye mass of the field $\chi$, 
$m^{-1}$ ({\it cf.} the case of Abelian-projected theories, 
where the r\^ole of $T_g$ was played by the inverse mass of the 
dual gauge boson(s), generated by the Higgs mechanism). 
The coefficient function 
${\cal D}_1^{\rm full}
\left(x^2\right)$ will be derived later on. 

Let us now proceed to the problem of string representation of 3D 
compact QED. To this end, we shall 
consider an expression for the Wilson loop in this theory 
and try to find a mechanism realizing the independence of this 
object of the form of a certain string world-sheet 
$\Sigma$, bounded by the contour $C$. 
By virtue of the Stokes theorem, the Wilson loop 
can be rewritten in the following form

$$
\left<W(C)\right>=\left<\exp\left(\frac{i}{2}\int
\limits_{\Sigma}^{}d\sigma_{\mu\nu}{\cal F}_{\mu\nu}\right)
\right>_{A_\mu,\rho}=
\left<\exp\left(i\int\limits_{\Sigma}^{}d\sigma_\mu {\cal H}_\mu\right)
\right>_{A_\mu,\rho}=
$$

\begin{equation}
\label{wqed}
=\left<W(C)\right>_{A_\mu}\left<\exp\left(\frac{i}{2}\int d^3x\rho
({\bf x})\eta({\bf x})\right)\right>_\rho,
\end{equation}
where the free photon contribution reads 

\begin{equation}
\label{wphot}
\left<W(C)\right>_{A_\mu}=
\left<\exp\left(i\oint\limits_{C}^{}A_\mu dx_\mu\right)\right>_{A_\mu}=
\exp\left(-\frac{e^2}{8\pi} 
\oint\limits_{C}^{}dx_\mu\oint
\limits_{C}^{}dy_\mu\frac{1}{|{\bf x}-{\bf y}|}\right).
\end{equation}
In Eq.~(\ref{wqed}), $d\sigma_\mu\equiv\frac12\varepsilon_{\mu\nu\lambda}
d\sigma_{\nu\lambda}$, and 
$\eta({\bf x})=\frac{\partial}{\partial x_\mu}
\int\limits_{\Sigma}^{}d\sigma_\mu({\bf y})
\frac{1}{|{\bf x}-{\bf y}|}$ stands for the 
solid angle under which the surface $\Sigma$ 
shows up to an observer at the point ${\bf x}$. 
Notice that due to the Gauss law, 
in the case when $\Sigma$ is a closed surface surrounding 
the point ${\bf x}$, 
$\eta({\bf x})=4\pi$, which is the standard result for the 
total solid angle in 3D. 

Equation~(\ref{wqed}) seems to contain some discrepancy, since its L.H.S. 
depends only on the contour $C$, whereas the R.H.S. depends 
on an arbitrary 
surface $\Sigma$, bounded by $C$. However, this actually occurs to be 
not a discrepancy, but a 
key point in the construction of the desired string representation. 
The resolution of the apparent paradox lies in the observation that 
during the derivation of the effective monopole potential~(\ref{pot}), 
we have accounted only for the one, namely real, branch of the solution 
to the saddle-point equation~(\ref{speq}).   
Actually, however, one should 
sum up over all the complex-valued branches of the integrand of the 
effective 
potential~(\ref{pot}) at every space point ${\bf x}$. This requires 
to replace $V_{\rm real}[\rho]$ by the total expression 

\begin{equation}
\label{issimo}
V[\rho]=\sum\limits_{n=-\infty}^{+\infty}
\int d^3x\left\{\rho\left[\ln\left[\frac{\rho}{2\zeta}+
\sqrt{1+\left(\frac{\rho}{2\zeta}\right)^2}\right]+2\pi in\right]
- 2\zeta
\sqrt{1+\left(\frac{\rho}{2\zeta}\right)^2}
\right\},  
\end{equation}
which, in particular, being substituted for $V_{\rm real}[\rho]$ 
into Eq.~(\ref{zdens}) yields the 
full partition function, mentioned above.
As far as the Wilson loop, given by the expression 

$$
\left<W(C)\right>=
\left<W(C)\right>_{A_\mu}\times
$$

\begin{equation}
\label{strrepr}
\times\int D\rho
\exp\left\{-\left[\frac{\pi}{2e^2}\int d^3xd^3y
\rho({\bf x})\frac{1}{|{\bf x}-{\bf y}|}
\rho({\bf y})+V[\rho]
-\frac{i}{2}\int d^3x\rho
({\bf x})\eta({\bf x})
\right]\right\}, 
\end{equation} 
is concerned, such a summation over the branches of $V[\rho]$ 
thus restores the independence of this object 
of the choice of the world-sheet. (Notice  
that from now on we omit an inessential normalization factor, 
implying everywhere the normalization $\left<W(0)\right>=1$.)

It is worth noting that the obtained string 
representation~(\ref{strrepr}) has been for the first time 
derived in another, more 
indirect, way in Ref.~\cite{confstr}. It is therefore instructive 
to establish a correspondence between our approach and 
the one of that paper.

The main idea of Ref.~\cite{confstr} was to calculate 
the Wilson loop starting with the direct definition of this average 
in the sense of the partition function~(\ref{zmon}) of the monopole gas. 
The corresponding expression has the form

$$
\left<W(C)\right>_{\rm mon.}=1+
$$

$$
+\sum\limits_{N=1}^{\infty}\sum\limits_{q_a=\pm 1}^{}
\frac{\zeta^N}{N!}\prod\limits_{i=1}^{N}\int d^3z_i
\exp\left[-\frac{\pi}{2e^2}\int d^3xd^3y \rho_{\rm gas}({\bf x})
\frac{1}{|{\bf x}-{\bf y}|}\rho_{\rm gas}({\bf y})+
\frac{i}{2}\int d^3x\rho_{\rm gas}
({\bf x})\eta({\bf x})
\right]=
$$ 

$$
=\int D\chi\exp\left\{-\int d^3x
\left[\frac12\left(\partial_\mu\chi\right)^2-2\zeta
\cos\left(g\chi+\frac{\eta}{2}\right)\right]\right\}=
$$

\begin{equation}
\label{wvarphi}
=\int D\varphi\exp\left\{-\int d^3x\left[\frac{e^2}{8\pi^2}
\left(\partial_\mu\varphi-\frac12\partial_\mu\eta\right)^2-
2\zeta\cos\varphi\right]\right\},
\end{equation}
where $\varphi\equiv g\chi+\frac{\eta}{2}$. 

Next, one can prove 
the following equality

$$
\exp\left[-\frac{e^2}{8\pi}\oint\limits_{C}^{}dx_\mu\oint
\limits_{C}^{}dy_\mu\frac{1}{|{\bf x}-{\bf y}|}-
\frac{e^2}{8\pi^2}\int d^3x
\left(\partial_\mu\varphi-\frac12\partial_\mu\eta\right)^2\right]= 
$$

\begin{equation}
\label{ramond}
=\int Dh_{\mu\nu}\exp\left[-\int d^3x\left(i\varphi
\varepsilon_{\mu\nu\lambda}\partial_\mu h_{\nu\lambda}+
g^2h_{\mu\nu}^2-2\pi ih_{\mu\nu}\Sigma_{\mu\nu}\right)\right],
\end{equation}
which makes it possible to represent 
the contribution of the 
kinetic term on the R.H.S. of Eq.~(\ref{wvarphi}) and the free 
photon contribution~(\ref{wphot}) 
to the Wilson loop as an integral over the 
Kalb-Ramond field. The only nontrivial point necessary to prove the 
equality~(\ref{ramond}) is an expression for the 
derivative of the solid angle. 
One has

\begin{equation}
\label{parteta}
\partial_\lambda\eta({\bf x})=
\int\limits_{\Sigma}^{}\left(d\sigma_\mu({\bf y})
\frac{\partial}{\partial y_\lambda}- 
d\sigma_\lambda({\bf y})
\frac{\partial}{\partial y_\mu}\right)\frac{\partial}{\partial y_\mu}
\frac{1}{|{\bf x}-{\bf y}|}+\int\limits_{\Sigma}^{}
d\sigma_\lambda({\bf y})\Delta
\frac{1}{|{\bf x}-{\bf y}|}.
\end{equation}
Applying to the first integral on the R.H.S. of Eq.~(\ref{parteta}) 
Stokes theorem in the operator form,  
$d\sigma_\mu\frac{\partial}{\partial y_\lambda}-
d\sigma_\lambda\frac{\partial}{\partial y_\mu}\to 
\varepsilon_{\mu\lambda\nu}dy_\nu$,    
one finally obtains  

$$
\partial_\lambda\eta({\bf x})=
\varepsilon_{\lambda\mu\nu}\frac{\partial}{\partial x_\mu}
\oint\limits_{C}^{}dy_\nu\frac{1}{|{\bf x}-{\bf y}|}-
4\pi\int\limits_{\Sigma}^{}d\sigma_\lambda({\bf y})
\delta({\bf x}-{\bf y}).
$$ 
Making use of this result and carrying out the Gaussian integral 
over the field $h_{\mu\nu}$, one can 
demonstrate that both sides of Eq.~(\ref{ramond}) are equal to 

$$
\exp\left\{-\frac{e^2}{2}\left[\frac{1}{4\pi^2}\int d^3x
\left(\partial_\mu\varphi\right)^2+\frac{1}{\pi}\int\limits_{\Sigma}^{}
d\sigma_\mu\partial_\mu\varphi+\int\limits_{\Sigma}^{}d\sigma_\mu
({\bf x})\int\limits_{\Sigma}^{}d\sigma_\mu
({\bf y})\delta({\bf x}-{\bf y})
\right]\right\}, 
$$
thus proving the validity of this equation. 

Substituting now Eq.~(\ref{ramond}) into Eq.~(\ref{wvarphi}), it 
is straightforward  
to carry out the integral over the field $\varphi$. Since this 
field has 
no more kinetic term, such an integration can be performed 
in the saddle-point approximation. 
The corresponding saddle-point 
equation has the same form as Eq.~(\ref{speq}) 
with the 
replacement $\rho\to\varepsilon_{\mu\nu\lambda}\partial_\mu 
h_{\nu\lambda}$. As a result, we obtain 
the following expression for the full Wilson loop

$$
\left<W(C)\right>=\left<W(C)\right>_{A_\mu}
\left<W(C)\right>_{\rm mon.}=
$$

\begin{equation}
\label{wtot}
=\int Dh_{\mu\nu}\exp\left\{-\int d^3x\left(g^2h_{\mu\nu}^2
+V\left[\varepsilon_{\mu\nu\lambda}
\partial_\mu h_{\nu\lambda}\right]\right)+2\pi i
\int\limits_{\Sigma}^{}d\sigma_{\mu\nu}h_{\mu\nu}\right\},
\end{equation}
where the world-sheet independence of the R.H.S. is again provided 
by the summation over the branches of the multivalued action, which 
is now the action of the Kalb-Ramond field. 

Comparing now Eqs.~(\ref{strrepr}) and~(\ref{wtot}), we see that 
the Kalb-Ramond field is indeed related to the monopole density 
via the equation $\varepsilon_{\mu\nu\lambda}\partial_\mu 
h_{\nu\lambda}=\rho$. Thus, a conclusion following 
from the representation of the 
full Wilson loop in terms of the integral over the Kalb-Ramond field 
is that this field is simply related to the sum of the photon 
and monopole field strength tensors as
$h_{\mu\nu}=\frac{1}{4\pi}
{\cal F}_{\mu\nu}$. In the 
formal language, such a 
decomposition of the Kalb-Ramond field is just the essence of the 
Hodge decomposition theorem. 

Let us now consider the low-energy limit of Eq.~(\ref{wtot}) 
and again restrict ourselves to the real branch of the effective 
potential, {\it i.e.}, replace $V\left[
\varepsilon_{\mu\nu\lambda}\partial_\mu h_{\nu\lambda}\right]$ by 
$V_{\rm real}\left[
\varepsilon_{\mu\nu\lambda}\partial_\mu h_{\nu\lambda}\right]$.
This yields the following expression for the Wilson loop 

\begin{equation}
\label{small}
\left<W(C)\right>_{\rm low-energy}=\int Dh_{\mu\nu}
\exp\left\{-\int d^3x\left[\frac{1}{6\zeta}H_{\mu\nu\lambda}^2+
g^2h_{\mu\nu}^2-2\pi ih_{\mu\nu}\Sigma_{\mu\nu}\right]\right\}.
\end{equation}
Notice, that the mass of the 
Kalb-Ramond field resulting from this equation is equal to the Debye 
mass $m$ of the field $\chi$ from Eq.~(\ref{zcos}).

One can now see that Eq.~(\ref{small})
is quite similar to the 3D version of 
Eq.~(\ref{mon1}) (with the $A_\mu$-field 
gauged away) 
we had in the DAHM case.
However, the important difference from DAHM 
is that restricting ourselves to the 
real branch of the potential, 
we have violated 
the surface independence 
of the R.H.S. of Eq.~(\ref{small}).
This problem is similar to the one we met within the 
MFC in Subsection 1.3,  
where in the expression for the Wilson loop, written via the non-Abelian 
Stokes theorem and cumulant expansion, all the 
cumulants higher than the bilocal one were disregarded. 
There, the surface independence was restored by 
replacing $\Sigma$ by the surface of the minimal area, 
$\Sigma_{\rm min.}=\Sigma_{\rm min.}[C],$  
bounded by the contour $C$. 
Let us follow this recipe, after which the quantity 

\begin{equation}
\label{slowen}
\left.S_{\rm str.}=
-\ln\left<W(C)\right>_{\rm low-energy}\right|_{\Sigma\to\Sigma_{\rm min.}}
\end{equation} 
can be considered as a low-energy string 
effective action of 3D compact QED. 

The integration over the Kalb-Ramond field in Eq.~(\ref{small})
is now 
almost the same as the one of Appendix C and yields 

$$
\left.\left<W(C)\right>_{\rm low-energy}\right|_{\Sigma
\to\Sigma_{\rm min.}}=
\exp\left\{-\frac18
\int\limits_{\Sigma_{\rm min.}}^{}
d\sigma_{\lambda\nu}({\bf x})
\int\limits_{\Sigma_{\rm min.}}^{}
d\sigma_{\mu\rho}({\bf y})
\left<{\cal F}_{\lambda\nu}({\bf x}) 
{\cal F}_{\mu\rho}({\bf y})
\right>_{A_\mu,\rho}\right\}, 
$$
which is consistent with the result following directly from the 
cumulant expansion of Eq.~(\ref{wqed}).
Here, the 
bilocal correlator is defined by Eq.~(\ref{cora}) with the 
function ${\cal D}^{\rm mon.}$ given by Eq.~(\ref{dmon}) and 
${\cal D}_1^{\rm full}={\cal D}_1^{\rm phot.}+{\cal D}_1^{\rm mon.}$, 
where the photon and monopole contributions read 

$${\cal D}_1^{\rm phot.}\left(x^2\right)=
\frac{e^2}{2\pi|{\bf x}|^3}$$
and

\begin{equation}
\label{d1mon}
{\cal D}_1^{\rm mon.}\left(x^2\right)=\frac{e^2}{4\pi x^2}
\left(m+\frac{1}{|{\bf x}|}\right){\rm e}^{-m|{\bf x}|},
\end{equation}
respectively.
Since the 
approximation $\rho\ll\zeta$, in which Eq.~(\ref{dmon}) has been derived, 
is just the low-energy limit, in which 
Eq.~(\ref{small}) follows from Eq.~(\ref{wtot}), 
coincidence of 
the function ${\cal D}^{\rm mon.}$, following from the 
propagator of the Kalb-Ramond field, with the one of 
Eq.~(\ref{dmon})  
confirms the consistency of our calculations. 

Notice that by performing an expansion of the nonlocal string 
effective action~(\ref{slowen}) 
in powers of the derivatives {\it w.r.t.} the world-sheet 
coordinates $\xi^a$'s, one gets the following 
string tension of the Nambu-Goto term

\begin{equation}
\label{sigqed}
\sigma=\pi^2\frac{\sqrt{2\zeta}}{g}. 
\end{equation}
Similarly to the expression~(\ref{sigahm}) 
for the string tension in the 
Abelian-projected $SU(2)$-gluodynamics, which is 
nonanalytic in the magnetic coupling constant,  
Eq.~(\ref{sigqed}) is nonanalytic in the electric coupling constant.  
This observation manifests the 
nonperturbative nature of string representation of these 
theories. 

Note that the inverse bare coupling constant of the rigidity term 
corresponding to the string effective action~(\ref{slowen})  
can be also calculated by virtue of the results of Section 2, and 
the result reads $\frac{1}{\alpha_0}=-\frac{\pi^2}{8\sqrt{2\zeta}g^3}$. 
This expression is also nonanalytic in $e$, and its negative sign 
supports the stability of strings. However, as it has been 
demonstrated in Section 2, at the 
surface of the minimal area, $\Sigma_{\rm min.}[C]$, at which 
string effective action~(\ref{slowen}) is defined, the rigidity 
term vanishes. 

We see that the long- and short distance 
asymptotic behaviours of the functions~(\ref{dmon}) and~(\ref{d1mon}) 
display the properties analogous to those 
of the corresponding 
functions in QCD within MFC~\cite{di, digiac}.
Namely, at large distances   
both of the 
functions~(\ref{dmon}) and~(\ref{d1mon}) decrease exponentially 
with the correlation length $m^{-1}$, and at such distances 
${\cal D}_1^{\rm mon.}\ll {\cal D}^{\rm mon.}$ due to the preexponential 
factor. In the same time, in the opposite case 
$|{\bf x}|\ll m^{-1}$, the function ${\cal D}_1^{\rm mon.}$ is much 
larger than the function ${\cal D}^{\rm mon.}$, which also  
parallels the MFC results. Notice, however, that the short-distance 
similarity takes place only to the lowest order of perturbation theory 
in QCD, 
where its specific non-Abelian properties are not important. 
Note also that the above described 
asymptotic behaviours of the 
functions ${\cal D}^{\rm mon.}$ and ${\cal D}_1^{\rm mon.}$ clearly 
match those of the corresponding functions, which parametrize the 
bilocal correlator of the dual field strength tensors in 
the Abelian-projected $SU(2)$- and $SU(3)$-gluodynamics 
({\it cf.} the previous Section). 

\subsubsection{4D-case}

In the present Subsection, we shall generalize the above considerations 
to the case of 4D compact QED. 
Note that in this case, contrary to what happens in 3D, monopoles 
are no more point-like particles at rest, but are rather represented 
by the closed loops. The disorder effects in the ensemble of such 
loops, leading to the confinement of an external electrically charged 
test particle, become strong enough only in the so-called 
strong coupling limit, {\it i.e.}, at sufficiently large values 
of the electric coupling constant. This situation differs from  
the case of 3D compact QED, where 
confinement takes place at arbitrary values of the 
coupling constant (see Ref.~\cite{polbook} for an extended discussion). 

The action we start with 
is again of the form 
of AHM without the Higgs field, extended by external monopoles
({\it cf.} Eq.~(\ref{zofrho})) 

\begin{equation}
\label{sgas}
S_{\rm gas}=
\frac14\int d^4x\left(F_{\mu\nu}+F_{\mu\nu}^{\rm gas}\right)^2.
\end{equation} 
Here, the monopole field strength tensor describes the 
violation of the Bianchi identities by the collective current of $N$ 
monopoles, which form a dilute gas

\begin{equation}
\label{viol}
\partial_\mu\tilde F_{\mu\nu}^{\rm gas}(x)=j_\nu^{\rm gas}(x)\equiv
g\sum\limits_{a=1}^{N}q_a\oint dz_\nu^a\delta(x-x^a(\tau)).
\end{equation}
In Eq.~(\ref{viol}), we have parametrized the trajectory of the 
$a$-th monopole by the vector $x_\mu^a(\tau)=y_\mu^a+z_\mu^a(\tau)$, 
where $y_\mu^a=\int\limits_{0}^{1}d\tau x_\mu^a(\tau)$ denotes the 
position of the trajectory, whereas the vector $z_\mu^a(\tau)$ 
corresponds to its shape, both of which should be eventually averaged 
over. The procedure of a derivation of the statistical weight 
entering the partition function of the 
grand canonical ensemble of monopoles is analogous to the corresponding 
3D one, described in the previous Subsection. Firstly, one can 
linearize the term $\frac14\int d^4x F_{\mu\nu}^2$ in the 
action~(\ref{sgas}) by introducing an integration over an auxiliary 
antisymmetric tensor field $B_{\mu\nu}$ as follows 

$${\cal Z}\left[j_\mu^{\rm gas}\right]\equiv\int DA_\mu 
{\rm e}^{-S_{\rm gas}}=$$

$$=\int DA_\mu DB_{\mu\nu}\exp\left\{
-\int d^4x\left[\frac14B_{\mu\nu}^2+\frac{i}{2}\tilde B_{\mu\nu}
F_{\mu\nu}+\frac12F_{\mu\nu}F_{\mu\nu}^{\rm gas}+\frac14\left(
F_{\mu\nu}^{\rm gas}\right)^2\right]\right\}.
$$
Next, the constraint $\partial_\mu\left(\tilde B_{\mu\nu}-
iF_{\mu\nu}^{\rm gas}\right)=0$ emerging after the integration 
over the field $A_\mu$ can be resolved by setting $B_{\mu\nu}=
i\tilde F_{\mu\nu}^{\rm gas}+\hat F_{\mu\nu}$, where 
$\hat F_{\mu\nu}=\partial_\mu B_\nu-\partial_\nu B_\mu$. 
This finally yields 

\begin{equation}
\label{zjgas}
{\cal Z}\left[j_\mu^{\rm gas}\right]=\int DB_\mu
\exp\left\{-\int d^4x\left[\frac14\hat F_{\mu\nu}^2-
iB_\mu j_\mu^{\rm gas}\right]\right\}.
\end{equation}
This expression should now be averaged over the 
monopole gas in the following sense 

\begin{equation}
\label{calo}
\left<{\cal O}\left[j_\mu^{\rm gas}\right]\right>_{\rm gas}=
\prod\limits_{i=1}^{N}\int d^4y^i Dz^i\mu\left[z^i\right] 
\sum\limits_{q_a=\pm 1}^{}{\cal O}\left[j_\mu^{\rm gas}\right].
\end{equation}
Here, $\mu\left[z^i\right]$ is a certain rotation- and translation 
invariant measure of integration over the shapes of monopole loops, 
whose concrete form will not be specified here (For example, one 
can choose it in the form of the properly normalized measure of 
an ensemble of random loops, representing trajectories of scalar 
particles, {\it i.e.}, 

$$\int Dz^i\mu\left[z^i\right]{\cal O}\left[z^i\right]=
{\cal N}\int\limits_{0}^{+\infty}\frac{ds_i}{s_i}\int 
\limits_{u(0)=u(s_i)}^{}Du(s_i')\exp\left(-\frac14\int\limits_{0}^{s_i}
\dot u^2(s_i')ds_i'\right){\cal O}\left[u(s_i')\right],$$
where the vector $u_\mu(s_i')$ parametrizes the same contour as 
the vector $z_\mu^i(\tau)$.).

One can now write down the contribution 
of $N$ monopoles 
to the partition function of their grand canonical ensemble. 
Owing to Eq.~(\ref{calo}) it reads

$$
{\cal Z}\left[B_\mu\right]=
1+\sum\limits_{N=1}^{\infty}\frac{\zeta^N}{N!}\left<
\exp\left(i\int d^4x B_\mu j_\mu^{\rm gas}\right)
\right>_{\rm gas}=$$

\begin{equation}
\label{partaver}
=1+\sum\limits_{N=1}^{\infty}\frac{\left(2\zeta\right)^N}{N!}
\left\{\int d^4y\int Dz\mu[z]\cos\left(
g\oint dz_\mu B_\mu(x)\right)\right\}^N.
\end{equation}
Here, 
$\zeta\propto {\rm e}^{-S_0}$ again stands for the fugacity (Boltzmann 
factor of a single monopole) 
of dimension $({\rm mass})^4$ with the action of a single 
monopole given by 
$S_0=\frac{{\rm const.}}{e^2}$. Note that the constant in this 
formula is proportional to the length of the monopole loop.

In order to evaluate the path-integral over $z_\mu$'s in 
Eq.~(\ref{partaver}), 
let us employ
the dilute gas approximation, which 
requires 
that typical distances between the monopole loops are much larger than 
their sizes. This means that generally $\left|y^a\right|\gg\left|z^a
\right|$, where $\left|y\right|\equiv\sqrt{y_\mu^2}$. 
Let us denote characteristic distances $\left|y\right|$  
by $L$, characteristic sizes of monopole loops 
$\left(=\int\limits_{0}^{1}d\tau 
\sqrt{\dot z^2}\right)$ by $a$, and perform the Taylor expansion 
of $B_\mu(x)$ up to the first order in $a/L$ (which is the first one 
yielding a nonvanishing contribution to the integral 
$\oint dz_\mu B_\mu(x)$   
on the R.H.S. of Eq.~(\ref{partaver})), 

\begin{equation}
\label{taylorB}
B_\mu(x)=B_\mu(y)+z_\nu \partial_\nu B_\mu(y)+O\left(\left(\frac{a}{L}
\right)^2\right).
\end{equation}
Then, the substitution of this 
expansion into Eq.~(\ref{partaver}) yields 

$$
\int Dz\mu[z]\cos\left(
g\oint dz_\mu B_\mu(x)\right)\simeq\int Dz\mu[z] 
\cos\left(gP_{\mu\nu}[z]\partial_\nu B_\mu(y)\right)=$$

\begin{equation}
\label{nsion} 
=\sum\limits_{n=0}^{\infty}\frac{(-1)^n}{(2n)!}g^{2n}\partial_{\nu_1}
B_{\mu_1}(y)\cdots \partial_{\nu_{2n}}B_{\mu_{2n}}(y)
\int Dz\mu[z]P_{\mu_1\nu_1}[z]\cdots P_{\mu_{2n}\nu_{2n}}[z],
\end{equation}
where 
$P_{\mu\nu}[z]\equiv\oint dz_\mu z_\nu$ 
stands for the tensor area associated with the contour parametrized by 
$z_\mu(\tau)$~\footnote{One can check that for the plane contour, 
$P_{\mu\nu}=-P_{\nu\mu}=-S$, $\mu<\nu$, where $S$ is the area inside 
the contour.}. Due to the rotation- and 
translation invariance of the measure $\mu[z]$, the average of the 
product of the tensor areas can be written in the form 

\begin{equation}
\label{prodareas}
\int Dz\mu[z]P_{\mu_1\nu_1}[z]\cdots P_{\mu_{2n}\nu_{2n}}[z]=
\frac{\left(a^2\right)^{2n}}{(2n-1)!!}\left[\hat 1_{\mu_1\nu_1, 
\mu_2\nu_2}\cdots \hat 1_{\mu_{2n-1}\nu_{2n-1},\mu_{2n}\nu_{2n}}+{\,} 
{\rm permutations}{\,}\right].
\end{equation}
Here,  
$\hat 1_{\mu\nu,\lambda\rho}=\frac12\left(\delta_{\mu\lambda}
\delta_{\nu\rho}-\delta_{\mu\rho}\delta_{\nu\lambda}\right)$ 
({\it cf.} Appendix C),   
and the normalization factor $(2n-1)!!$ is explicitly extracted 
out since the sum in square brackets on the R.H.S. of 
Eq.~(\ref{prodareas}) 
contains $(2n-1)!!$ terms. 
Substituting now Eq.~(\ref{prodareas}) into 
Eq.~(\ref{nsion}) and contracting the indices, it is worth noting the
following. Due to the Hodge decomposition theorem, the field 
$B_\mu$ can always be represented as $\partial_\mu\varphi+
\varepsilon_{\mu\nu\lambda\rho}\partial_\nu h_{\lambda\rho}$, where 
$\varphi$ and $h_{\lambda\rho}$ stand for a scalar and an 
antisymmetric tensors, respectively. Owing to the conservation of 
the current $j_\mu^{\rm gas}$, the contribution of the first of them 
to the partition function~(\ref{zjgas}) vanishes. The remaining 
part of the $B_\mu$-field automatically obeys the condition 
$\partial_\mu B_\mu=0$. Taking this into account, estimating 
$\left(\partial_\mu B_\nu\right)^2$ as $|B_\mu|^2/L^2$,  
and denoting $\frac{ga^2}{L\sqrt{2}}\left(\ll a\right)$ by 
$\Lambda^{-1}$, where $\Lambda$ 
acts as a natural UV momentum cutoff,  
we finally obtain 

$$
\int Dz\mu[z]\cos\left(
g\oint dz_\mu B_\mu(x)\right)\simeq
\cos\left(\frac{\left|B_\mu(y)\right|}{\Lambda}\right).$$
This yields the desired expression for the 
partition function of the grand canonical ensemble of 
monopoles

\begin{equation}
\label{grandfour}
{\cal Z}_{\rm mon.}=\int DB_\mu\exp\left\{-\int d^4x\left[
\frac14\hat F_{\mu\nu}^2-2\zeta
\cos\left(\frac{\left|B_\mu\right|}{\Lambda}\right)\right]
\right\}.
\end{equation}
The Debye mass of the $B_\mu$-field following from the expansion 
of the cosine in the action on the R.H.S. of Eq.~(\ref{grandfour})
reads $m=\frac{\sqrt{2\zeta}}{\Lambda}$.

Let us now turn ourselves to a derivation of the representation 
for the partition function~(\ref{grandfour}) in terms of the 
monopole currents similar to the representation of its 3D
analogue~(\ref{zcos}) in terms of the monopole densities, 
given by Eqs.~(\ref{zdens}) and~(\ref{issimo}). Such a representation will 
enable us to obtain the bilocal correlator of the 
field strength tensors and to construct the string representation for 
the Wilson loop of an external electrically charged test particle in 
the 4D-case under study.  
To proceed with note that, analogously to the 3D-case, 
the integration over the field $B_\mu$ 
in Eq.~(\ref{zjgas}) leads to the Coulomb interaction between the 
monopole currents, 

$${\cal Z}\left[j_\mu^{\rm gas}\right]=\exp\left[
-\frac{1}{8\pi^2}\int d^4xd^4x'j_\mu^{\rm gas}(x)
\frac{1}{(x-x')^2}j_\mu^{\rm gas}(x')\right].$$
Averaging this expression over the grand canonical ensemble of 
monopoles we obtain 

$$
{\cal Z}_{\rm mon.}=1+\sum\limits_{N=1}^{\infty}\frac{\zeta^N}{N!}
\left<\int Dj_\mu\delta\left(j_\mu-j_\mu^{\rm gas}
\right){\cal Z}\left[j_\mu\right]
\right>_{\rm gas}=
$$

\begin{equation}
\label{newrepr}
=\int Dj_\mu D\lambda_\mu\exp\left[
-\frac{1}{8\pi^2}
\int d^4x d^4x' j_\mu(x)\frac{1}{(x-x')^2}
j_\mu(x')-i\int d^4x \lambda_\mu j_\mu+
2\zeta\int d^4x\cos\left(\frac{\left|\lambda_\mu\right|}{\Lambda}
\right)\right],
\end{equation}
where the term fixing the Fock-Schwinger gauge for the 
Lagrange multiplier $\lambda_\mu$ is assumed to be included 
into the integration measure. Notice that $Dj_\mu$ here is the standard 
integration measure over the vector field, 
which is of the same form as $D\lambda_\mu$.  

Clearly in order to integrate now over 
$\lambda_\mu$, one should solve the saddle-point equation

\begin{equation}
\label{newsaddle}
\frac{\lambda_\mu}{\left|\lambda_\mu\right|}
\sin\left(\frac{\left|\lambda_\mu\right|}{\Lambda}\right)=
-\frac{i\Lambda}{2\zeta} 
j_\mu.
\end{equation} 
This can be done by noting that its L.H.S. is a vector in the 
direction $\lambda_\mu$, which means that it can be equal to the 
R.H.S. only provided that 
the direction of the vector $\lambda_\mu$ coincides 
with the direction of the vector $j_\mu$. Therefore, it is reasonable 
to seek for a solution to Eq.~(\ref{newsaddle}) 
in the form $\lambda_\mu=\left|\lambda_\mu\right|
\frac{j_\mu}{\left|j_\mu\right|}$. Then, 
Eq.~(\ref{newsaddle}) reduces to the scalar 
equation $\sin\left(\frac{\left|\lambda_\mu\right|}{\Lambda}\right)=
-\frac{i\Lambda\left|j_\mu\right|}{2\zeta}$. Straightforward 
solution of the 
latter one yields the desired current representation  

\begin{equation}
\label{newmonrepr}
{\cal Z}_{\rm mon.}=\int Dj_\mu\exp\left\{-\left[
\frac{1}{8\pi^2}    
\int d^4x d^4x' j_\mu(x)\frac{1}{(x-x')^2}
j_\mu(x')+V\left[j_\mu\right]\right]\right\},
\end{equation}
where the effective
complex-valued potential of monopole currents reads 

$$
V\left[j_\mu\right]=$$

\begin{equation}
\label{newmonpot}
=\sum\limits_{n=-\infty}^{+\infty}
\int d^4x\left\{\Lambda \left|j_\mu\right|\left[\ln\left[
\frac{\Lambda}{2\zeta}\left|
j_\mu\right|+\sqrt{1+\left(
\frac{\Lambda}{2\zeta}\left|
j_\mu\right|\right)^2}\right]+2\pi in\right]-2\zeta
\sqrt{1+\left(
\frac{\Lambda}{2\zeta}\left|
j_\mu\right|\right)^2}
\right\}.
\end{equation}

Analogously to the 3D-case, 
by virtue of the representation~(\ref{newmonrepr}), (\ref{newmonpot}) 
it is now possible to construct the string representation for the 
Wilson loop of an external electrically charged test particle. 
Once being written via the Stokes theorem, the Wilson loop reads 

\begin{equation}
\label{fullwils}
\left<W(C)\right>=\left<\exp\left(\frac{i}{2}\int
\limits_{\Sigma}^{}d\sigma_{\mu\nu}{\cal F}_{\mu\nu}
\right)\right>_{A_\mu, j_\mu}=\left<W(C)\right>_{A_\mu}\left<
\exp\left(\frac{i}{2}\int\limits_{\Sigma}^{}d\sigma_{\mu\nu}
h_{\mu\nu}\right)\right>_{j_\mu},
\end{equation}
where ${\cal F}_{\mu\nu}=eF_{\mu\nu}+h_{\mu\nu}$ is the full 
field strength tensor, extended by the Kalb-Ramond field $h_{\mu\nu}$,
which obeys the equation $\partial_\mu\tilde h_{\mu\nu}=j_\nu$. 
Next, on the R.H.S. of Eq.~(\ref{fullwils}), the $A_\mu$-average 
is defined as 

$$\left<\ldots\right>_{A_\mu}=
\frac{\int DA_\mu\left(\ldots\right)\exp\left(
-\frac14\int d^4xF_{\mu\nu}^2\right)}{\int DA_\mu\exp\left(
-\frac14\int d^4xF_{\mu\nu}^2\right)},$$
whereas the $j_\mu$-average is defined by means of the partition 
function~(\ref{newmonrepr})-(\ref{newmonpot}). Among the two 
multipliers on the R.H.S. of Eq.~(\ref{fullwils}), the first one 
is the standard ``perimeter'' (Gaussian) contribution to the 
Wilson loop, brought about by the free photons, {\it i.e.}, 

$$\left<W(C)\right>_{A_\mu}=\exp\left(-\frac{e^2}{8\pi^2}
\oint\limits_{C}^{}dx_\mu\oint\limits_{C}^{}dy_\mu\frac{1}{(x-y)^2}
\right).$$
As far as the second one is concerned, expressing $h_{\mu\nu}$ via 
$j_\mu$, it is possible to rewrite it directly as  

\begin{equation}
\label{jwils}
\left<\exp\left(-i\int d^4xj_\mu\eta_\mu\right)\right>_{j_\mu},
\end{equation}
where 

\begin{equation}
\label{foursolid}
\eta_\mu(x)=\frac{1}{8\pi^2}\varepsilon_{\mu\nu\lambda\rho}
\frac{\partial}{\partial x_\nu}\int\limits_{\Sigma}^{}
d\sigma_{\lambda\rho}(x(\xi))\frac{1}{(x-x(\xi))^2}
\end{equation}
stands for the 4D solid angle, under which the surface $\Sigma$  
is seen by an observer at the 
point $x$ 
(If $\Sigma$ is a closed surface surrounding the point $x$  
than by virtue of the Gauss law, 
$d\tilde\sigma_{\mu\nu}\to dS_\mu\partial_\nu-dS_\nu\partial_\mu$, 
one can check that 
for the conserved current $j_\mu$, $\int d^4x j_\mu\eta_\mu=
\int dS_\mu j_\mu$, as it should be. Here, $dS_\mu$ stands for the 
oriented element of the hypersurface bounded by $\Sigma$.). 
Again, an apparent $\Sigma$-dependence 
of Eq.~(\ref{jwils}) actually drops out due to the summation over branches 
of the multivalued potential~(\ref{newmonpot}). This is the essence 
of the string representation of the Wilson loop in the 4D monopole gas.

Let us now consider the low-energy limit of the Wilson loop, {\it i.e.},  
the limit $\Lambda\left|j_\mu\right|\ll\zeta$, and investigate the 
(stable) minimum of the real branch of the potential. This corresponds to 
extracting the term with $n=0$ from the whole sum 
in Eq.~(\ref{newmonpot}). Then, since we have 
restricted ourselves to the only one branch of the potential, the  
$\Sigma$-independence of the Wilson loop is spoiled. 
In order to restore it, 
let us choose $\Sigma$ to be the surface of the minimal area 
$\Sigma=\Sigma_{\rm min.}\left[C\right]$. Then the Wilson 
loop takes the form 

$$
\left<W(C)\right>_{\rm low-energy}=\left<W(C)\right>_{A_\mu}\times
$$

\begin{equation}
\label{veryweak}
\times\int Dj_\mu\exp\left\{-\left[\frac{1}{8\pi^2} 
\int d^4x d^4x' j_\mu(x)\frac{1}{(x-x')^2}
j_\mu(x')+\frac{\Lambda^2}{4\zeta}\int d^4xj_\mu^2+
i\int d^4x j_\mu
\eta_\mu\right]\right\}, 
\end{equation}
where now $\eta_\mu$ is defined by Eq.~(\ref{foursolid}) 
with the replacement 
$\Sigma\to\Sigma_{\rm min.}$. 
Recalling the expression for $j_\mu$ via $h_{\mu\nu}$, 
Eq.~(\ref{veryweak}) can be written as follows

\begin{equation}
\label{weakviah}
\left<W(C)\right>_{\rm low-energy}=\left<W(C)\right>_{A_\mu}
\int Dh_{\mu\nu}\exp\left[-\int d^4x\left(\frac{\Lambda^2}{24\zeta}
H_{\mu\nu\lambda}^2+\frac14 h_{\mu\nu}^2\right)+\frac{i}{2}
\int\limits_{\Sigma_{\rm min.}}^{} 
d\sigma_{\mu\nu}h_{\mu\nu}\right].
\end{equation}
Note that the mass of the Kalb-Ramond field following 
from the quadratic part of the action standing in the exponent 
on the R.H.S. of 
Eq.~(\ref{weakviah}) is equal to the Debye mass $m$ of the field $B_\mu$
following from Eq.~(\ref{grandfour}). 

Integrating over the Kalb-Ramond field we obtain 

$$\left<W(C)\right>_{\rm low-energy}=\exp\left\{
-\frac18\int\limits_{\Sigma_{\rm min.}}^{}d\sigma_{\lambda\nu}(x)
\int\limits_{\Sigma_{\rm min.}}^{}d\sigma_{\mu\rho}(y)
\left<{\cal F}_{\lambda\nu}(x){\cal F}_{\mu\rho}(y)
\right>_{A_\mu, j_\mu}\right\}.$$
Here, owing to the Appendix C, 
the bilocal correlator, once being parametrized as  

$$\left<{\cal F}_{\lambda\nu}(x) 
{\cal F}_{\mu\rho}(0)
\right>_{A_\mu, j_\mu}=
\Biggl(\delta_{\lambda\mu}\delta_{\nu\rho}-\delta_{\lambda\rho}
\delta_{\nu\mu}\Biggr){\cal D}^{\rm mon.}\left(x^2\right)+$$

$$
+\frac12\Biggl[\partial_\lambda
\Biggl(x_\mu\delta_{\nu\rho}-x_\rho\delta_{\nu\mu}\Biggr)
+\partial_\nu\Biggl(x_\rho\delta_{\lambda\mu}-x_\mu\delta_{\lambda\rho}
\Biggr)\Biggr]{\cal D}_1^{\rm full}\left(x^2\right)  
$$
with ${\cal D}_1^{\rm full}={\cal D}_1^{\rm phot.}+
{\cal D}_1^{\rm mon.}$,  
yields for the coefficient functions ${\cal D}^{\rm mon.}$ and 
${\cal D}_1^{\rm mon.}$ the expressions given by Eqs.~(\ref{dvadtri}) 
and~(\ref{dvadchetyr}), respectively, with the Higgs mass 
of the dual gauge boson replaced by the 
Debye one. Consequently, similarly to the case of the Abelian-projected 
theories, the large- (and short) distance 
asymptotic behaviours of the functions ${\cal D}^{\rm mon.}$ 
and ${\cal D}_1^{\rm mon.}$ match those of the corresponding 
functions in QCD within the MFC (see discussion after 
Eq.~(\ref{dvadchetyr})). We also see that 
in the case of 4D compact QED under study, 
the r\^ole of the correlation 
length of the vacuum $T_g$ is played by the inverse 
Debye mass of the $B_\mu$-field, $m^{-1}$.
As far as the function 
${\cal D}_1^{\rm phot.}\left(x^2\right)$ is concerned, 
it turns out to be equal to $\frac{e^2}{8\pi^2|x|^4}$ 
({\it cf.} Eq.~(\ref{oge})). All this demonstrates the relevance of 
concepts of the MFC to the description of confinement in compact QED.

\subsection{A Method of Description of the String World-Sheet 
Excitations}

The results of the previous Sections tell us that strings in QCD 
within the MFC and 
the approach based on the method of Abelian projections, 
as well as in 
compact QED, can be with a good accuracy described 
by the same action of the  
massive Kalb-Ramond field interacting with the string 
world-sheet. This makes it reasonable to develop a unified mechanism 
of description of the world-sheet excitations in all these theories. 
This can be done on the basis of 
the background-field method proposed in Ref.~\cite{braat} 
for the nonlinear sigma models. 

Let us for concreteness work with 
the action standing in the exponent on the R.H.S. of Eq.~(\ref{odinnad}) 
for the case when there are no external quarks, {\it i.e.}, the string 
world-sheets are closed. This action has the form 

\begin{equation}
\label{strn}
S=\int d^4x\left(
\frac{1}{12\eta^2}H_{\mu\nu\lambda}^2+g_m^2h_{\mu\nu}^2-
i\pi h_{\mu\nu}
\Sigma_{\mu\nu}\right),
\end{equation}
({\it cf.} 
Eq.~(\ref{mon1})
where the $A_\mu$-field has been absorbed by fixing the gauge of 
the field $h_{\mu\nu}$).
In order to develop the background-field method, one should define a 
geodesics passing through the background world-sheet 
$y_\mu (\xi)$ and the 
excited one $x_\mu (\xi)=y_\mu (\xi)+z_\mu (\xi)$, where $z_\mu 
(\xi)$ stands for the world-sheet fluctuation. Such a geodesics 
has the form 
$\rho_\mu(\xi,s)=y_\mu (\xi) +sz_\mu (\xi)$, where $s$ denotes the 
arc-length parameter, $0\le s\le 1$. The expansion of the string 
effective action~(\ref{strn})  
in powers of quantum fluctuations $z_\mu (\xi)$'s can be performed 
by virtue of the 
arc-dependent term describing the interaction of the Kalb-Ramond 
field with the string, which reads  

$$S[\rho(\xi,s)]=-i\pi\int d^2\xi h_{\mu\nu}[\rho(\xi,s)]
\varepsilon^{ab}\left(\partial_a\rho_\mu (\xi,s)\right)\left(
\partial_b\rho_\nu (\xi,s)\right).$$
Then the term containing $n$ quantum fluctuations has the form 

$$S^{(n)}=\left.\frac{1}{n!}\frac{d^n}{ds^n}S[\rho(\xi,s)]
\right|_{s=0}.$$
In particular, we obtain

\begin{equation}
\label{fl15}
S^{(0)}=-i\pi\int d\sigma_{\mu\nu}(y(\xi))h_{\mu\nu}[y(\xi)],
\end{equation}

\begin{equation}
\label{fl16}
S^{(1)}=-i\pi\int d\sigma_{\mu\nu}(y(\xi))z_\lambda(\xi)H_{\mu\nu\lambda}
[y(\xi)], 
\end{equation}
and

\begin{equation}
\label{fl17}
S^{(2)}=-i\pi\int d^2\xi z_\nu(\xi)\varepsilon^{ab}\left(\partial_a 
y_\mu(\xi)\right)\Biggl(\left(\partial_b z_\lambda(\xi)\right)
H_{\nu\mu\lambda}[y(\xi)]+\frac{1}{2}z_\alpha (\xi)\left(
\partial_b y_\lambda(\xi)\right)\partial_\alpha H_{\nu\mu\lambda}
[y(\xi)]\Biggr),
\end{equation}
where, owing to the closeness of $\Sigma$,  
during the derivation of Eqs.~(\ref{fl16}) and~(\ref{fl17}) 
several full derivative terms were omitted. 

Notice that as it was discussed in Ref.~\cite{braat}, 
the terms~(\ref{fl15})-(\ref{fl17}) are 
necessary and sufficient to determine all one-loop UV 
divergencies in the theory~(\ref{strn}). That is why in what 
follows we shall restrict ourselves to the derivation of the 
effective action, quadratic in quantum fluctuations $z_\mu(\xi)$'s. 

In order to get such an action, 
we shall first carry out the integral

$$\int Dh_{\mu\nu}\exp\Biggl[-\int d^4x\Biggl(\frac{1}{12\eta^2}
H_{\mu\nu\lambda}^2+g_m^2h_{\mu\nu}^2\Biggr)-S^{(0)}-S^{(1)}
\Biggr].$$
It turns out to be equal to 

$$\exp\Biggl[-(\pi\eta)^2\int d\sigma_{\mu\nu}(y(\xi))
\int d\sigma_{\mu\nu}(y(\xi'))
D_m^{(4)}(y(\xi)-y(\xi'))+$$

\begin{equation}
\label{fl18}
+2z_\alpha(\xi)\frac{\partial}{\partial y_\alpha(\xi)}
D_m^{(4)}(y(\xi)-y(\xi'))
+z_\alpha(\xi)z_\beta(\xi')\frac{\partial^2}{\partial 
y_\alpha(\xi)\partial y_\beta(\xi')}
D_m^{(4)}(y(\xi)-y(\xi'))
\Biggr)\Biggr].
\end{equation}
Here, we have again taken into account that the world-sheets under study are 
closed, so that the boundary terms vanish. 

Secondly, one should substitute the saddle-point value 
of the integral

$$\int Dh_{\mu\nu}\exp\Biggl[-\int d^4x\Biggl(\frac{1}{12\eta^2}
H_{\mu\nu\lambda}^2+g_m^2h_{\mu\nu}^2\Biggr)-
S^{(0)} \Biggr] $$ 
into Eq.~(\ref{fl17}). This saddle-point value reads 

\begin{equation}
\label{newequation1}
h^{{\rm s.p.}}_{\mu\nu}[y(\xi)]=\frac{im^3}{8\pi g_m^2}\int 
d\sigma_{\mu\nu}(y(\xi'))\frac{K_1\left(m\left|y(\xi)-
y(\xi')\right|\right)}{\left|y(\xi)-y(\xi')\right|},
\end{equation}
where $m=2g_m\eta$ is again the mass of the Kalb-Ramond field.
Upon the substitution of Eq.~(\ref{newequation1}) 
into Eq.~(\ref{fl17}) and  
accounting for Eq.~(\ref{fl18}), 
we finally get the following 
action quadratic in quantum 
fluctuations

$$S_{{\rm quadr.}}=\frac{(\pi\eta)^2}{2}\int d\sigma_{\mu\nu}(y)
\int d\sigma_{\mu\nu}(y')\Biggl\{2z_\alpha(\xi)(y'-y)_\alpha
{\cal D}_1\left((y-y')^2\right)+
$$

$$
+z_\alpha(\xi)z_\beta(\xi')\Biggl[\delta_{\alpha\beta}
{\cal D}_1\left((y-y')^2\right)-\frac{(y-y')_\alpha 
(y-y')_\beta}{(y-y')^2}\Biggl(3{\cal D}_1\left((y-y')^2\right)+
$$

$$
+\frac{m^3}{8\pi^2\left|y-y'\right|}\left(3K_1\left(m\left|y-y'\right|
\right)+K_3\left(m\left|y-y'\right|\right)\right)\Biggr)\Biggr]
\Biggr\}-
$$ 

$$
-(\pi\eta)^2\Biggl[\int d^2\xi\varepsilon^{ab}
\left(\partial_a y_\mu\right)
z_\nu(\xi)\partial_b z_\lambda(\xi)+\frac12\int d\sigma_{\mu\lambda}
(y)z_\nu(\xi)z_\alpha(\xi)\frac{\partial}{\partial y_\alpha}\Biggr]
\times
$$

$$\times\Biggl\{\int d\sigma_{\mu\lambda}(y')(y-y')_\nu+
(\nu\to\mu,{\,}\mu\to\lambda,{\,}\lambda\to\nu)+
(\lambda\to\mu,{\,}\nu\to\lambda,{\,}\mu\to\nu)\Biggr\}
{\cal D}_1\left((y-y')^2\right),
$$
where $y\equiv y(\xi)$, $y'\equiv y(\xi')$, and the function 
${\cal D}_1$ is defined by Eq.~(\ref{dvadchetyr}). It is remarkable 
that though the interactions between the points lying on the background 
world-sheet are completely described via the function ${\cal D}$, the 
dynamics of the 
world-sheet fluctuations is governed by the 
function ${\cal D}_1$, which in the case of open world-sheets 
is responsible for the 
perimeter-type interactions. This phenomenon can be interpreted 
as an interpolation between the world-sheet and world-line dynamics, 
which is absent on the background level. 

\subsection{Abelian-Projected Theories as Ensembles of Vortex Loops 
and Modified Expressions for the Field Strength Correlators}

In our investigations of Abelian-projected theories performed in the 
previous Section, the interaction of topological defects (Abrikosov 
type electric vortices in 3D and Nielsen-Olesen type strings in 4D) 
has not been taken into account, {\it i.e.}, these objects have 
been treated as individual ones. Owing to that, string representations 
of the partition functions of Abelian-projected theories as well 
as field strength correlators in these theories were insensitive to the 
properties of the ensemble of strings as a whole and depended 
actually on a certain fixed string configuration only ({\it e.g.} a 
single string). On the other hand, it is known~\cite{pit} that in the  
case of zero temperature under study Abrikosov vortices in the 
Ginzburg-Landau theory form bound states, consisting of a vortex and 
an antivortex, which are usually referred 
to as vortex dipoles. Such vortex dipoles are short living (virtual) 
objects, whose typical sizes are much smaller than 
the typical distances between them. This means that similarly 
to monopoles in compact QED, vortex dipoles form a dilute gas.
In the 2D-case, the summation over 
the grand canonical ensemble of such dipoles in the dilute gas 
approximation  has been performed 
in Ref.~\cite{schap}, and the result has the form of the 2D 
sine-Gordon theory of the scalar field similar to Eq.~(\ref{zcos}), 
albeit with an additional mass term of the field $\chi$.
The aim of the present Subsection is to perform analogous summations  
in the 3D- and 4D-cases of the Abelian-projected $SU(2)$- and 
$SU(3)$-gluodynamics. Besides the representations in terms of the 
effective sine-Gordon theories, the related direct 
representations in terms 
of the integrals over the vortex dipoles (small vortex loops built out 
of two Nielsen-Olesen type electric strings in the 
4D-case) similar to Eqs.~(\ref{zdens}), (\ref{issimo}) 
((\ref{newmonrepr}), (\ref{newmonpot}) in 4D)
will be derived, which will enable us to evaluate correlation 
functions of vortex dipoles (loops). Such correlation functions 
will further be applied to the more precise determination of 
the correlators of field strength tensors in Abelian-projected theories.
In particular, it will be shown that the effect of Debye screening 
in the gas of vortex dipoles (loops) leads to the modification 
of the correlation length of the vacuum, at which these correlators 
decrease. Namely, it changes from the inverse mass of the dual gauge 
boson, which it acquires due to the Higgs mechanism, 
to its inverse full mass, which accounts also 
for the above mentioned Debye screening effect.

Let us start our analysis with the dual Ginzburg-Landau theory 
(3D Abelian-projected $SU(2)$-gluodynamics) in the London limit, 
whose partition function is given by Eq.~(\ref{ginland}). As it has 
been demonstrated in the previous Section, the path-integral duality 
transformation casts this partition function into the form~(\ref{threed}).
After gauging away the field $\varphi$ by the gauge transformation 
${\bf h}\to {\bf h}-\frac{1}{g_m\sqrt{2}}\nabla\varphi$ the 
partition function takes the form  

\begin{equation}
\label{newthree}
{\cal Z}=\int D{\bf x}(\tau)D{\bf h}\exp\left\{
-\int d^3x\left[\frac{1}{4\eta^2}H_{\mu\nu}^2+2g_m^2{\bf h}^2-
2\pi ih_\mu\delta_\mu\right]\right\},
\end{equation}
where we have denoted for brevity 
$H_{\mu\nu}=\partial_\mu h_\nu-\partial_\nu h_\mu$. 
Note that according to the Hodge decomposition theorem, 
the field $h_\mu$ can always be represented as $\partial_\mu\phi+
\varepsilon_{\mu\nu\lambda}\partial_\nu\psi_\lambda$. Due to the 
closeness of vortices, the field $\phi$ decouples from $\delta_\mu$.
Moreover, the $\phi$-field 
decouples from the field $\psi_\mu$ as well and yields only an 
inessential determinant factor, which is not of our interest.
Therefore, this field can be disregarded from the very beginning,
which means that the ${\bf h}$-field obeys 
the equation $\nabla {\bf h}=0$.

To proceed from individual 
vortices to the grand canonical 
ensemble of vortex dipoles, one should replace $\delta_\mu$ in 
Eq.~(\ref{newthree}) by the following expression describing 
the gas of $N$ vortex dipoles

\begin{equation}
\label{vorgas}
\delta_\mu^{\rm gas}({\bf x})=
\sum\limits_{a=1}^{N}n_a\oint dz_\mu^a(\tau)
\delta\left({\bf x}-{\bf x}^a(\tau)\right).
\end{equation}
Here, $n_a$'s stand for winding numbers, 
and we have decomposed 
the vector ${\bf x}^a(\tau)$ as ${\bf x}^a(\tau)={\bf y}^a+
{\bf z}^a(\tau)$, where ${\bf y}^a=\int\limits_{0}^{1}d\tau
{\bf x}^a(\tau)$ denotes the position of the $a$-th vortex dipole. 
In what follows, we shall restrict ourselves to the vortices 
possessing the minimal winding numbers, $n_a=\pm 1$. That is because 
the energy of a single vortex is known to be a quadratic function 
of the flux~\cite{pit}, and therefore the existence 
of two vortices of a unit flux is more energetically 
favorable than the existence of one vortex of the double flux. 
Besides that, as it has been mentioned above, 
we shall work in the dilute gas approximation, 
according to which 
characteristic distances $|{\bf y}|$, which we shall denote by $L$, 
are much larger than the characteristic sizes of vortex dipoles, 
$\int\limits_{0}^{1}d\tau\sqrt{\dot{\bf z}^2}$, which we shall denote 
by $a$.

Within these two approximations, by 
substituting Eq.~(\ref{vorgas}) into Eq.~(\ref{newthree}) 
one can proceed with 
the summation over the grand canonical ensemble of vortex dipoles. 
This procedure can be performed with quite general form of the 
measure of integration over the shapes of vortex dipoles, which 
should obey only the requirements of rotation- and 
translation invariance. Clearly, such a summation 
essentially parallels the summation over the grand canonical ensemble 
of monopoles in 4D compact QED, described in 
Subsection 4.1. 
As a result, we arrive at the following expression 
for the grand canonical partition function

\begin{equation}
\label{moregrand}
{\cal Z}=\int D{\bf h}
\exp\left\{-\int d^3x\left[\frac{1}{4\eta^2}H_{\mu\nu}^2+
2g_m^2{\bf h}^2-2\zeta\cos\left(\frac{|{\bf h}|}{\Lambda}
\right)\right]\right\}.
\end{equation}
Here, $\zeta\propto {\rm e}^{-S_0}$ again denotes the 
fugacity of a vortex dipole, which 
has the dimension $({\rm mass})^3$, with $S_0$ standing for the 
dipole's action~\footnote{This action can be estimated as $\sigma a$,
where $\sigma$ is the energy of the Abrikosov vortex 
per its unit length, {\it i.e.}, the 3D analogue of the string tension. 
With the logarithmic accuracy one has~\cite{pit} $\sigma\propto\eta^2\ln
\frac{\sqrt{\lambda}}{g_m}$ ({\it cf.} Eq.~(\ref{sigahm})), where 
$\lambda$ is the magnetic Higgs field coupling constant.},   
and $\Lambda=\frac{L}{\sqrt{2}\pi a^2}$ 
being the UV momentum cutoff. Thus, the 
summation over the grand canonical ensemble of vortex dipoles, 
built out of electric Abrikosov vortices, 
with the most general 
form of the measure of integration over their shapes yields in the 
dilute gas approximation the 
effective sine-Gordon theory~(\ref{moregrand}). In particular, 
this way of treating the gas of vortex dipoles 
leads to increasing of the mass of the  
field ${\bf h}$. Namely, expanding the cosine in 
Eq.~(\ref{moregrand}) we 
get the square of the full mass, $M^2=m^2+m_D^2\equiv Q^2\eta^2$, 
where $m=2g_m\eta$ is 
the usual mass of this field 
(equal to the mass of the dual gauge boson), 
and $m_D=\frac{\eta}{\Lambda}\sqrt{2
\zeta}$ is the additional contribution coming  
from the Debye screening. We have also introduced the full magnetic 
charge $Q=\sqrt{4g_m^2+\frac{2\zeta}{\Lambda^2}}$.

Our next aim is to derive the representation 
for the partition function~(\ref{moregrand}) directly in the 
form of an integral over the vortex dipoles. 
This can be done by making 
use of the following equality 

$$
\exp\left\{-\int d^3x\left[\frac{1}{4\eta^2}H_{\mu\nu}^2+
2g_m^2{\bf h}^2\right]\right\}=$$

$$
=\int D{\bf j}\exp\left\{-\left[\frac{\pi\eta^2}{2}
\int d^3xd^3y
j_\mu({\bf x})\frac{{\rm e}^{-m|{\bf x}-{\bf y}|}}{|{\bf x}-
{\bf y}|}j_\mu({\bf y})+2\pi i\int d^3x h_\mu j_\mu\right]\right\},$$
in whose derivation it has been used that $\nabla {\bf h}=0$.
Substituting it into Eq.~(\ref{moregrand}), 
one can straightforwardly resolve 
the resulting saddle-point equation for the field ${\bf h}$, 
$\frac{h_\mu}{|{\bf h}|}\sin\left(\frac{|{\bf h}|}{\Lambda}\right)=
-\frac{i\pi\Lambda}{\zeta}j_\mu$, which yields the desired vortex 
representation 

\begin{equation}
\label{moremore}
{\cal Z}=
\int D{\bf j}\exp\left\{-\left[\frac{\pi\eta^2}{2}
\int d^3xd^3y
j_\mu({\bf x})\frac{{\rm e}^{-m|{\bf x}-{\bf y}|}}{|{\bf x}-
{\bf y}|}j_\mu({\bf y})+V\left[2\pi j_\mu\right]\right]\right\}.
\end{equation}
Here, the complex-valued potential of the vortex dipoles 
is given by Eq.~(\ref{newmonpot}) with the replacement 
$d^4x\to d^3x$.

The obtained representation~(\ref{moremore}) can be now applied 
to the calculation of correlators of vortex dipoles. 
Indeed, it is possible to  
demonstrate that if we introduce into Eq.~(\ref{newthree}) 
(with $\delta_\mu$ 
replaced by $\delta_\mu^{\rm gas}$) a unity of the form 

\begin{equation}
\label{ljeq}
1=\int D{\bf j}\delta\left(j_\mu-
\delta_\mu^{\rm gas}
\right)=\int D{\bf j} D{\bf l}\exp\left[
-2\pi i\int d^3x l_\mu\left(j_\mu-\delta_\mu^{\rm gas}\right)
\right]
\end{equation}
and integrate out all the fields except ${\bf j}$, the result will 
coincide with Eq.~(\ref{moremore}). 
This is the reason why the correlators 
of ${\bf j}$'s are nothing else, but the correlators of the vortex 
dipoles. 
Such correlators can be calculated in the low-energy limit, {\it i.e.},
when $\Lambda|{\bf j}|\ll\zeta$. Moreover, we shall perform an 
additional approximation by restricting ourselves to 
the real branch of the effective potential of vortex dipoles, which 
corresponds to extracting from the whole sum standing on the R.H.S. of 
Eq.~(\ref{newmonpot}) the term with $n=0$. 
Within these approximations, we arrive at the following expression for the 
generating functional of the correlators of ${\bf j}$'s

$${\cal Z}[{\bf J}]\equiv\frac{
\int D{\bf j} 
\exp\left\{-\left[\frac{\pi\eta^2}{2}
\int d^3xd^3yj_\mu({\bf x})
\frac{{\rm e}^{-m|{\bf x}-{\bf y}|}}{|{\bf x}-
{\bf y}|}j_\mu({\bf y})+\int d^3x\left(-2\zeta+\frac{\pi^2\Lambda^2
{\bf j}^2}{\zeta}+J_\mu j_\mu\right)\right]\right\}} 
{\int D{\bf j} 
\exp\left\{-\left[\frac{\pi\eta^2}{2}
\int d^3xd^3yj_\mu({\bf x})
\frac{{\rm e}^{-m|{\bf x}-{\bf y}|}}{|{\bf x}-
{\bf y}|}j_\mu({\bf y})+\int d^3x\left(-2\zeta+\frac{\pi^2\Lambda^2
{\bf j}^2}{\zeta}\right)\right]\right\}}=$$

\begin{equation}
\label{Iphi}
=\exp\left[-\int d^3x\int d^3yJ_\mu({\bf x}){\cal K}({\bf x}-{\bf y})
J_\mu({\bf y})\right],
\end{equation}
where ${\cal K}({\bf x})\equiv\frac{m_D^2}{32\pi^3\eta^2}(\partial^2-m^2)
\frac{{\rm e}^{-M|{\bf x}|}}{|{\bf x}|}$. Next, since due to 
Eq.~(\ref{ljeq}) $\partial_\mu j_\mu=\partial_\mu\delta_\mu^{\rm gas}=0$,
the Hodge decomposition theorem requires that $j_\mu$ should have the 
form $j_\mu=\varepsilon_{\mu\nu\lambda}\partial_\nu\varphi_\lambda$.
Then owing to the same theorem, 
the coupling $\int d^3x J_\mu j_\mu$ is nonvanishing only provided that 
$J_\mu$ can be represented as $\varepsilon_{\mu\nu\lambda}
\partial_\nu I_\lambda$. This coupling then takes the form 
$\int d^3xI_\mu T_{\mu\nu}\varphi_\nu$, where $T_{\mu\nu}({\bf x})\equiv
\partial_\mu^x\partial_\nu^x-\delta_{\mu\nu}\partial^{x{\,}2}$. 
On the other hand, 
once being substituted into the R.H.S. of Eq.~(\ref{Iphi}), such a  
representation for $J_\mu$ yields 

$${\cal Z}[{\bf J}]=\exp\left[
-\int d^3x\int d^3y I_\mu({\bf x}) I_\nu({\bf y})T_{\mu\nu}({\bf x}) 
{\cal K}({\bf x}-{\bf y})\right].$$
Thus, varying ${\cal Z}[{\bf J}]$ twice {\it w.r.t.} ${\bf I}$, we get

$$
T_{\mu\nu}({\bf x})T_{\lambda\rho}({\bf y})\left<\varphi_\nu({\bf x})
\varphi_\rho({\bf y})\right>=-2T_{\mu\lambda}({\bf x}){\cal K}({\bf x}-
{\bf y}).$$
Due to the rotation- and translation 
invariance of space, it is further natural to 
write down for the correlator $\left<\varphi_\nu({\bf x})
\varphi_\rho({\bf y})\right>$ the following {\it ansatz}:
$\delta_{\nu\rho}f({\bf x}-{\bf y})$. This yields 

\begin{equation}
\label{feqnew}
f({\bf x})=-\frac{1}{2\pi}\int d^3y\frac{{\cal K}({\bf y})}{|{\bf x}-
{\bf y}|}.
\end{equation}
The desired correlator of ${\bf j}$'s reads

$$\left<j_\mu({\bf x})j_\nu({\bf y})\right>=
\varepsilon_{\mu\alpha\beta}\varepsilon_{\nu\rho\sigma}
\partial_\alpha^x\partial_\rho^y\left<\varphi_\beta({\bf x})
\varphi_\sigma({\bf y})\right>.$$
Taking into account the above obtained results, we eventually have 

$$\left<j_\mu({\bf x})j_\nu({\bf 0})\right>=T_{\mu\nu}({\bf x})f({\bf x})=
-\frac{m_D^2}{64\pi^4\eta^2}T_{\mu\nu}({\bf x})(\partial^{x{\,}2}-m^2)
\int d^3y\frac{{\rm e}^{-M|{\bf y}|}}{|{\bf x}-{\bf y}||{\bf y}|},$$
where in a derivation of the last equality the definition of ${\cal K}$
and Eq.~(\ref{feqnew}) have been employed. The remained integral can 
most easily be calculated by dividing the integration region over 
$|{\bf y}|$ into two parts, $[0,|{\bf x}|]$, $[|{\bf x}|,+\infty)$, 
and expanding $\frac{1}{|{\bf x}-{\bf y}|}$ in Legendre polynomials 
$P_n$'s on both of them. Then, the integration over the azimuthal
angle singles out from the whole series only the zeroth term, 
$\int\limits_{-1}^{+1}P_n(\cos\theta)d\cos\theta=2\delta_{n0}$.
After that, the integration over $|{\bf y}|$ at both intervals 
is straightforward and yields 
 
$$\int d^3y\frac{{\rm e}^{-M|{\bf y}|}}{|{\bf x}-{\bf y}||{\bf y}|}=
\frac{4\pi}{M^2|{\bf x}|}\left(1-{\rm e}^{-M|{\bf x}|}\right).$$
Note that an alternative method of calculation of this integral
(which is the 3D analogue of the respective 4D calculation, to be 
discussed below) 
can be found in Ref.~\cite{ijnew}. Finally, we arrive at the following 
expression for the bilocal correlator of vortex dipoles:

$$\left<j_\mu({\bf x})j_\nu({\bf 0})\right>=\frac{1}{\pi}
\left(\frac{m_D}{4\pi\eta M}\right)^2T_{\mu\nu}({\bf x})
\frac{1}{|{\bf x}|}\left(m^2+m_D^2{\rm e}^{-M|{\bf x}|}\right).$$
Clearly, when $m_D\to 0$, this correlator vanishes, which means that
it is the Debye screening in the ensemble of vortex dipoles, which
is responsible for their correlations.

Let us now proceed to the 4D-case of small electric 
vortex loops in the DAHM. This model is of more interest 
for us, since it is the 4D-case, where we have calculated 
field strength correlators in the previous Section, and thus
the final aim of further investigations will be to find 
modifications of these correlators due to the interactions 
in the ensemble of vortex loops.
To study the grand canonical ensemble of such loops, 
it is necessary to replace $\Sigma_{\mu\nu}$ in Eq.~(\ref{strn}) 
by the following expression 

\begin{equation}
\label{13new}
\Sigma_{\mu\nu}^{\rm gas}(x)=\sum\limits_{a=1}^{N}n_a\int 
d\sigma_{\mu\nu}(x^a(\xi))\delta(x-x^a(\xi)),
\end{equation}
where below in this Section we shall set $\xi\in [0,1]\times [0,1]$. 
After that, the summation over the grand canonical ensemble of 
vortex loops is similar to the summation over the ensemble of vortex 
dipoles in the dual Ginzburg-Landau theory. Referring the reader 
for the details to Appendix D, we shall present here the result of 
this procedure, which has the form

\begin{equation}
\label{14new}
{\cal Z}=\int Dh_{\mu\nu}\exp
\left\{-\int d^4x\left[\frac{1}{12\eta^2}H_{\mu\nu\lambda}^2+
g_m^2h_{\mu\nu}^2-2\zeta\cos\left(\frac{\left|h_{\mu\nu}
\right|}{\Lambda^2}\right)\right]\right\}.
\end{equation} 
Here $\left|h_{\mu\nu}\right|\equiv\sqrt{h_{\mu\nu}^2}$, and 
the fugacity $\zeta$ (Boltzmann factor of a single vortex loop) 
has now the dimension $({\rm mass})^4$. We have also introduced 
a new UV momentum cutoff $\Lambda$ equal to $\sqrt{\frac{L}{\pi a^3}}$ 
with $L$ and $a$ denoting the characteristic distances between 
vortex loops and their typical sizes, respectively.
The square of the 
full mass of the field $h_{\mu\nu}$ following from Eq.~(\ref{14new}) 
reads $M^2=m^2+m_D^2\equiv Q^2\eta^2$. Here,  
$m=2g_m\eta$ is the usual Higgs contribution, 
$m_D=\frac{2\eta\sqrt{\zeta}}{\Lambda^2}$ is the Debye contribution, and 
$Q=2\sqrt{g_m^2+\frac{\zeta}{\Lambda^4}}$ 
is the full magnetic charge.

The representation for the partition function~(\ref{14new})
in terms of the vortex loops can be obtained 
by virtue of the following equality

$$
\exp
\left\{-\int d^4x\left[\frac{1}{12\eta^2}H_{\mu\nu\lambda}^2+
g_m^2h_{\mu\nu}^2\right]\right\}=$$

\begin{equation}
\label{sovsemnew}
=\int DS_{\mu\nu}\exp\left\{-\left[(\pi\eta)^2
\int d^4x\int d^4y S_{\mu\nu}(x)D_m^{(4)}(x-y)
S_{\mu\nu}(y)+i\pi\int d^4x h_{\mu\nu}S_{\mu\nu}\right]\right\}.
\end{equation}
It can be derived by taking into account that $\partial_\mu h_{\mu\nu}=0$.
Indeed, owing to the Hodge decomposition theorem, the Kalb-Ramond
field can always be represented as follows: $h_{\mu\nu}=\partial_\mu
\varphi_\nu-\partial_\nu\varphi_\mu+\varepsilon_{\mu\nu\lambda\rho}
\partial_\lambda\psi_\rho$. Clearly, the field $\varphi_\mu$ decouples 
not only from $\Sigma_{\mu\nu}^{\rm gas}$ (due to the conservation 
of the latter one), but also from $\psi_\mu$. The $\varphi_\mu$-field
thus yields only an inessential determinant factor, which 
is not of our interest. Therefore this field can be disregarded, 
which proves the above statement.

Substituting Eq.~(\ref{sovsemnew}) 
into Eq.~(\ref{14new}), we can integrate the field 
$h_{\mu\nu}$ out, which yields the desired representation 
for the partition function~(\ref{14new}), 

\begin{equation}
\label{15new}
{\cal Z}=\int DS_{\mu\nu}\exp\left\{
-\left[(\pi\eta)^2
\int d^4x\int d^4y S_{\mu\nu}(x)D_m^{(4)}(x-y)
S_{\mu\nu}(y)
+V\left[\pi\Lambda S_{\mu\nu}\right]\right]\right\}
\end{equation}
with the effective potential $V$ given by Eq.~(\ref{newmonpot}). 

Similarly to the 3D case, correlation functions of $S_{\mu\nu}$'s, 
calculated by virtue of the partition function~(\ref{15new}), are nothing 
else, but the correlation functions of vortex loops in the gas. This can 
be seen by mentioning that if we insert into the partition function 
${\cal Z}=\int Dx_\mu(\xi)Dh_{\mu\nu}{\rm e}^{-S}$ with the action 
$S$ given by Eq.~(\ref{strn}) and 
$\Sigma_{\mu\nu}$ replaced by $\Sigma_{\mu\nu}^{\rm gas}$ the following  
unity 

\begin{equation}
\label{ssigmauni}
1=\int DS_{\mu\nu}\delta\left(S_{\mu\nu}-
\Sigma_{\mu\nu}^{\rm gas}\right)=\int DS_{\mu\nu}
Dl_{\mu\nu}\exp\left[-i\pi\int d^4x l_{\mu\nu}\left(S_{\mu\nu}-
\Sigma_{\mu\nu}^{\rm gas}\right)\right]
\end{equation}
and integrate out all the fields except $S_{\mu\nu}$, the result 
will coincide with Eq.~(\ref{15new}). Such correlation functions    
of the vortex loops 
can be most easily calculated in the low-energy limit, $\Lambda^2
\left|S_{\mu\nu}\right|\ll\zeta$, by restricting oneself to the 
vicinity of the minimum 
of the real branch of the potential~(\ref{newmonpot}), 
where it reduces just to a quadratic functional of $S_{\mu\nu}$.
To calculate these correlation functions, 
let us derive an expression for the respective 
generating functional. It reads

$${\cal Z}[J_{\mu\nu}]=$$

$$=\frac{
\int DS_{\mu\nu} 
\exp\left\{
-\left[(\pi\eta)^2
\int d^4x\int d^4y S_{\mu\nu}(x)D_m^{(4)}(x-y)
S_{\mu\nu}(y)+\int d^4x\left(\frac{\pi^2\Lambda^4}{4\zeta}S_{\mu\nu}^2
+J_{\mu\nu}S_{\mu\nu}\right)\right]\right\}}{\int DS_{\mu\nu}
\exp\left\{
-\left[(\pi\eta)^2
\int d^4x\int d^4y S_{\mu\nu}(x)D_m^{(4)}(x-y)
S_{\mu\nu}(y)
+\frac{\pi^2\Lambda^4}{4\zeta}\int d^4x S_{\mu\nu}^2
\right]\right\}}=$$

\begin{equation}
\label{ss4d}
=\exp\left[-\int d^4x\int d^4yJ_{\mu\nu}(x){\cal G}(x-y)J_{\mu\nu}(y)
\right],
\end{equation}
where 

\begin{equation}
\label{calGM}
{\cal G}(x)\equiv\frac{\zeta}{\pi^2\Lambda^4}(\partial^2-m^2)
D_M^{(4)}(x).
\end{equation}
Next, since $\partial_\mu\Sigma_{\mu\nu}^{\rm gas}=0$, 
Eq.~(\ref{ssigmauni}) requires that $\partial_\mu S_{\mu\nu}=0$ 
as well. Then, the Hodge decomposition theorem leads to the 
following representation for $S_{\mu\nu}$: $S_{\mu\nu}=
\varepsilon_{\mu\nu\lambda\rho}\partial_\lambda\varphi_\rho$.
Owing to this fact and the same theorem, the coupling $\int d^4x
J_{\mu\nu}S_{\mu\nu}$ will be nonvanishing only provided that 
$J_{\mu\nu}=\varepsilon_{\mu\nu\lambda\rho}\partial_\lambda
I_\rho$. This coupling then reads $2\int d^4x I_\mu T_{\mu\nu}
\varphi_\nu$. On the other hand, substituting the above representation 
for $J_{\mu\nu}$ into the R.H.S. of Eq.~(\ref{ss4d}), we have 

$${\cal Z}[J_{\mu\nu}]=\exp\left[-2\int d^4x\int d^4y I_\mu(x)
I_\nu(y)T_{\mu\nu}(x){\cal G}(x-y)\right].$$
Thus, varying ${\cal Z}[J_{\mu\nu}]$ twice {\it w.r.t.} $J_{\mu\nu}$
and setting then $J_{\mu\nu}=0$, we get

$$T_{\mu\nu}(x)T_{\lambda\rho}(y)\left<\varphi_\nu(x)
\varphi_\rho(y)\right>=-T_{\mu\lambda}(x){\cal G}(x-y).$$
Again, due to the rotation- and translation invariance 
of space-time, it is natural to seek for 
$\left<\varphi_\nu(x)\varphi_\rho(y)\right>$
in the form of the following {\it ansatz}: $\delta_{\nu\rho}
g(x-y)$. This yields the equation $\partial^2g={\cal G}$, whose 
solution reads 

$$g(x)=-\frac{\zeta}{\pi^2\Lambda^4}(\partial^{x{\,}2}-m^2)
\int d^4yD_0^{(4)}(x-y)D_M^{(4)}(y),$$
where $D_0^{(4)}(x)\equiv D_m^{(4)}(x)$ at $m=0$, {\it i.e.},
it is just $\frac{1}{4\pi^2x^2}$. The last integral can obviously be 
rewritten as 

\begin{equation}
\label{xzinte}
\int d^4z D_0^{(4)}(z)D_M^{(4)}(z-x).
\end{equation}
As we will see
below, it will be necessary to know the more general expression,
namely that for the integral 

\begin{equation}
\label{mMint}
\int d^4z D_m^{(4)}(z)D_M^{(4)}(z-x).
\end{equation}
Its calculation is outlined in Appendix E, and the result 
reads

\begin{equation}
\label{Mmres}
\frac{1}{m_D^2}\left(D_m^{(4)}(x)-D_M^{(4)}(x)\right).
\end{equation}
Note that as it obviously follows from Eq.~(\ref{mMint}), 
since this result depends on $x$ only as $|x|$,
it should be symmetric {\it w.r.t.} interchange $m\leftrightarrow M$.
One can see that this really holds: Eq.~(\ref{Mmres}) is invariant
under this interchange, since during it $m_D^2$ changes its sign.
 
Setting now in Eq.~(\ref{Mmres}) $m=0$, we get 
$\frac{1}{m_D^2}
\left(D_0^{(4)}(x)-D_{m_D}^{(4)}(x)\right)$,~\footnote{
Note that 
this result can also be obtained directly by making use of the method 
presented in Appendix E, which was done in Ref.~\cite{bohmplb}.} 
which yields for the 
desired integral~(\ref{xzinte}) the same result with the 
substitution $m_D\to M$. 
Thus, the final expression for the function $g$ reads

\begin{equation}
\label{newg}
g(x)=\frac{\zeta}{(\pi M\Lambda^2)^2}(\partial^2-m^2)\left(
D_M^{(4)}(x)-D_0^{(4)}(x)\right).
\end{equation}
The desired correlator of $S_{\mu\nu}$'s 
has the form 

$$\left<S_{\mu\nu}(x)S_{\lambda\rho}(y)\right>=
\varepsilon_{\mu\nu\alpha\beta}\varepsilon_{\lambda\rho\gamma\sigma}
\partial_\alpha^x\partial_\gamma^y\left<\varphi_\beta(x)\varphi_\sigma(y)
\right>$$
and therefore  

$$
\left<S_{\mu\nu}(x)S_{\lambda\rho}(0)\right>=
-\varepsilon_{\mu\nu\alpha\beta}\varepsilon_{\lambda\rho\gamma\beta}
\partial_\alpha^x\partial_\gamma^x g(x)=$$

\begin{equation}
\label{SScor}
=\left(\delta_{\lambda\nu}\delta_{\mu\rho}-\delta_{\nu\rho}
\delta_{\mu\lambda}\right){\cal G}(x)+\left(\delta_{\mu\lambda}
\partial_\rho\partial_\nu+\delta_{\nu\rho}\partial_\mu\partial_\lambda-
\delta_{\mu\rho}\partial_\lambda\partial_\nu-\delta_{\lambda\nu}
\partial_\mu\partial_\rho\right)g(x),
\end{equation}
where it has been used that $\partial^2g(x)={\cal G}(x)$.

This result can immediately be applied to the calculation of the
string contribution to the bilocal cumulant~(\ref{dvaddva}).
Indeed, applying to the average on the R.H.S. of Eq.~(\ref{otherhand}) 
the cumulant expansion in the bilocal approximation, one gets: 

$${\cal Z}\simeq\exp\Biggl\{-\int d^4x\int d^4y D_m^{(4)}(x-y)\left[
(4\pi\eta)^2\Sigma_{\mu\nu}^E(x)\Sigma_{\mu\nu}^E(y)+\frac12
j_\mu^E(x)j_\mu^E(y)\right]+$$

$$+32(\pi\eta)^4\int d^4xd^4yd^4zd^4uD_m^{(4)}(x-z)D_m^{(4)}(y-u)
\Sigma_{\mu\nu}^E(x)\Sigma_{\lambda\rho}^E(y)\left<\left<\Sigma_{\mu\nu}(z)
\Sigma_{\lambda\rho}\right>\right>_{x_\mu(\xi)}\Biggr\}.$$
Comparing this expression with Eq.~(\ref{Zonehand}), we see that 
owing to Eq.~(\ref{SScor}), 
the additional 
string contribution to the cumulant~(\ref{dvaddva}) has the form 

$$\Delta\left<\left<f_{\mu\nu}(x)f_{\lambda\rho}(y)\right>
\right>_{a_\mu, j_\mu^M}=\left(4\pi g_m\eta^2\right)^2\int d^4zd^4u
D_m^{(4)}(x-z)D_m^{(4)}(y-u)\times$$

$$\times\left\{\left(\delta_{\mu\lambda}\delta_{\nu\rho}-
\delta_{\mu\rho}\delta_{\nu\lambda}\right){\cal G}(z-u)+
\left[\delta_{\mu\rho}\partial_\lambda^z\partial_\nu^z+
\delta_{\nu\lambda}\partial_\mu^z\partial_\rho^z-\delta_{\mu\lambda}
\partial_\rho^z\partial_\nu^z-\delta_{\nu\rho}\partial_\mu^z
\partial_\lambda^z\right]g(z-u)\right\}.$$
Comparing this equation further with Eq.~(\ref{dvaddva}) 
and taking into account that 

$$(x-y)_\mu{\cal D}_1\left((x-y)^2\right)=
-\frac12\partial_\mu^x G\left((x-y)^2\right),$$ 
where the function $G$ 
is defined by Eq.~(\ref{g}) with the replacement $D_1\to{\cal D}_1$,
we arrive at the following system of equations, which determine 
string contributions to the dunctions ${\cal D}$ and $G$:

\begin{equation}
\label{deltaD}
\Delta{\cal D}\left((x-y)^2\right)=\left(4\pi g_m\eta^2\right)^2
\int d^4z d^4uD_m^{(4)}(x-z)D_m^{(4)}(y-u){\cal G}(z-u),
\end{equation}

\begin{equation}
\label{deltaG}
\Delta G\left((x-y)^2\right)=\left(8\pi g_m\eta^2\right)^2
\int d^4z d^4uD_m^{(4)}(x-z)D_m^{(4)}(y-u)g(z-u).
\end{equation}
Inserting now Eq.~(\ref{calGM}) into Eq.~(\ref{deltaD}), we get

$$\Delta{\cal D}\left((x-y)^2\right)=-\frac{\left(4g_m\eta^2\right)^2
\zeta}{\Lambda^4}\int d^4uD^{(4)}_m(y-u)D^{(4)}_M(x-u).$$ 
By virtue 
of the result of Appendix E, we have

$$\Delta{\cal D}\left(x^2\right)=\frac{m^2}{4\pi^2}\left[
\frac{M}{|x|}K_1(M|x|)-\frac{m}{|x|}K_1(m|x|)\right].$$
Adding this result to Eq.~(\ref{dvadtri}), 
we finally obtain for the finction ${\cal D}$ the following full result:

\begin{equation}
\label{Dtot}
{\cal D}^{\rm full}\left(x^2\right)=\frac{m^2M}{4\pi^2}
\frac{K_1(M|x|)}{|x|}.
\end{equation}
Analogously, inserting Eq.~(\ref{newg}) into Eq.~(\ref{deltaG}),
we have 

$$\Delta G\left((x-y)^2\right)=\zeta\left(\frac{8g_m\eta^2}{\Lambda^2M}
\right)^2\int d^4uD_m^{(4)}(y-u)\left[D_0^{(4)}(x-u)-D_M^{(4)}(x-u)
\right],$$ 
or further by virtue of Appendix E, 

$$\Delta G\left(x^2\right)=\left(\frac{m_D}{\pi M|x|}\right)^2+
\left(\frac{2m}{M}\right)^2D_M^{(4)}(x)-4D_m^{(4)}(x).$$ 
Together with Eq.~(\ref{dvadchetyr}), 
this yields the following full result for 
the function ${\cal D}_1$:

\begin{equation}
\label{D1tot}
{\cal D}_1^{\rm full}\left(x^2\right)=\frac{m_D^2}{\pi^2M^2|x|^4}+
\frac{m^2}{2\pi^2Mx^2}\left[\frac{K_1(M|x|)}{|x|}+\frac{M}{2}
\left(K_0(M|x|)+K_2(M|x|)\right)\right].
\end{equation}

It is worth noting that the functions $\Delta{\cal D}$ and 
$\Delta{\cal D}_1$ contained the terms exactly equal to 
Eqs.~(\ref{dvadtri}) and (\ref{dvadchetyr}), respectively, but with the 
opposite sign,
which just canceled out in the full 
functions~(\ref{Dtot}) and (\ref{D1tot}).
We also see that, as it should be, the functions~(\ref{Dtot}) and 
(\ref{D1tot}) go over into Eqs.~(\ref{dvadtri}) and (\ref{dvadchetyr}),
respectively, when $m_D\to 0$, {\it i.e.}, when one neglects the effect of 
screening in the ensemble of vortex loops. An obvious important 
consequence of the obtained Eqs.~(\ref{Dtot}) and (\ref{D1tot}) is that
the correlation length of the vacuum, $T_g$, becomes modified 
from $1/m$ (according to Eqs.~(\ref{dvadtri}) and (\ref{dvadchetyr}))
to $1/M$. (It is worth emphasizing once more that this effect is just 
due to the Debye screening of magnetic charge of the dual vector boson 
in the ensemble of electrically charged vortex loops, which makes 
this particle more heavy, namely enlarges its mass from $m$ to $M$.)
Indeed, it is straightforward to see that at $|x|\gg\frac{1}{M}$,

$${\cal D}^{\rm full}\longrightarrow\frac{(mM)^2}{4\sqrt{2}
\pi^{\frac32}}\frac{{\rm e}^{-M|x|}}{(M|x|)^{\frac32}}$$
and

$${\cal D}_1^{\rm full}\longrightarrow
\frac{m_D^2}{\pi^2M^2|x|^4}+\frac{(mM)^2}{2\sqrt{2}\pi^{\frac32}}
\frac{{\rm e}^{-M|x|}}{(M|x|)^{\frac52}}.$$
It is also remarkable  
that the leading term of the large-distance asymptotics 
of the function ${\cal D}_1^{\rm full}$ is a pure power-like one, rather 
than that of the function ${\cal D}_1$, given by Eq.~(\ref{dvadshest}).
Another nontrivial result is that the screening does not change the 
short-distance asymptotic behaviours of the functions~(\ref{dvadtri}) 
and (\ref{dvadchetyr}), {\it i.e.}, the short-distance asymptotics of 
the functions~(\ref{Dtot}) and (\ref{D1tot}) are given by 
Eqs.~(\ref{dvadsem}) and (\ref{dvadvosem}), respectively.

Let us now turn ourselves to the case of the Abelian-projected 
$SU(3)$-gluodynamics. In the case when Abrikosov-Nielsen-Olesen type 
electric strings in this theory are considered as noninteracting objects, 
the expression for the partition function is given by Eq.~(\ref{suz6}), 
which can be shortly written as 

$$
{\cal Z}=\int Dx_\mu^1(\xi) Dx_\mu^2(\xi)\times
$$

\begin{equation}
\label{5su3}
\times\exp\left\{
-g_m\eta^3\sqrt{\frac32}\int d^4xd^4y\left[\Sigma_{\mu\nu}^1(x)
\Sigma_{\mu\nu}^1(y)+\Sigma_{\mu\nu}^1(x)\Sigma_{\mu\nu}^2(y)+
\Sigma_{\mu\nu}^2(x)\Sigma_{\mu\nu}^2(y)\right]
\frac{K_1(m_B|x-y|)}{|x-y|}\right\}.
\end{equation}
In order to proceed from the individual strings to the grand canonical 
ensemble of interacting vortex loops, one should replace 
$\Sigma_{\mu\nu}^a(x)$, where from now on $a=1,2$, in Eq.~(\ref{5su3}) 
by 

$$
\Sigma_{\mu\nu}^{a{\,}{\rm gas}}(x)=\sum\limits_{k=1}^{N}n_k^{a}
\int d\sigma_{\mu\nu}\left(x_k^a(\xi)\right)\delta\left(
x-x_k^a(\xi)\right).
$$
Here, $n_k^a$'s stand for winding numbers, which 
we shall again set to be equal $\pm 1$. 
Performing such a replacement, one can see the crucial difference 
of the grand canonical ensemble of small vortex loops in the model 
under study from that in the Abelian-projected 
$SU(2)$-gluodynamics, studied above. Namely, the system 
has now the form of two interacting gases consisting of the vortex loops 
of two kinds, while in the $SU(2)$-case the gas was built out of 
vortex loops of the only one kind.

Analogously to that case, we shall 
treat such a grand canonical ensemble of vortex loops in the dilute 
gas approximation. According to it, characteristic sizes of loops are 
much smaller than characteristic distances between them, which in 
particular means that the vortex loops are short living objects.
Then the summation over this grand canonical ensemble
can be most easily performed by inserting the 
unity 

\begin{equation}
\label{auxsu3}
1=\int DS_{\mu\nu}^a\delta\left(S_{\mu\nu}^a-
\Sigma_{\mu\nu}^{a{\,}{\rm gas}}\right)
\end{equation} 
into the R.H.S. of Eq.~(\ref{5su3})
(with $\Sigma_{\mu\nu}^a$ replaced by $\Sigma_{\mu\nu}^{a{\,}{\rm gas}}$) 
and representing the $\delta$-functions as the integrals over Lagrange 
multipliers. Then, the contribution of $N$ vortex loops of each kind 
to the grand canonical ensemble takes the following form  

$$
{\cal Z}\left[\Sigma_{\mu\nu}^{a{\,}{\rm gas}}\right]
=\int DS_{\mu\nu}^a D\lambda_{\mu\nu}^a\times$$

$$
\times\exp\left\{
-g_m\eta^3\sqrt{\frac32}\int d^4xd^4y\left[S_{\mu\nu}^1(x)
S_{\mu\nu}^1(y)+S_{\mu\nu}^1(x)S_{\mu\nu}^2(y)+
S_{\mu\nu}^2(x)S_{\mu\nu}^2(y)\right]
\frac{K_1(m_B|x-y|)}{|x-y|}-\right.$$

\begin{equation}
\label{6su3}
\left.-i\int d^4x\lambda_{\mu\nu}^a\left(S_{\mu\nu}^a-
\Sigma_{\mu\nu}^{a{\,}{\rm gas}}\right)\right\}.
\end{equation}
After that, the desired summation is straightforward, since it technically 
parallels the one of Abelian-projected $SU(2)$-gluodynamics. 
We have 

$$
\left\{1+\sum\limits_{N=1}^{\infty}\frac{\zeta^N}{N!}\left(
\prod\limits_{i=1}^{N}\int d^4y_i^1\int Dz_i^1(\xi)\mu
\left[z_i^1\right]\right)\times\right.$$

$$\left.\times\sum\limits_{n_k^1=\pm 1}^{}
\exp\left[i\sum\limits_{k=1}^{N}n_k^1\int d\sigma_{\mu\nu}
\left(z_k^1(\xi)\right)\lambda_{\mu\nu}^1\left(x_k^1(\xi)\right)
\right]\right\}\times$$

$$\times
\left\{ {\rm the}~ {\rm same}~ {\rm term}~ {\rm with}~ {\rm the}~ 
{\rm replacement}~ {\rm of}~ {\rm indices}~ 1\to 2
\right\}=$$

\begin{equation}
\label{7su3}
=\exp\left\{2\zeta\int d^4y\left[\cos\left(\frac{\left|
\lambda_{\mu\nu}^1(y)\right|}{\Lambda^2}\right)+
\cos\left(\frac{\left|
\lambda_{\mu\nu}^2(y)\right|}{\Lambda^2}\right)\right]\right\}.
\end{equation}
Here, the world-sheet coordinate of the $k$-th vortex loop 
of the $a$-th type~\footnote{For brevity, we omit the Lorentz index.} 
$x_k^a(\xi)$ has been decomposed 
as $x_k^a(\xi)=y_k^a+z_k^a(\xi)$, where the vector 
$y_k^a\equiv\int d^2\xi x_k^a(\xi)$ describes the position 
of the vortex loop, whereas the vector $z_k^a(\xi)$ describes 
its shape. Next, on the L.H.S. of Eq.~(\ref{7su3}), $\mu\left[z_i^a
\right]$ again stands for a certain rotation- and translation invariant 
measure of integration over the shapes of vortex 
loops, and $\zeta\propto {\rm e}^{-S_0}$ denotes the fugacity 
(Boltzmann factor of a single vortex loop~\footnote{It is natural 
to assume that the vortex loops of different kinds have the same 
fugacity, since different $\theta_a^{\rm sing.}$'s enter
the initial partition function~(\ref{suz2}) in the same way.}) 
of dimension $({\rm mass})^4$ 
with $S_0$ being the action of a single loop. In Eq.~(\ref{7su3}), 
we have also introduced the UV momentum cutoff $\Lambda\equiv\sqrt{
\frac{L}{a^3}}$ $\left(\gg a^{-1}\right)$, 
where $a$ is a typical size of the 
vortex loop, and $L$ is a typical distance between loops, so that 
in the dilute gas approximation under study $a\ll L$. Finally in 
Eq.~(\ref{7su3}), we have denoted $\left|\lambda_{\mu\nu}^a\right|\equiv
\sqrt{\left(\lambda_{\mu\nu}^a\right)^2}$. 
 
Next, it is possible to integrate out the Lagrange multipliers 
by solving the saddle-point equation following from Eqs.~(\ref{6su3}) 
and~(\ref{7su3}), 

$$\frac{\lambda_{\mu\nu}^a}{\left|\lambda_{\mu\nu}^a\right|}\sin\left(
\frac{\left|\lambda_{\mu\nu}^a\right|}{\Lambda^2}\right)=
-\frac{i\Lambda^2}{2\zeta}S_{\mu\nu}^a.$$
After that, we 
arrive at the following representation for the partition function 
of the grand canonical ensemble 

$$
{\cal Z}_{\rm grand}=
\int DS_{\mu\nu}^a
\exp\left\{-\left[
g_m\eta^3\sqrt{\frac32}\int d^4xd^4y\left[S_{\mu\nu}^1(x)
S_{\mu\nu}^1(y)+S_{\mu\nu}^1(x)S_{\mu\nu}^2(y)+
S_{\mu\nu}^2(x)S_{\mu\nu}^2(y)\right]\times\right.\right.$$

\begin{equation}
\label{8su3}
\left.\left.\times
\frac{K_1(m_B|x-y|)}{|x-y|}+V\left[\Lambda S_{\mu\nu}^1\right]+
V\left[\Lambda S_{\mu\nu}^2\right]\right]\right\},
\end{equation}
which owing to Eq.~(\ref{auxsu3}) is natural to be referred to as 
the representation in terms of the vortex loops. In Eq.~(\ref{8su3}), 
the effective potential of 
vortex loops $V\left[\Lambda S_{\mu\nu}^a\right]$ is given by 
Eq.~(\ref{newmonpot}).

It is further instructive to illustrate the difference of such a 
partition function of two interacting gases of vortex loops from the 
case of Abelian-projected $SU(2)$-gluodynamics by studying a related 
representation in terms of a certain effective sine-Gordon 
theory. This can be done by introducing the new integration variables 
${\cal S}_{\mu\nu}^1=\frac{\sqrt{3}}{2}\left(S_{\mu\nu}^1+S_{\mu\nu}^2
\right)$ and ${\cal S}_{\mu\nu}^2=\frac12\left(S_{\mu\nu}^1-S_{\mu\nu}^2
\right)$, which diagonalize the quadratic form in square brackets 
on the R.H.S. of Eq.~(\ref{6su3}). Then Eqs.~(\ref{6su3}) 
and~(\ref{7su3}) yield

$${\cal Z}_{\rm grand}=\int D{\cal S}_{\mu\nu}^a
D\lambda_{\mu\nu}^a
\exp\left\{
-g_m\eta^3\sqrt{\frac32}\int d^4xd^4y
{\cal S}_{\mu\nu}^a(x)\frac{K_1(m_B|x-y|)}{|x-y|}{\cal S}_{\mu\nu}^a(y)+
\right.$$

\begin{equation}
\label{10su3}
\left.+2\zeta\int d^4x\left[\cos\left(\frac{\left|
\lambda_{\mu\nu}^1(x)\right|}{\Lambda^2}\right)+
\cos\left(\frac{\left|
\lambda_{\mu\nu}^2(x)\right|}{\Lambda^2}\right)\right]
-i\int d^4xh_{\mu\nu}^a{\cal S}_{\mu\nu}^a\right\},
\end{equation}
where we have denoted
$h_{\mu\nu}^1=\frac{1}{\sqrt{3}}
\left(\lambda_{\mu\nu}^1+\lambda_{\mu\nu}^2\right)$ and 
$h_{\mu\nu}^2=\lambda_{\mu\nu}^1-\lambda_{\mu\nu}^2$. The partition 
function of the desired sine-Gordon theory can be obtained 
from Eq.~(\ref{10su3}) by making use of the following 
equality

$$
\int  D{\cal S}_{\mu\nu}^a
\exp\left\{
-\left[g_m\eta^3\sqrt{\frac32}\int d^4xd^4y
{\cal S}_{\mu\nu}^a(x)\frac{K_1(m_B|x-y|)}{|x-y|}{\cal S}_{\mu\nu}^a(y)+
i\int d^4x h_{\mu\nu}^a{\cal S}_{\mu\nu}^a\right]\right\}=
$$

\begin{equation}
\label{calSh}
=\exp\left\{-\frac{1}{2\pi^2}\int d^4x\left[\frac{1}{12\eta^2}
\left(H_{\mu\nu\lambda}^a\right)^2+\frac32 
g_m^2\left(h_{\mu\nu}^a\right)^2
\right]\right\}
\end{equation}
({\it cf.} the R.H.S. with the quadratic part of the action of the 
Kalb-Ramond fields on the R.H.S. of Eq.~(\ref{su3}) with the fields 
$a_\mu^a$ gauged away). It can easily be proved
by noting that due to the Hodge decomposition theorem and
the equation $\partial_\mu{\cal S}_{\mu\nu}^a=0$ (which follows 
from Eq.~(\ref{auxsu3}) and conservation of 
$\Sigma_{\mu\nu}^{a{\,}{\rm gas}}$), $\partial_\mu h_{\mu\nu}^a=0$. 
Substituting further Eq.~(\ref{calSh}) into Eq.~(\ref{10su3})
and performing the rescaling $\frac{h_{\mu\nu}^a}{\pi\sqrt{2}}
\to h_{\mu\nu}^a$, 
we arrive at the following representation for the partition function 
of the grand canonical ensemble of vortex loops in terms of the 
local sine-Gordon theory, equivalent to the 
nonlocal theory~(\ref{8su3}): 

$${\cal Z}_{\rm grand}=\int Dh_{\mu\nu}^a\exp\left\{
-\int d^4x\left\{
\frac{1}{12\eta^2}
\left(H_{\mu\nu\lambda}^a\right)^2+\frac32 
g_m^2\left(h_{\mu\nu}^a\right)^2
-\right.\right.$$

\begin{equation}
\label{11su3}
\left.\left.-2\zeta\left[\cos\left(\frac{\pi}{\Lambda^2\sqrt{2}}
\left|\sqrt{3}
h_{\mu\nu}^1+h_{\mu\nu}^2\right|\right)+ 
\cos\left(\frac{\pi}{\Lambda^2\sqrt{2}}\left|\sqrt{3}
h_{\mu\nu}^1-h_{\mu\nu}^2\right|\right)\right]\right\}\right\}.
\end{equation}
As we now see, an essential property of the obtained sine-Gordon theory, 
which distinguishes it from an analogous theory describing the 
grand canonical ensemble of vortex loops in the 
Abelian-projected $SU(2)$-gluodynamics, is the presence 
of two interacting Kalb-Ramond fields, while in the $SU(2)$-case there 
was only one self-interacting field. Notice that upon the expansion 
of the cosines on the R.H.S. of Eq.~(\ref{11su3}), 
it is straightforward to see 
that only the interaction terms of the type $\left(h_{\mu\nu}^1
\right)^{2n}\left(h_{\mu\nu}^2\right)^{2k}$ survive. In another words, 
despite of the mixing of the Kalb-Ramond 
fields in the arguments of the cosines, no terms linear in any of 
these fields appear in the action. In particular, the full masses of 
both Kalb-Ramond fields, $M_1$ and $M_2$, 
can be read off from Eq.~(\ref{11su3}) by 
expanding the cosines up to the quadratic terms. The result reads 
$M_a^2=m_B^2+m_a^2\equiv Q_a^2\eta^2$, 
where $m_1=\frac{2\pi\eta}{\Lambda^2}
\sqrt{3\zeta}$, $m_2=\frac{2\pi\eta}{\Lambda^2}\sqrt{\zeta}$
are the Debye masses, and we have introduced the magnetic charges 
$Q_1=\sqrt{6g_m^2+\frac{12\pi^2\zeta}{\Lambda^4}}$, 
$Q_2=\sqrt{6g_m^2+\frac{4\pi^2\zeta}{\Lambda^4}}$.  

Equation~(\ref{8su3}) can now be used for the evaluation 
of correlators of vortex loops, which due to Eq.~(\ref{auxsu3}), 
are nothing else but the correlators of $S_{\mu\nu}^a$'s. Those are 
calculable in the low-energy limit, $\Lambda^2\left|S_{\mu\nu}^a
\right|\ll\zeta$, by considering 
the real branch of the potential~(\ref{newmonpot}), {\it i.e.}, 
extracting from the sum over branches of this potential the term 
with $n=0$. This branch has 
a simple parabolic form, and in the vicinity of its minimum 
(corresponding to the low-energy limit) 
the generating functional
for correlators of $S_{\mu\nu}^a$'s reads

$${\cal Z}\left[J_{\mu\nu}^a\right]=
\int D{\cal S}_{\mu\nu}^a
\exp\left\{
-\left[g_m\eta^3\sqrt{\frac32}\int d^4xd^4y
{\cal S}_{\mu\nu}^a(x)\frac{K_1(m_B|x-y|)}{|x-y|}
{\cal S}_{\mu\nu}^a(y)+
\right.\right.
$$

$$
\left.\left.
+\frac{\Lambda^4}{2\zeta}\int d^4x\left[\frac13\left({\cal S}_{\mu\nu}^1
\right)^2+\left({\cal S}_{\mu\nu}^2\right)^2\right]+\int d^4x
\left[{\cal S}_{\mu\nu}^1\frac{J_{\mu\nu}^{+}}{\sqrt{3}}
+{\cal S}_{\mu\nu}^2J_{\mu\nu}^{-}\right]
\right]\right\},$$
where $J_{\mu\nu}^a$ is a source of $S_{\mu\nu}^a$, and 
$J_{\mu\nu}^{\pm}\equiv J_{\mu\nu}^1\pm J_{\mu\nu}^2$. Such two Gaussian 
integrals can be calculated by virtue of the following equality 

$$\int 
D{\cal S}_{\mu\nu}
\exp\left\{
-\left[g_m\eta^3\sqrt{\frac32}\int d^4xd^4y
{\cal S}_{\mu\nu}(x)\frac{K_1(m_B|x-y|)}{|x-y|}{\cal S}_{\mu\nu}(y)
+\int d^4x\left(\frac{\Lambda^4}{2\zeta}
{\cal S}_{\mu\nu}^2+
J_{\mu\nu}{\cal S}_{\mu\nu}\right)\right]\right\}=$$

$$=\exp\left\{-\frac{M_2\zeta}{8\pi^2\Lambda^4}\int d^4xd^4y
J_{\mu\nu}(x)J_{\mu\nu}(y)\left(\partial_x^2-m_B^2\right)
\frac{K_1(M_2|x-y|)}{|x-y|}\right\},$$
and the result reads
  
$${\cal Z}\left[J_{\mu\nu}^a\right]=$$

$$=\exp\left\{-\int d^4x\int d^4y
\left[J_{\mu\nu}^{+}(x)J_{\mu\nu}^{+}(y){\cal G}_1(x-y)+
J_{\mu\nu}^{-}(x)J_{\mu\nu}^{-}(y){\cal G}_2(x-y)\right]\right\},$$
where ${\cal G}_a(x)\equiv\frac{\zeta}{2\Lambda^4}
\left(\partial^2-m_B^2\right)D_{M_a}^{(4)}(x)$.
Owing to the conservation of $S_{\mu\nu}^a$'s and the 
Hodge decomposition theorem, we again have $S_{\mu\nu}^a=
\varepsilon_{\mu\nu\lambda\rho}\partial_\lambda\varphi_\rho^a$ and
$J_{\mu\nu}^a=\varepsilon_{\mu\nu\lambda\rho}\partial_\lambda I_\rho^a$, 
which yields

$${\cal Z}\left[J_{\mu\nu}^a\right]=\exp\Biggl\{-2\int d^4x\int d^4y
\Biggl[I_\mu^a(x)I_\nu^a(y)T_{\mu\nu}(x)\left({\cal G}_1(x-y)+
{\cal G}_2(x-y)\right)+$$

$$+2I_\mu^1(x)I_\nu^2(y)T_{\mu\nu}(x)
\left({\cal G}_1(x-y)-{\cal G}_2(x-y)\right)\Biggr]\Biggr\}.$$
On the other hand, the coupling $\int d^4xJ_{\mu\nu}^aS_{\mu\nu}^a$
takes the form $2\int d^4xI_\mu^aT_{\mu\nu}\varphi_\nu^a$, and
varying ${\cal Z}\left[J_{\mu\nu}^a\right]$ twice {\it w.r.t.} 
$I_\mu^a$'s
we arrive at the following system of equations: 

$$
T_{\mu\nu}(x)T_{\lambda\rho}(y)\left<\varphi_\nu^1(x)
\varphi_\rho^1(y)\right>=
T_{\mu\nu}(x)T_{\lambda\rho}(y)\left<\varphi_\nu^2(x)
\varphi_\rho^2(y)\right>=
-T_{\mu\lambda}(x)\left({\cal G}_1(x-y)+{\cal G}_2(x-y)\right),$$

$$
T_{\mu\nu}(x)T_{\lambda\rho}(y)\left<\varphi_\nu^1(x)
\varphi_\rho^2(y)\right>=
-T_{\mu\lambda}(x)\left({\cal G}_1(x-y)-{\cal G}_2(x-y)\right).$$
Adopting for the correlators of $\varphi_\mu^a$'s the following 
{\it ans\"atze},

$$
\left<\varphi_\nu^1(x)
\varphi_\rho^1(0)\right>=
\left<\varphi_\nu^2(x)
\varphi_\rho^2(0)\right>=\delta_{\nu\rho}f_{+}(x),~~
\left<\varphi_\nu^1(x)
\varphi_\rho^2(y)\right>=\delta_{\nu\rho}f_{-}(x),$$
we get:

$$f_{\pm}(x)=\frac{\zeta}{2\Lambda^4}\left(\partial^2-m_B^2\right)
\left[\frac{1}{M_1^2}\left(D_{M_1}^{(4)}(x)-D_0^{(4)}(x)\right)\pm
\frac{1}{M_2^2}\left(D_{M_2}^{(4)}(x)-D_0^{(4)}(x)\right)\right].$$
This result makes the choice of notations ``$f_{\pm}(x)$'' quite 
natural. Finally, the correlators of vortex loops read

$$\left<S_{\mu\nu}^1(x)S_{\lambda\rho}^1(0)\right>=
\left<S_{\mu\nu}^2(x)S_{\lambda\rho}^2(0)\right>=
\left(\delta_{\nu\lambda}\delta_{\mu\rho}-\delta_{\nu\rho}
\delta_{\mu\lambda}\right)\left({\cal G}_1(x)+{\cal G}_2(x)\right)+$$

$$+\left(\delta_{\mu\lambda}\partial_\rho\partial_\nu+
\delta_{\nu\rho}\partial_\mu\partial_\lambda-
\delta_{\mu\rho}\partial_\lambda\partial_\nu-
\delta_{\nu\lambda}\partial_\mu\partial_\rho\right)f_{+}(x),$$

$$\left<S_{\mu\nu}^1(x)S_{\lambda\rho}^2(0)\right>=
\left(\delta_{\nu\lambda}\delta_{\mu\rho}-\delta_{\nu\rho}
\delta_{\mu\lambda}\right)\left({\cal G}_1(x)-{\cal G}_2(x)\right)+$$

$$+\left(\delta_{\mu\lambda}\partial_\rho\partial_\nu+
\delta_{\nu\rho}\partial_\mu\partial_\lambda-
\delta_{\mu\rho}\partial_\lambda\partial_\nu-
\delta_{\nu\lambda}\partial_\mu\partial_\rho\right)f_{-}(x).$$

This result can now immediately be applied to the calculation of
the modification of the correlator~(\ref{colorcorrel}) due 
to the screening in the ensemble of vortex loops. 
Indeed, applying to the average over world-sheets, standing 
on the R.H.S. of Eq.~(\ref{Zexact}) the cumulant expansion in the 
bilocal approximation, we have due to Eq.~(\ref{Zapr}): 

$$\Delta\left<\left<f_{\mu\nu}^i(x)f_{\lambda\rho}^i(y)\right>
\right>_{{\bf a}_\mu, {\bf j}_\mu^M}=-24\pi^2g_m^2\eta^4s_a^{(c)}
s_b^{(c)}\int d^4zd^4u D_{m_B}^{(4)}(x-z)D_{m_B}^{(4)}(y-u)
\left<S_{\mu\nu}^a(z)S_{\lambda\rho}^b(u)\right>.$$
Taking further into account the equalities

$$\left<S_{\mu\nu}^1(x)S_{\lambda\rho}^3(y)\right>=
\left<S_{\mu\nu}^2(x)S_{\lambda\rho}^3(y)\right>=
\left<S_{\mu\nu}^1(x)S_{\lambda\rho}^2(y)\right>,$$
and the facts that for every $c$, $s_a^{(c)}s_a^{(c)}=2$,
$s_1^{(c)}s_2^{(c)}+s_1^{(c)}s_3^{(c)}+s_2^{(c)}s_3^{(c)}=-1$,
we can write

$$s_a^{(c)}s_b^{(c)}\left<S_{\mu\nu}^a(z)S_{\lambda\rho}^b(u)\right>=
2\left(\left<S_{\mu\nu}^1(z)S_{\lambda\rho}^1(u)\right>-
\left<S_{\mu\nu}^1(z)S_{\lambda\rho}^2(u)\right>\right).$$
This leads to:

$$\Delta\hat D\left((x-y)^2\right)=48\pi^2g_m^2\eta^4\int d^4zd^4u
D_{m_B}^{(4)}(x-z)D_{m_B}^{(4)}(y-u){\cal G}_2(z-u),$$

$$\Delta\hat G\left((x-y)^2\right)=96\pi^2g_m^2\eta^4\int d^4zd^4u
D_{m_B}^{(4)}(x-z)D_{m_B}^{(4)}(y-u)\left(f_{+}(z-u)-f_{-}(z-u)\right),$$
where $\hat G$ is given by Eq.~(\ref{g}) with the replacement 
$D_1\to\hat D_1$. Carrying now the integrals analogously to how it was
done above in the $SU(2)$-case, we get

$$\Delta\hat D\left(x^2\right)=m_B^2\left(D_{M_2}^{(4)}(x)
-D_{m_B}^{(4)}(x)\right),$$

$$\Delta\hat G\left(x^2\right)=4\left[\left(\frac{m_2}{M_2}\right)^2
D_0^{(4)}(x)-D_{m_B}^{(4)}(x)+\left(\frac{m}{M_2}\right)^2
D_{M_2}^{(4)}(x)\right].$$
Together with the old expressions for the functions $\hat D$ and 
$\hat D_1$ (given by Eqs.~(\ref{dvadtri}) and 
(\ref{dvadchetyr}), respectively, with $m\to m_B$),  
which did not account for the screening effect in the 
ensemble of vortex loops~\footnote{
It is remarkable that these expressions again become exactly canceled
by the corresponding terms in $\Delta\hat D$ and $\Delta\hat D_1$.}, 
we finally obtain that $\hat D^{\rm full}$
and $\hat D_1^{\rm full}$ are given by Eqs.~(\ref{Dtot}) and 
(\ref{D1tot}), respectively, with the replacements 
$m\to m_B$, $m_D\to m_2$, and $M\to M_2$. 
Therefore the whole discussion, following after Eq.~(\ref{D1tot}),  
remains the same modulo these replacements. In particular, 
when $m_2$ vanishes,
{\it i.e.}, one disregards the effect of screening, the old
expressions for the functions $\hat D$ and $\hat D_1$ are recovered.

\section{Conclusion and Outlook}

The main problem addressed in the present review was an 
attempt of an analytical description of confinement in QCD and 
other gauge theories. As a guiding principle for our investigations 
served the so-called Wilson's picture of confinement, according to 
which this phenomenon can be described in terms of some effective 
theory of strings, joining coloured objects to each other and 
preventing them from moving apart to macroscopic distances.  
In this review, we have proceeded with a derivation 
of such string theories corresponding to various gauge ones, 
including QCD, {\it i.e.}, with the solution of the 
problem of string representation of gauge theories. 
We have started our analysis 
with the nonlocal string effective action, arising within the 
so-called Method of Field Correlators of QCD, where the interaction 
between the string world-sheet elements is mediated by the 
phenomenological background gluon propagator. By performing the 
derivative expansion of this action, we have derived the first few 
terms of the string Lagrangian. The first two 
nontrivial of them turned out 
to be the Nambu-Goto and rigidity terms with the coupling constants 
expressed completely via the gluonic condensate and correlation 
length of the QCD vacuum. The signs of these constants ensure 
the stability of strings in the so-obtained effective string theory. 
After that, 
we have investigated the problem of crumpling of the 
string world-sheet by a derivation of the 
topological string term in the instanton gas model of the gluodynamics 
vacuum. Next, by making use of perturbation theory in the 
nonperturbative QCD vacuum, we have calculated perturbative corrections 
to the obtained string effective action. Those led to a new form of 
the nonlocal string effective action with the 
propagator between the elements of the world-sheet being the one 
of a perturbative gluon in the confining background. 
By the derivative expansion of this action, we got a 
correction to the rigidity term coupling constant, whereas the string 
tension of the Nambu-Goto term occurs to get no corrections due to 
perturbative gluonic exchanges. Finally, we have derived the 
Hamiltonian of QCD string with spinless quarks at the ends, associated 
with the obtained string effective action including the rigidity term. 
In the particular case of vanishing orbital momentum of the system, 
this Hamiltonian reduces to that 
of the so-called relativistic quark model, albeit with some 
modifications due to the rigidity term, 
which might have some influence to   
the dynamics of the QCD string with quarks.   
All these topics have been elaborated on in Section 2, and form the 
essence of the string representation of QCD within the 
Method of Field Correlators.

In Section 3, we have addressed the problem of string representation 
of Abelian-projected theories. In this way, we have started with 
the string representation for the partition function of the simplest 
model of this kind, namely the Abelian-projected $SU(2)$-QCD, which 
is argued to be 
the dual Abelian Higgs Model with external electrically charged 
particles. The advantage of this approach to the string representation of 
QCD {\it w.r.t.} the one based on the Method of Field Correlators 
is a possibility 
to get an integration over string world-sheets, resulting from the 
integration over the singular part of the phase of the Higgs field. 
After the string representation for the partition function in the 
London limit, we have derived such a representation for the bilocal
cumulant of the field strength tensors. 
>From this, in the confining regime for 
a test external quark, we got the expressions for the two functions,
which parametrize the bilocal cumulant within the Method of Field
Correlators. 
The obtained results demonstrate that the large-distance 
asymptotic behaviour of the bilocal field strength cumulant 
matches the one of the corresponding gauge-invariant cumulant in QCD, 
predicted by the Method of Field Correlators and measured in the 
lattice experiments. In particular, the r\^ole of the correlation 
length of the vacuum, introduced in the Method of Field Correlators 
as a phenomenological input, turned out to be played in the 
Abelian-projected $SU(2)$-gluodynamics by the inverse mass of the 
dual gauge boson. 
These results support the method 
of Abelian projection on the one hand and give a new field-theoretical 
status to the Method of Field Correlators on the other hand. 
After that, we have discussed 
the other fundamental nonperturbative phenomenon of QCD, the chiral 
symmetry breaking, from the point of view of the Abelian-projected 
theories and the Method of Field Correlators. In particular, 
in the $SU(2)$-case under study, we have quoted the derivation 
of the relation between the chiral and gluonic condensates, following 
from these two approaches.

Next, by making use of the Abelian projection method, 
we have addressed the problem of string 
representation of the $SU(3)$-gluodynamics. 
Namely, we have casted the related dual model, 
containing three types of the dual Higgs fields, into the string form.
Consequently, the latter one 
turned out to contain three types of strings, among which, however,  
only two were actually independent. As a result, 
we have found that both the ensemble of strings as a whole and 
individual strings display confining properties in the sense that 
these two distinct 
types of strings (self)interact via the exchanges of the massive dual 
gauge bosons. We have further also derived 
string representation for the Abelian-projected
$SU(3)$-QCD with quarks, from which there was again obtained 
the related string representation for the bilocal cumulants 
of field strength tensors. In the confining regime for a test quark, 
it coincides for diagonal cumulants 
with the one of the $SU(2)$-case modulo the modification of 
mass of the dual vector boson. This means that 
only {\it w.r.t.} those cumulants, which are  
built out of gluonic fields referring to the same generator 
of the Cartan subalgebra 
the vacuum of the model under study exhibits 
a nonvanishing correlation length. The latter one again turned out
to be equal to the inverse mass of the dual gauge bosons, which they 
acquire by means of the Higgs mechanism.

In conclusion of this topic, we have derived 
another useful representation for the partition functions of the 
Abelian-projected theories in the form of the integral over monopole 
currents. Besides the part quadratic in these currents, which represents 
the Biot-Savart interaction and the kinetic energy of monopole 
Cooper pairs, the obtained Lagrangians contain also the 
terms describing the long-range interaction of magnetic currents 
with the string world-sheets. If one localizes the (quantum) monopole 
currents along the classical trajectories, these terms take the form 
of the Gauss linking numbers between the currents and world-sheets. 
This means that the latter objects 
may be viewed as solenoids, which scatter the former ones ({\it cf.} the 
Aharonov-Bohm effect).

In Section 4, we have studied another model, allowing for an analytical 
description of confinement, 
which is the compact QED. In this way, 
we have successively investigated the 3D- and 4D-cases 
and in both of them  
derived the string representations for the Wilson loop of 
an external electrically charged test particle.
The essence of these representations is a certain mechanism 
realizing the independence of the Wilson loop of the shape 
of an arbitrary surface bounded by its contour. In this way, it has 
been argued that this mechanism is based on the summation over the 
branches of a certain multivalued effective potential, which in 
the 3D-case is the potential of the monopole densities, whereas 
in the 4D-case it is the potential of the monopole currents. 
In the former case, we have established 
a correspondence of this 
approach to another recently found one, the so-called confining 
string theory. After that, in both cases, 
we have calculated the bilocal cumulants 
of the field strength tensors in the low-energy limit of the model 
under study. Those also turned out to be in line with the general 
concepts of the Method of Field Correlators and therefore match the 
corresponding results known from the lattice measurements in QCD and 
found analytically for the effective 
Abelian-projected theories in Section 3. 
Note only that contrary to the Abelian-projected theories, 
in compact QED monopoles are not condensed (in the sense that the dual 
Higgs field does not exist), 
but form a gas, and consequently the dual 
gauge field acquires a nonvanishing mass due to the Debye 
screening in such a gas, rather than due to the Higgs 
mechanism. Correspondingly, 
the correlation length of the vacuum in compact QED 
is equal to the inverse Debye mass of the dual gauge boson.

On the basis of the found correspondence between 
string representations of various gauge theories and the 
above discussed similarities in the 
large-distance asymptotic 
behaviours of the bilocal cumulants in QCD, 
Abelian-projected theories, and (low-energy limit of) compact QED,  
we have 
elaborated on a unified method of 
description of the string world-sheet excitations in the respective 
string theories. This method, which employed
the techniques of nonlinear 
sigma models, enabled us to derive 
the effective action, quadratic in the 
world-sheet fluctuations.

Finally, we have applied the methods of summation over the 
grand canonical ensemble of monopoles in compact QED to the 
summation over the grand canonical ensemble of topological 
defects (vortex dipoles built out 
of electric Abrikosov vortices in 3D and vortex loops built out 
of electric Nielsen-Olesen strings in 4D) in Abelian-projected theories. 
In this way, 
we have evaluated correlation functions of these defects in the 
low-energy limit. This enabled us to improve on the calculation 
of the bilocal field strength correlators in Abelian-projected 
theories by a derivation of the corresponding string contributions.
In this way, it has been demonstrated that the true correlation 
length of the vacuum in these theories is given not simply by the 
inverse mass of the dual vector bosons, acquired by them due to the 
Higgs mechanism, 
but rather by their full mass,
which accounts also for screening in the gas of topological defects.
Besides that, it turned out that the effect of screening 
leads also to the appearance of a power-like term in the large-distance
asymptotics of one of the two functions, which parametrize the 
bilocal cumulant.

To summarize, the correspondence between various field-theoretical 
models, considered in the present review, as well as the approaches 
to their string representation may be established by the 
Table 1, presented at the end of this Section.  
In conclusion, the proposed nonperturbative techniques provide us 
with some new information on the mechanisms of confinement in QCD and 
other gauge theories and shed some light on the structure of the vacua
of these theories. 
They also demonstrate the relevance of the Method of Field Correlators 
to the description of this phenomenon and yield several prescriptions 
for the construction of the adequate string theories from the 
corresponding gauge ones.

Further investigations of the problems addressed in the present 
review are planned to follow in at least two directions.
First of them is a derivation of the string representation for 
the last up to now known model allowing for an analytical description 
of confinement, which is the low-energy $SU(2)$ ${\cal N}=2$ 
supersymmetric Yang-Mills theory (the so-called Seiberg-Witten 
theory)~\cite{seib}. The second line of investigations is devoted 
to a better understanding of interrelation between confinement 
and chiral symmetry breaking. In this way, it is planned 
to develop further 
the approach proposed in Ref.~\cite{chiral} by virtue of  
bosonization of the equations obtained there. This  
turned out to be a rather hard topic due to the non-translation-invariant 
character of the corresponding interaction kernel. However, it is this 
kernel, which is responsible for the confining effects in the system, 
and therefore it should not probably be reduced to some more simple 
translation-invariant one.  
Investigations of 
this problem together with an application of other well elaborated 
methods known from the theory of NJL models might finally enable one 
to understand completely the interrelations between the phenomena 
of confinement and chiral symmetry breaking in QCD. 

\newpage

\noindent
{\bf Table 1. Correspondence between various field-theoretical models and 
approaches to their string representation}
\vspace{3mm}

\begin{tabular}{||p{38mm}||p{38mm}|p{38mm}|p{38mm}||}
\hline
Model & QCD within MFC & Abelian-projected theories & Compact QED\\
\hline
\hline
Mechanism of the string representation & No integral over string 
world-sheets. String effective action is defined  
only {\it w.r.t.} $\Sigma_{\rm min.}$ &
$\int D\theta^{\rm sing.}\to \int Dx_\mu(\xi)$ & $\Sigma$-independence 
of $\left<W(C)\right>$ is realized by the summation over branches 
in $V[\rho]$ in 3D or $V[j_\mu]$ in 4D\\
\hline
Mechanism of the mass generation & Due to stochastic background 
fields & Higgs mechanism & Debye screening in the monopole gas\\
\hline
Type of propagator between the elements of the world-sheet(s) & 
Nonperturbative gluon propagator ($D$-function) 
or propagator of perturbative gluon in the nonperturbative 
background & Propagators of the Kalb-Ramond fields & Propagator 
of the Kalb-Ramond field\\
\hline
Parameter of the expansion of the resulting nonlocal interaction 
between the elements of the world-sheet(s) & 
Correlation length of the QCD vacuum, $T_g$ & Inverse mass of the dual 
gauge bosons, which they acquire both due the Higgs mechanism and due to the 
Debye screening in the ensemble of topological defects & 
Inverse Debye mass of the dual gauge boson\\
\hline
\end{tabular}

\newpage

\section{Acknowledgments}

The author is grateful to Profs. D. Ebert and Yu.A. Simonov 
for many helpful discussions and correspondence.
He is also indebted to Prof. A. Di Giacomo for fruitful
discussions and cordial hospitality extended to him during 
his stay at the Department of Physics of the University of Pisa,
where this work has been completed. 
Besides that, he would like
to acknowledge for useful discussions Profs. H.G. Dosch,
R.N. Faustov, H. Kleinert, Yu.M. Makeenko, M. M\"uller-Preussker, 
M.I. Polikarpov, H. Reinhardt, M.G. Schmidt,  
T. Suzuki, and A. Wipf. He is also greatful to 
Drs. E.T. Akhmedov, G. Bali, N. Brambilla, M. Campostrini, 
M.N. Chernodub, C. Diamantini, H. Dorn,  
V.O. Galkin, E.-M. Ilgenfritz, E. Meggiolaro, 
C. Preitschopf, C. Schubert, V.I. Shevchenko, C.A. Trugenberger, and A. Vairo 
for valuable discussions and correspondence.
Useful discussions with the participants of the  
seminars at the theoretical high energy physics groups of 
ITEP, Moscow, Humboldt and Freie Universities of Berlin, Universities 
of Heidelberg, Jena, Pisa, and T\"ubingen are greatfully appreciated. 
The author would also like to thank the staffs of the Quantum Field 
Theory Departments of the Institute of Physics of the Humboldt University 
of Berlin, where the present work has been started, and of the 
Department of Physics of the University of Pisa, where it has been 
completed,
for kind hospitality and technical support. 
Besides that, he is greatful to INFN and the Graduiertenkolleg 
{\it ``Strukturuntersuchungen, Pr\"azisionstests und Erweiterungen des 
Standard-Modells der Elementarteilchenphysik''} 
of the Humboldt University of Berlin 
for financial support. And last but not least, 
he would like to 
thank his wife and parents for all their love, support, and sacrifices.

\newpage

\section{Appendices}

\subsection*{A. Details of a Derivation of the Hamiltonian of 
the Straight-Line QCD String with Quarks} 

Assuming that a meson as a whole moves with a constant speed
(which is true for a free meson), {\it i.e.}, $\ddot R=0$, and bringing
together quark kinetic terms~(\ref{kin}) and the pure 
string action~(\ref{noq}),
we arrive at the following action of the QCD string with quarks 
 
$$S_{\rm tot.}=
\int\limits_0^T d\tau\Biggl\{\frac{m_1^2}{2\mu_1}+\frac{m_2^2}{2\mu_2}+
\frac{\mu_1}{2}+\frac{\mu_2}{2}+
\frac{1}{2}\left(\mu_1+\mu_2+\int\limits_0^1 d\beta \nu\right)\dot R^2+$$

$$
+\left(\mu_1(1-\zeta_1)-\mu_2\zeta_1+\int\limits_0^1 
d\beta (\beta-\zeta_1)
\nu\right)\left(\dot R\dot r\right)
-\int\limits_0^1 d\beta\nu\eta \left(\dot R r\right)+
\int\limits_0^1 d\beta (\zeta_1-\beta)\eta\nu (\dot r r)+$$

$$+\frac{1}{2}
\left(\mu_1(1-\zeta_1)^2+\mu_2\zeta_1^2+\int\limits_0^1 d\beta (\beta-
\zeta_1)^2\nu\right)\dot r^2+\frac{1}{2}\int\limits_0^1 d\beta\left(
\frac{\sigma^2}{\nu}+\eta^2\nu\right)r^2+$$

$$+\frac{1}{\alpha_0}
\Biggl[\zeta_2(\mu_1(\zeta_1-1)+\mu_2\zeta_1)\dot r^2-\zeta_2(\mu_1+\mu_2)
\left(\dot R \dot r\right)
+\int\limits_0^1 d\beta\nu\Biggl(\zeta_2(\zeta_1-\beta)
\dot r^2-\zeta_2\left(\dot R \dot r\right)
+\zeta_2\eta(\dot r r)+$$

$$+\frac{1}{2}(\beta-
\zeta_1)^2\left[\ddot{\bf r}, {\bf r}\right]^2
+\frac{1}{2}\dot R^2\dot r^2-\frac{1}{2}
\left(\dot R\dot r\right)^2+
(\beta-\zeta_1)\left(\left(\ddot r \dot R\right)
(\dot r r)-(\ddot r\dot r)
\left(\dot R r\right)\right)
+\frac{1}{2}\biggl(\biggl(\frac{\sigma}{\nu}\biggr)^2+\eta^2
\biggr)\left[\dot{\bf r}, {\bf r}\right]^2+$$

$$
+\eta\biggl((\beta-\zeta_1)\left((\ddot r r)
(\dot r r)-(\ddot r \dot r) r^2\right)
+\left(\dot R r\right)\dot r^2-\left(\dot R \dot r\right)(\dot r
r)\biggr)\Biggr)\Biggr]\Biggr\},\eqno(A.1)
$$
where we have performed a rescaling $z_\mu\to\bar \alpha \sqrt{\frac{h}
{\sigma T^2}}z_\mu$, $\bar z_\mu\to\bar\alpha\sqrt{\frac{h}{\sigma T^2}}
\bar z_\mu$, $\nu\to\frac{\sigma T^2}{\bar\alpha^2 h}\nu$.
Integrating then over $\eta$, one gets in the zeroth 
order in $\frac{1}{\alpha_0}$

$$\eta_{\rm extr.}=
\frac{(\dot r r)}{r^2}\biggl(\beta-\frac{\mu_1}{\mu_1+\mu_2}
\biggr),$$ 
which together with the condition $\dot R\dot r=0$ yields

$$\zeta_1^{\rm extr.}=\frac{\mu_1+\int
\limits_0^1 d\beta\beta\nu}{\mu_1+\mu_2+\int\limits_0^1 d\beta\nu},~ 
\zeta_2^{\rm extr.}=
\frac{(\dot r r)^2}{r^2}\frac{\frac{\mu_1}{\mu_1+\mu_2}
\int\limits_0^1 d\beta\nu-\int\limits_0^1 d\beta\beta\nu}{\mu_1+\mu_2+
\int\limits_0^1 d\beta\nu}.$$

Finally, in order to obtain the desired Hamiltonian, 
we shall perform the usual
canonical transformation from $\dot{\bf R}$ to the total momentum
${\bf P}$, which in the Minkowski space-time reads 

$$\int D{\bf R} 
\exp\biggl[i\int L\left(\dot{\bf R},\ldots\right)d\tau\biggr]=
\int D{\bf R} D{\bf P} \exp\biggl[i\int\left({\bf P}\dot{\bf R}- 
H\left(
{\bf P},\ldots\right)\right)
d\tau\biggr]$$
with $H\left({\bf P},\ldots\right)={\bf P}\dot{\bf R}- 
L\left(\dot{\bf R},\ldots\right)$, and
choose the meson rest frame as 

$${\bf P}=\frac{\partial L\left(\dot{\bf R},\ldots\right)}{\partial 
\dot{\bf R}}=0.$$ 
After performing the transformation
from $\dot{\bf r}$ to ${\bf p}$ we arrive at the 
Hamiltonian~(\ref{fullham}) of the main text 
with the coefficient functions 
$a_k$, which read as follows

$$a_1=\frac{\sigma^2}{2}\int\limits_0^1\frac{d\beta}{\nu},~ 
a_2=3\frac{\dot{\tilde\mu}}{\tilde\mu}B-\dot B,$$

$$a_3=\frac{1}{2(\mu_1+\mu_2)(\mu_1+\mu_2+\nu_0)}
\Biggl[\frac{\nu_0(\mu_1\nu_0-\nu_1(\mu_1+\mu_2))^2}{(\mu_1+\mu_2)
(\mu_1+\mu_2+\nu_0)}-\nu_1(\mu_1+\mu_2)(\nu_1-2\mu_2)-\mu_1\nu_0
(\mu_1+2\mu_2)\Biggr],$$

$$a_4=\frac{\dot{\tilde\mu}}{\tilde\mu}B-\dot B,$$

$$a_5=\nu_2+\frac{\nu_1^2+2\mu_1\nu_0-2\mu_2\nu_1}{\mu_1+
\mu_2+\nu_0}+\frac{1}{\mu_1+\mu_2}\Biggl[\frac{1}{\mu_1+\mu_2}
\Biggl(\frac{(\mu_1\nu_0-\nu_1(\mu_1+\mu_2))^2(3\nu_0+2(\mu_1+\mu_2))}
{(\mu_1+\mu_2+\nu_0)^2}+$$

$$+\mu_1(\mu_1\nu_0-2\nu_1(\mu_1+\mu_2))\Biggr)-
\frac{\mu_1^2\nu_0}{\mu_1+\mu_2+\nu_0}\Biggr],$$
and 

$$B\equiv\frac{\nu_1(\mu_1+\mu_2)(\nu_1-2\mu_2)+\mu_1\nu_0(\mu_1+
2\mu_2)}{(\mu_1+\mu_2)(\mu_1+\mu_2+\nu_0)}.$$
Notice that during the derivation of $H^{(1)}$, we have chosen the origin 
at the centre of masses of the initial state,
so that $\dot{\bf R}\dot{\bf r}\ll 1$, and the term 
$-\frac{1}{2\alpha_0}\int\limits_0^T d\tau\nu_0\left(\dot{\bf R}\dot
{\bf r}\right)^2$
on the R.H.S. of Eq.~(A.1) has been disregarded.

\subsection*{B. Path-Integral Duality Transformation}

In this Appendix, we shall outline some details of a derivation of 
Eq.~(\ref{odinnad}) from the main text. 
Firstly, one can linearize the term $\frac{\eta^2}{2}
\left(\partial_\mu\theta-2g_mB_\mu
\right)^2$ in the exponent on the R.H.S. of Eq.~(\ref{vosem}) 
and carry out the integral over 
$\theta^{{\rm reg.}}$ as follows 

$$\int D\theta^{{\rm reg.}}\exp\left\{-\frac{\eta^2}{2}
\int d^4x \left(\partial_\mu\theta-2g_mB_\mu
\right)^2\right\}=$$

$$=\int DC_\mu D\theta^{{\rm reg.}}
\exp\left\{\int d^4x\left[-\frac{1}{2\eta^2}C_\mu^2+iC_\mu
\left(\partial_\mu\theta-2g_mB_\mu
\right)\right]\right\}=$$

$$
=\int DC_\mu\delta\left(\partial_\mu C_\mu\right)
\exp\left\{\int d^4x\left[-\frac{1}{2\eta^2}C_\mu^2+iC_\mu
\left(\partial_\mu\theta^{{\rm sing.}}-2g_mB_\mu
\right)\right]\right\}.\eqno(B.1) 
$$
The constraint $\partial_\mu C_\mu=0$ can be uniquely resolved by 
representing $C_\mu$ in the form $C_\mu=\partial_\nu\tilde
h_{\mu\nu}$, where 
$h_{\mu\nu}$ stands for an antisymmetric tensor field. 
Notice, that the number 
of degrees of freedom during such a substitution  
is conserved, since both of the fields 
$C_\mu$ and $h_{\mu\nu}$ have three independent components.

Then, taking 
into account the relation~(\ref{devyat}) 
between $\theta^{{\rm sing.}}$ and 
$\Sigma_{\mu\nu}$, we get 
from Eq.~(B.1) 

$$\int D\theta^{{\rm sing.}}
D\theta^{{\rm reg.}}\exp\left\{-\frac{\eta^2}{2}
\int d^4x \left(\partial_\mu\theta-2g_mB_\mu
\right)^2\right\}=$$

$$
=\int Dx_\mu (\xi) Dh_{\mu\nu}
\exp\left\{\int d^4x 
\left[-\frac{1}{12\eta^2}H_{\mu\nu\lambda}^2+i\pi h_{\mu\nu}
\Sigma_{\mu\nu}-ig_m\varepsilon_{\mu\nu\lambda\rho}B_\mu
\partial_\nu h_{\lambda\rho}\right]\right\}.\eqno(B.2) 
$$
In a derivation of Eq.~(B.2), 
we have replaced $D\theta^{{\rm sing.}}$ by 
$Dx_\mu(\xi)$ (since the surface $\Sigma$, parametrized by 
$x_\mu(\xi)$, is just the surface, at which the field 
$\theta$ is singular) and, for 
simplicity, discarded the Jacobian arising during 
such a change of the integration variable~\footnote{For the case when 
the surface $\Sigma$ has a spherical topology, this Jacobian 
has been evaluated in Ref.~\cite{zubkov}.}. 
In the literature, the above described sequence of 
transformations 
of integration variables is   
usually referred to as a 
``path-integral duality transformation''. In particular, it 
has been applied in Ref.~\cite{lee1} 
to the model with a global $U(1)$-symmetry.   

It is further convenient 
to rewrite  
$\exp\left[-\frac14\int d^4x\left( 
F_{\mu\nu}+F_{\mu\nu}^E\right)^2\right]$ as 

$$\int DG_{\mu\nu}\exp\left\{\int d^4x
\left[-G_{\mu\nu}^2+i\left(\tilde F_{\mu\nu}+\tilde F_{\mu\nu}^E\right)
G_{\mu\nu}\right]\right\},$$
after which the $B_\mu$-integration yields 

$$\int DG_{\mu\nu}\exp\left[\int d^4x\left(-G_{\mu\nu}^2
+iG_{\mu\nu}\tilde F_{\mu\nu}^E\right)\right]
\delta\left(\varepsilon_{\mu\nu\lambda\rho}\partial_\nu
\left(G_{\lambda\rho}-g_mh_{\lambda\rho}\right)\right)=$$

$$
=\int DA_\mu\exp\left\{\int d^4x\left[-\left(
g_mh_{\mu\nu}
+\partial_\mu A_\nu-\partial_\nu A_\mu\right)^2+ig_mh_{\mu\nu}\tilde
F_{\mu\nu}^E-2iA_\mu j_\mu^E\right]\right\}.\eqno(B.3)
$$
Here $A_\mu$ is the electric field, dual to the 
dual gauge field $B_\mu$. The dependence on this electric field 
actually drops out upon the hypergauge transformation
(see {\it e.g.} Ref.~\cite{orl}) 
$h_{\mu\nu}\to h_{\mu\nu}-\frac{1}{g_m}\left(
\partial_\mu A_\nu-\partial_\nu A_\mu\right)$. In particular, the 
interaction of the $A_\mu$-field with the electric current
cancels out during such a transformation. The outcome of this 
transformation together with Eq.~(B.2) yield Eq.~(\ref{odinnad}) 
of the main text.

\subsection*{C. Integration over the Kalb-Ramond Field}

Let us carry out the following integration over the Kalb-Ramond field  

$$
{\cal Z}=\int Dh_{\mu\nu}\exp\Biggl[
-\int d^4x\Biggl(\frac{1}
{12\eta^2}
H_{\mu\nu\lambda}^2+g_m^2h_{\mu\nu}^2+
i\pi h_{\mu\nu}\hat\Sigma_{\mu\nu}
\Biggr)\Biggr].\eqno(C.1) 
$$
To this end, it is necessary to substitute the saddle-point value of 
such a functional integral  
back into the integrand. The saddle-point equation in the momentum 
representation reads 

$$\frac{1}{2\eta^2}\left(p^2h_{\nu\lambda}^{\rm extr.}(p)+p_\lambda 
p_\mu h_{\mu\nu}^{\rm extr.}(p)+p_\mu p_\nu h_{\lambda\mu}^{\rm extr.}(p)
\right)+2g_m^2h_{\nu\lambda}^{\rm extr.}(p)=-i\pi
\hat\Sigma_{\nu\lambda}(p).$$
This equation can be most easily solved by rewriting it in the 
following way

$$
\left(p^2 \hat P_{\lambda\nu, \alpha\beta}+m^2 \hat 1_{\lambda\nu, 
\alpha\beta}\right)h_{\alpha\beta}^{\rm extr.}(p)=-2\pi i\eta^2
\hat\Sigma_{\lambda
\nu}(p),\eqno(C.2) 
$$
where we have introduced the projection operators~\cite{petlqed} 

$$\hat P_{\mu\nu, \lambda\rho}\equiv\frac12\left({\cal P}_{\mu\lambda}
{\cal P}_{\nu\rho}-{\cal P}_{\mu\rho} {\cal P}_{\nu\lambda}\right)~
\mbox{and}~ 
\hat 1_{\mu\nu, \lambda\rho}\equiv\frac12 \left(\delta_{\mu\lambda}
\delta_{\nu\rho}-\delta_{\mu\rho}\delta_{\nu\lambda}\right)$$
with ${\cal P}_{\mu\nu}\equiv\delta_{\mu\nu}-\frac{p_\mu p_\nu}{p^2}$. 
These projection operators obey the following relations 

$$
\hat 1_{\mu\nu, \lambda\rho}=-\hat 1_{\nu\mu, \lambda\rho}=
-\hat 1_{\mu\nu, \rho\lambda}=\hat 1_{\lambda\rho, \mu\nu},~ 
\hat 1_{\mu\nu, \lambda\rho} \hat 1_{\lambda\rho, \alpha\beta}=
\hat 1_{\mu\nu, \alpha\beta}\eqno(C.3) 
$$
(the same relations hold for $\hat P_{\mu\nu, \lambda\rho}$), and 

$$
\hat P_{\mu\nu, \lambda\rho}\left(\hat 1-\hat P\right)_{\lambda
\rho, \alpha\beta}=0.\eqno(C.4) 
$$
By virtue of the properties~(C.3) and~(C.4), 
the solution to the saddle-point equation~(C.2) reads

$$h_{\lambda\nu}^{\rm extr.}(p)=-\frac{2\pi i\eta^2}{p^2+m^2}\left[
\hat 1+\frac{p^2}{m^2}\left(\hat 1-\hat P\right)\right]_{\lambda
\nu, \alpha\beta}\hat\Sigma_{\alpha\beta}(p),$$
which, once being substituted back into the 
partition function~(C.1), 
yields 
for it the following expression

$$
{\cal Z}=\exp\Biggl\{-\pi^2\eta^2\int\frac{d^4p}{(2\pi)^4}
\frac{1}{p^2+m^2}\left[\hat 1+\frac{p^2}{m^2}\left(\hat 1-\hat P
\right)\right]_{\mu\nu, \alpha\beta}\hat\Sigma_{\mu\nu}(-p)
\hat\Sigma_{\alpha\beta}(p)
\Biggr\}.\eqno(C.5) 
$$

Let us now prove that the term 
proportional to the projection operator 
$\left(\hat 1-\hat P\right)$ on the 
R.H.S. of Eq.~(C.5) yields in the 
coordinate representation the 
boundary term. One has 

$$
p^2(\hat 1-\hat P)_{\mu\nu, \alpha\beta}=\frac12
(\delta_{\nu\beta}p_\mu p_\alpha+\delta_{\mu\alpha}
p_\nu p_\beta-\delta_{\nu\alpha}p_\mu p_\beta-\delta_{\mu
\beta}p_\nu p_\alpha).\eqno(C.6)
$$
Then, by making use of Eq.~(C.6), the term 

$$-\pi^2\eta^2\int\frac{d^4p}{(2\pi)^4}\frac{1}{p^2+m^2}
\frac{p^2}{m^2}\left(\hat 1-\hat P\right)_{\mu\nu, \alpha\beta}
\int d^4x\int d^4y
{\rm e}^{ip(y-x)}\hat\Sigma_{\mu\nu}(x)\hat\Sigma_{\alpha\beta}(y)$$
under study, after carrying out the integration over $p$, reads

$$-2\left(\frac{\pi\eta}{m}\right)^2\int d^4x\int d^4y(\partial_\mu
\hat\Sigma_{\mu\nu}(x))(\partial_\lambda\hat\Sigma_{\lambda\nu}(y))
D_m^{(4)}(x-y).$$
This is just the 
argument of the first exponential factor  
standing on the R.H.S. of Eq.~(\ref{pyatnad}), {\it i.e.}, 
the desired boundary term. 

\subsection*{D. Summation over the Grand Canonical Ensemble of 
Vortex Loops in the Dual Abelian Higgs Model}

In the present Appendix, we shall outline some steps of a derivation 
of Eq.~(\ref{14new}) of the main text. 
Let us first consider the infinitesimal world-sheet 
element of the $a$-th vortex loop, which has the form 

$$d\sigma_{\mu\nu}(x^a(\xi))=\varepsilon^{\alpha\beta}
(\partial_\alpha x_\mu^a(\xi))
(\partial_\beta x_\nu^a(\xi))d^2\xi.$$
Here, in order to distinguish from the index enumerating 
the vortex loop, we have denoted the indices referring to the 
world-sheet coordinate $\xi$ by $\alpha,\beta=1,2$.
Analogously to the 3D case, it is reasonable to introduce the 
centre-of-mass coordinate (position) 
of the world-sheet $y_\mu^a\equiv\int d^2\xi
x_\mu^a(\xi)$. 
The full world-sheet coordinate can be respectively decomposed as 
$x_\mu^a(\xi)=y_\mu^a+z_\mu^a(\xi)$ with the vector $z_\mu^a(\xi)$ 
describing the shape of the $a$-th vortex loop world-sheet. 
Then, substituting 
into Eq.~(\ref{strn}) instead of $\Sigma_{\mu\nu}$ the vorticity 
tensor current of the vortex loop gas~(\ref{13new}), we can perform the 
summation over the grand canonical ensemble of vortex loops as follows,

$$
1+\sum\limits_{N=1}^{\infty}\frac{\zeta^N}{N!}\left(\prod\limits_{i=1}^{N}
\int d^4y^i\int Dz_\rho^i(\xi)\mu\left[z^i\right]\right)
\sum\limits_{n_a=\pm 1}^{}\exp\left\{i\pi \sum\limits_{a=1}^{N}
n_a\int d\sigma_{\mu\nu}(z^a(\xi))h_{\mu\nu}(x^a(\xi))\right\}=$$

$$
=1+\sum\limits_{N=1}^{\infty}\frac{(2\zeta)^N}{N!}\left\{
\int d^4y \int Dz_\rho(\xi)\mu[z]\cos\left(\pi
\int d\sigma_{\mu\nu}(z(\xi))h_{\mu\nu}(x(\xi))\right)\right\}^N.$$
Here, $\mu$ is a certain rotation- and translation invariant measure 
of integration over shapes of the world-sheets of the vortex loops.
Employing now the dilute gas approximation, we can expand $h_{\mu\nu}$ 
up to the first order in $a/L$,  
where $a$ stands for the typical value of $\left|z^a(\xi)\right|$'s 
(sizes of vortex loops), which are much smaller than the typical 
value $L$ of $\left|y^a\right|$'s 
(distances between vortex loops)~\footnote{Similarly to the 3D-case, 
the approximation $a\ll L$ means that vortex loops are 
short living objects.}. This yields 

$$
\int Dz_\rho(\xi)\mu[z]
\cos\left(\pi
\int d\sigma_{\mu\nu}(z(\xi))h_{\mu\nu}(x(\xi))\right)\simeq 
\int Dz_\rho\mu[z]\cos\left(
\pi{\cal P}_{\mu\nu, \lambda}[z]\partial_\lambda
h_{\mu\nu}(y)\right)=$$

$$
=\sum\limits_{n=0}^{\infty}\frac{(-1)^n}{(2n)!}\pi^{2n}
\partial_{\lambda_1}h_{\mu_1\nu_1}(y)\cdots \partial_{\lambda_{2n}}
h_{\mu_{2n}\nu_{2n}}(y)\int Dz_\rho\mu[z]
{\cal P}_{\mu_1\nu_1, \lambda_1}[z]\cdots 
{\cal P}_{\mu_{2n}\nu_{2n}, \lambda_{2n}}[z],\eqno (D.1)$$
where 
${\cal P}_{\mu\nu, \lambda}[z]\equiv\int d\sigma_{\mu\nu}(z(\xi))
z_\lambda(\xi)$. Due to the rotation- and translation invariance 
of the measure $\mu[z]$, the last average has the form 

$$
\int Dz_\rho\mu[z]
{\cal P}_{\mu_1\nu_1, \lambda_1}[z]\cdots 
{\cal P}_{\mu_{2n}\nu_{2n}, \lambda_{2n}}[z]=$$

$$
=\frac{\left(a^3\right)^{2n}}{(2n-1)!!}\left[\hat 1_{\mu_1\nu_1, 
\mu_2\nu_2}\delta_{\lambda_1\lambda_2}
\cdots \hat 1_{\mu_{2n-1}\nu_{2n-1}, \mu_{2n}\nu_{2n}}
\delta_{\lambda_{2n-1}\lambda_{2n}}+{\,}
{\rm permutations}\right].$$
Substituting this expression into Eq.~(D.1) and 
estimating the derivative {\it w.r.t.} $y$ as $1/L$,
we finally obtain

$$
\int Dz_\rho(\xi)\mu[z]
\cos\left(\pi
\int d\sigma_{\mu\nu}(z(\xi))h_{\mu\nu}(x(\xi))\right)\simeq
\sum\limits_{n=0}^{\infty}\frac{(-1)^n}{(2n)!}\left(
\frac{\pi a^3}{L}\right)^{2n}\left|h_{\mu\nu}(y)\right|^{2n}=
\cos\left(\frac{\left|h_{\mu\nu}(y)\right|}{\Lambda^2}\right),$$
where we have introduced a new UV momentum cutoff 
$\Lambda\equiv\sqrt{\frac{L}{\pi a^3}}$, which is much larger 
than $1/a$. This yields the desired Eq.~(\ref{14new}).

\subsection*{E. Calculation of the Integral~(\ref{mMint})}

In this Appendix, we shall present some details of calculation 
of the integral~(\ref{mMint}). Firstly, owing to the definition 
of the functions $D_m^{(4)}$ and $D_M^{(4)}$, we have:

$$\int d^4z D_m^{(4)}(z)D_M^{(4)}(z-x)=\int\frac{d^4p}{(2\pi)^4}
\int\frac{d^4q}{(2\pi)^4}\int d^4z\frac{{\rm e}^{ipz}}{p^2+m^2}
\frac{{\rm e}^{iq(z-x)}}{q^2+M^2}=$$

$$=\int\frac{d^4p}{(2\pi)^4}\frac{{\rm e}^{ipx}}{(p^2+m^2)(p^2+M^2)}.$$
Next, this expression can be rewitten as

$$\int\frac{d^4p}{(2\pi)^4}\int\limits_{0}^{+\infty}d\alpha
\int\limits_{0}^{+\infty}d\beta {\rm e}^{ipx-\alpha(p^2+m^2)-
\beta(p^2+M^2)}=\frac{1}{(4\pi)^2}
\int\limits_{0}^{+\infty}d\alpha
\int\limits_{0}^{+\infty}d\beta\frac{{\rm e}^{-\alpha m^2-\beta M^2
-\frac{x^2}{4(\alpha+\beta)}}}{(\alpha+\beta)^2}.\eqno (E.1)$$
It is further convenient to introduce new integration variables
$a\in [0,+\infty)$ and $t\in [0,1]$ according to the formulae
$\alpha=at$ and $\beta=a(1-t)$. Then, the integration over $t$ 
yields for Eq.~(E.1) the following expression:

$$\frac{1}{(4\pi m_D)^2}\int\limits_{0}^{+\infty}
\frac{da}{a^2}{\rm e}^{-\frac{x^2}{4a}}\left({\rm e}^{-am^2}-
{\rm e}^{-aM^2}\right).$$
Such an integral can be carried out by virtue  
of the formula

$$\int\limits_{0}^{+\infty}x^{\nu-1}
{\rm e}^{-\frac{\beta}{x}-\gamma x}dx=
2\left(\frac{\beta}{\gamma}\right)^{\frac{\nu}{2}}K_\nu\left(2
\sqrt{\beta\gamma}\right),~ \Re\beta>0,~ \Re\gamma>0,$$
and the result has the form of Eq.~(\ref{Mmres}) from the main text.

\end{document}